\documentclass[a4,11pt]{article}
\usepackage{graphicx} 
\usepackage{float}
\usepackage{amsmath,amsthm}
\usepackage{amsfonts}
\usepackage{latexsym}
\usepackage{amssymb}
\usepackage{setspace}
\usepackage{comment}

\makeatletter
 
  \@addtoreset{equation}{section}
 \makeatother

\begin{document}
\title{Quadratic-exponential growth BSDEs with Jumps\\ and their Malliavin's Differentiability~\footnote{
{\it Forthcoming in Stochastic Processes and their Applications}.
All the contents expressed in this research are solely those of the author and do not represent any views or 
opinions of any institutions. The author is not responsible or liable in any manner for any losses and/or damages caused by the use of any contents in this research.
}
%\\ \Large{\it{--Implications of asymmetric CSA and suboptimal strategies--}}
}

\author{Masaaki Fujii\footnote{Quantitative Finance Course, Graduate School of Economics, The University of Tokyo. }
~~\&~~Akihiko Takahashi\footnote{Quantitative Finance Course, Graduate School of Economics, The University of Tokyo.} 
}
%\begin{center}
\date{ 
First version:  8 January, 2016\\This version: 5 September, 2017
}
%\end{center}
\maketitle

%%%%%%    TEXT START    %%%%%%

%%%%%%      Macros      %%%%%%
%nakamacro.tex(H120522;0730)
%\documentstyle[11pt]{article}
%\setlength{\textwidth}{10.5in}
%\setlength{\oddsidemargin}{0in}
%\setlength{\topmargin}{-0.52in}
%\setlength{\textheight}{9.0in}
%\setlength{\footskip}{0.7in}

\newtheorem{definition}{Definition}[section]
\newtheorem{assumption}{Assumption}[section]
\newtheorem{condition}{$[$ C}
\newtheorem{lemma}{Lemma}[section]
\newtheorem{proposition}{Proposition}[section]
\newtheorem{theorem}{Theorem}[section]
\newtheorem{remark}{Remark}[section]
\newtheorem{example}{Example}[section]
\newtheorem{corollary}{Corollary}[section]
%--------------------------------------------------------------------------
%BOLD FACES
\def\n{{\bf n}}
\def\A{{\bf A}}
\def\B{{\bf B}}
\def\C{{\bf C}}
\def\D{{\bf D}}
\def\E{{\bf E}}
\def\F{{\bf F}}
\def\G{{\bf G}}
\def\H{{\bf H}}
\def\I{{\bf I}}
\def\J{{\bf J}}
\def\K{{\bf K}}
\def\L{{\bf L}}
\def\M{{\bf M}}
\def\N{{\bf N}}
\def\O{{\bf O}}
\def\P{{\bf P}}
\def\Q{{\bf Q}}
\def\R{{\bf R}}
\def\S{{\bf S}}
\def\T{{\bf T}}
\def\U{{\bf U}}
\def\V{{\bf V}}
\def\W{{\bf W}}
\def\X{{\bf X}}
\def\Y{{\bf Y}}
\def\Z{{\bf Z}}
\def\cala{{\cal A}}
\def\calb{{\cal B}}
\def\calc{{\cal C}}
\def\cald{{\cal D}}
\def\cale{{\cal E}}
\def\calf{{\cal F}}
\def\calg{{\cal G}}
\def\calh{{\cal H}}
\def\cali{{\cal I}}
\def\calj{{\cal J}}
\def\calk{{\cal K}}
\def\call{{\cal L}}
\def\calm{{\cal M}}
\def\caln{{\cal N}}
\def\calo{{\cal O}}
\def\calp{{\cal P}}
\def\calq{{\cal Q}}
\def\calr{{\cal R}}
\def\cals{{\cal S}}
\def\calt{{\cal T}}
\def\calu{{\cal U}}
\def\calv{{\cal V}}
\def\calw{{\cal W}}
\def\calx{{\cal X}}
\def\caly{{\cal Y}}
\def\calz{{\cal Z}}
%
%YOKUTUKAUMONO
\def\sskip{\hspace{0.5cm}}
\def\simleq{ \raisebox{-.7ex}{\em $\stackrel{{\textstyle <}}{\sim}$} }
\def\leqsim{ \raisebox{-.7ex}{\em $\stackrel{{\textstyle <}}{\sim}$} }
\def\ep{\epsilon}
\def\half{\frac{1}{2}}
\def\iku{\rightarrow}
\def\Iku{\Rightarrow}
\def\ikup{\rightarrow^{p}}
\def\inclusion{\hookrightarrow}
\def\cadlag{c\`adl\`ag\ }
\def\up{\uparrow}
\def\down{\downarrow}
\def\doti{\Leftrightarrow}
\def\douti{\Leftrightarrow}
\def\dochi{\Leftrightarrow}
\def\douchi{\Leftrightarrow}%
%KAIGYOU,ARRAY
\def\yy{\\ && \nonumber \\}
\def\y{\vspace*{3mm}\\}
\def\nn{\nonumber}
\def\be{\begin{equation}}
\def\ee{\end{equation}}
\def\bea{\begin{eqnarray}}
\def\eea{\end{eqnarray}}
\def\beas{\begin{eqnarray*}}
\def\eeas{\end{eqnarray*}}
%
%KONO RONBUN DE TUKAU MONO
\def\hd{\hat{D}}
\def\hv{\hat{V}}
\def\hsd{{\hat{d}}}
\def\hx{\hat{X}}
\def\hsx{\hat{x}}
\def\bsx{\bar{x}}
\def\bsd{{\bar{d}}}
\def\bx{\bar{X}}
\def\ba{\bar{A}}
\def\bb{\bar{B}}
\def\bc{\bar{C}}
\def\bv{\bar{V}}
\def\balpha{\bar{\alpha}}
\def\bbalpha{\bar{\bar{\alpha}}}
\def\combi{\l(\begin{array}{c}\alpha\\ \beta \end{array}\r)}
\def\f{^{(1)}}
\def\s{^{(2)}}
\def\ss{^{(2)*}}
\def\l{\left}
\def\r{\right}
\def\a{\alpha}
\def\b{\beta}
\def\L{\Lambda}
%上に定義されたコマンドは数式モ−ドで用いる。
%--------------------------------------------------

%\newtheorem{definition}{Definition}[section]
%\newtheorem{assumption}{$[$ A}
%\newtheorem{condition}{$[$ C}
%\newtheorem{lemma}{Lemma}[section]
%\newtheorem{proposition}{Proposition}[section]
%\newtheorem{theorem}{Theorem}[section]
%\newtheorem{remark}{Remark}
%\newtheorem{example}{Example}
%\newtheorem{corollary}{Corollary}[section]

\def\E{{\bf E}}
\def\P{{\bf P}}
\def\Q{{\bf Q}}
\def\R{{\bf R}}

\def\cadlag{{c\`adl\`ag~}}

\def\calf{{\cal F}}
\def\calp{{\cal P}}
\def\calq{{\cal Q}}
\def\wtW{\widetilde{W}}
\def\wtB{\widetilde{B}}
\def\wtPsi{\widetilde{\Psi}}
\def\wt{\widetilde}
\def\mbb{\mathbb}
\def\ol{\overline}
\def\ul{\underline}

\def\hTheta{\hat{\Theta}}
\def\hPhi{\hat{\Phi}}
\def\L2nu{{\mbb{L}^2(\nu)}}
\def\esup{{\rm ess}\sup}

\def\LDis{\frac{\bigl.}{\bigr.}}
\def\ep{\epsilon}
\def\del{\delta}
\def\Del{\Delta}
\def\part{\partial}
\def\wh{\widehat}
\def\bsigma{\bar{\sigma}}
\def\yy{\\ && \nonumber \\}
\def\y{\vspace*{3mm}\\}
\def\nn{\nonumber}
\def\be{\begin{equation}}
\def\ee{\end{equation}}
\def\bea{\begin{eqnarray}}
\def\eea{\end{eqnarray}}
\def\beas{\begin{eqnarray*}}
\def\eeas{\end{eqnarray*}}
\def\l{\left}
\def\r{\right}

\def\bull{$\bullet~$}

\newcommand{\Slash}[1]{{\ooalign{\hfil/\hfil\crcr$#1$}}}
\vspace{-5mm}

%%%%%%%%%%%%%%%%%%%%%%%%%%%%%%
\begin{abstract}
%%%%%%%%%%%%%%%%%%%%%%%%%%%%%% 
We investigate a class of quadratic-exponential growth BSDEs with jumps. The quadratic structure introduced by Barrieu \& El Karoui (2013) yields the universal bounds on the possible solutions.
With local Lipschitz continuity and the so-called $A_\Gamma$-condition for the comparison principle to hold,
we prove the existence of a unique solution under the general quadratic-exponential structure.
We have also shown that the strong convergence occurs under more general (not necessarily monotone) sequence of drivers,
which is then applied to give the sufficient conditions for the Malliavin's
differentiability.
%%%%%%%%%%%%%%%%%%%%%%%%%%%%%%%
\end{abstract}
\vspace{2mm}
%%%%%%%%%%%%%%%%%%%%%%%%%%%%%%%%%$

%%%%%%%%%%%%%%%%%%%%%%%%%%%%%%%%%
\section{Introduction}
%%%%%%%%%%%%%%%%%%%%%%%%%%%%%%%%%
The backward stochastic differential equations (BSDEs) have been subjects of strong 
interest of many researchers since they were introduced by Bismut (1973)~\cite{Bismut}
and generalized later by Pardoux \& Peng (1990)~\cite{Pardoux-Peng}.
This is particularly because they provide a truly probabilistic approach to stochastic control problems, 
which has been soon recognized as a very powerful tool for both theoretical and numerical issues 
in many important applications.

More recently, there has appeared an acute interest in quadratic-growth BSDEs because of their
various fields of applications such as, risk sensitive control problems,
dynamic risk measures and indifference pricing in an incomplete market.
The first breakthrough was made by Kobylanski (2000)~\cite{Kobylanski}
in a Brownian filtration with a bounded terminal condition.
The result was then extended by Briand \& Hu (2006, 2008)~\cite{Briand-Hu-06, Briand-Hu-08} to unbounded solutions.
Direct convergence based on a fixed-point theorem was  proposed by 
Tevzadze (2008)~\cite{Tevzadze}. Various extensions/applications can be found in, for example,
Hu, Imkeller \& Muller (2005)~\cite{Hu-Imkeller}, 
Mania \& Tevzadze (2006)~\cite{Mania}, Morlais (2009)~\cite{Morlais-09},
Hu \& Schweizer (2011)~\cite{Hu-Schweizer}, Delbaen, Hu \& Richou (2011)~\cite{Delbaen}.

In contrast to the diffusion setup, the number of researches on quadratic BSDEs with jumps 
has been rather small. Morlais (2010)~\cite{Morlais}
deals with a particular BSDE appearing in the exponential utility optimization with jumps,
and Antonelli \& Mancini (2016)~\cite{Antonelli} studies the setup with local Lipschitz continuity with different assumptions.
Both of them adopt Kobylanski's approach making use of a weakly converging subsequence.  
Cohen \& Elliott (2015)~\cite{Cohen-Elliott} and also Kazi-Tani, Possamai \& Zhou (2015)~\cite{Kazi}
have adopted the fixed-point approach of Tevzadze~\cite{Tevzadze}. 
See also Becherer (2006)~\cite{Becherer} as an earlier attempt for utility optimization with different restrictions
on the driver.

Recently, Barrieu \& El Karoui (2013)~\cite{Barrieu-ElKaroui} have proposed a
 new approach based on the stability of quadratic semimartingales
by introducing a so-called quadratic structure condition.
They have shown the existence of a solution, without the uniqueness, 
under the minimal assumption  allowing the unbounded terminal condition in a continuous setup.
Their result has been extended to the exponential utility optimization in a market with counterparty default risks
by generalizing quadratic structure condition to a quadratic-exponential ($Q_{\exp}$) structure condition
in Ngoupeyou (2010)~\cite{Ngoupeyou} (See also Jeanblanc, Matoussi \& Ngoupeyou (2013)~\cite{Jeanblanc-Matoussi}
and El Karoui, Matoussi \& Ngoupeyou (2016)~\cite{ElKaroui-Matoussi}.).

%The convergence becomes more direct 
%but the method requires the second-order differentiability of the driver.

The current work, with local Lipschitz continuity and the so-called $A_\Gamma$-condition for the comparison principle to  hold,
proves the existence of a unique bounded solution under the {\it general} $Q_{\exp}$-structure condition.
Let us emphasize that the assumptions are more general than those used \cite{Cohen-Elliott, Kazi, Morlais, Antonelli}
where the existence of a unique solution is proved.
\cite{Cohen-Elliott, Kazi} additionally require the second-order differentiability of the driver.
\cite{Morlais, Antonelli} are using a special form of the driver,
in particular, it is bounded by a linear (not quadratic) function of $|z|$ from {\it below}, 
and the sign of the quadratic terms is prefixed. These features are inherited from the utility 
optimization problem in \cite{Morlais} and is explicitly assumed in \cite{Antonelli}. 
These assumptions play an important role for constructing a monotone sequence of drivers by simply truncating the quadratic terms.
In the current work, new regularization of the driver
inspired by \cite{Lepeltier-Martin, Cvitanic,  ElKaroui-Matoussi} provides a rather streamlined proof for the convergence
under the general $Q_{\exp}$-structure. 
Moreover, the uniqueness alone is proved without using the comparison principle by the new stability result.

The specific monotone sequence of drivers used in the proof for the existence is not useful 
for other purposes.
By generalizing Theorem 2.8~\cite{Kobylanski}, we prove the strong convergence under more
general (not necessary monotone) sequence of drivers. 
The result is then used to achieve the convergence of globally-Lipschitz BSDEs
constructed by a sequence of simply truncated drivers. 
The sufficient conditions for the Malliavin's differentiability of the $Q_{\exp}$-growth BSDEs 
are then obtained by exploiting the properties of locally Lipschitz BSDEs with $\mbb{H}^2_{BMO}$-coefficients.
This extends the work of Ankirchner, Imkeller \& Dos Reis (2007)~\cite{Imkeller-Reis}
on the Malliavin's differentiability in the diffusion setup.
The obtained representation theorem will be useful for 
the optimal hedging problems in financial applications, investigations 
on the path regularity necessary for numerical as well as analytical issues,
and also for the development of an 
asymptotic expansion for the quadratic BSDEs~\footnote{ 
Recently, we have proposed an analytic approximation method of the Lipschitz BSDEs with jumps in 
Fujii \& Takahashi (2015)~\cite{FT-BSDEJ}, which is based on 
the small-variance asymptotic expansion (See, Takahashi (2015)~\cite{T-review} as a general review.).
Its extension to the $Q_{\exp}$-growth BSDEs is now ready to be investigated 
using the new results obtained here, which will be pursued in 
a different opportunity.}.
\\

The organization of the paper is as follows:
Section 2 gives preliminaries including some important results on the BMO martingales.
Section 3 explains the setup of $Q_{\exp}$-growth BSDEs with jumps and gives the uniqueness result.
Section 4 proves the existence of a solution by using the monotone sequence and the comparison principle.
Sections 5 deals with the Malliavin's differentiability of the $Q_{\exp}$-growth BSDEs,
 which is then applied to a forward-backward system 
to obtain a representation theorem on the martingale components in Section 6.
Appendix A is a simple generalization of the results by Ankirchner, Imkeller \& Dos Reis (2007)~\cite{Imkeller-Reis}
and Briand \& Confortola (2008)~\cite{Briand-Confortola}
on the locally Lipschitz BSDEs with BMO coefficients to the setup with jumps.
Appendix B gives some results regarding the comparison principle.
Appendix C gives a detailed proof for the 
Malliavin's differentiability of the Lipschitz BSDEs with jumps, which generalizes the 
result of Delong \& Imkeller (2010)~\cite{Delong-Imkeller} and Delong (2013)~\cite{Delong}
to local (instead of global) Lipschitz continuity for the Malliavin derivative of the driver,
which becomes necessary to investigate a forward-backward system driven by a Markovian forward process.
Finally, Appendix D gives the technical details of the proof for Theorem~\ref{theorem-Qexp-MD} omitted in the main text.

%%%%%%%%%%%%%%%%%%%%%%%%%%%%%%%%%%%
\section{Preliminaries}
\label{sec-Preliminaries}
%%%%%%%%%%%%%%%%%%%%%%%%%%%%%%%%%%%
\subsection{General Setting}
%%%%%%%%%%%%%%%%%%%%%%%%%%%%%%%%%%
Let us first state the general setting to be used  throughout the paper.
$T>0$ is some bounded time horizon.
The space $(\Omega_W,\calf_W,\mbb{P}_W)$
is the usual canonical space for a $d$-dimensional Brownian motion equipped with the Wiener measure $\mbb{P}_W$.
We also denote  $(\Omega_\mu,\calf_\mu,\mbb{P}_\mu)$ as a product of 
canonical spaces $\Omega_\mu:=\Omega_\mu^1\times \cdots\times \Omega_\mu^k$,
$\calf_\mu:=\calf_\mu^1\times \cdots \times \calf_\mu^k$ and $\mbb{P}_\mu^1\times \cdots \times \mbb{P}_\mu^k$ with some constant $k\geq 1$,
on which each $\mu^i$ is a Poisson measure with a compensator $\nu^i(dz)dt$. Here, $\nu^i(dz)$ is a $\sigma$-finite 
measure on $\mbb{R}_0=\mbb{R}\backslash\{0\}$ satisfying $\int_{\mbb{R}_0} |z|^2 \nu^i(dz)<\infty$.
Throughout the paper, we work on the filtered probability space 
$(\Omega, \calf, \mbb{F}=(\calf_t)_{t\in[0,T]}, \mbb{P})$, where the  space $(\Omega,\calf,\mbb{P})$
is the product of the canonical spaces $(\Omega_W\times  \Omega_\mu, \calf_W\times\calf_\mu,\mbb{P}_W\times \mbb{P}_\mu)$,
and that the filtration $\mbb{F}=(\calf_t)_{t\in[0,T]}$ is the 
canonical filtration completed for $\mbb{P}$ and satisfying the usual conditions.
In this construction, $(W,\mu^1,\cdots,\mu^k)$ are independent.
We use a vector notation $\mu(\omega, dt,dz):=(\mu^1(\omega, dt,dz^1), \cdots, \mu^k(\omega, dt,dz^k))$
and denote the compensated Poisson measure as $\wt{\mu}:=\mu-\nu$.
We represent the $\mbb{F}$-predictable $\sigma$-field on $\Omega\times [0,T]$ by $\calp$.

%%%%%%%%%%%%%%%%%%%%%%%%%%%%%%%
\begin{remark}
%%%%%%%%%%%%%%%%%%%%%%%%%%%%%%%%
We have chosen the above setting mainly because that it is known to guarantee the
weak property of predictable representation and also because there exists an established Malliavin's differential rule.  
The contents up to Section~\ref{sec-existence-QexpBSDE} can be easily extendable to $\calp \otimes \cale$-measurable
random compensator $\nu_t(dx)$ as long as $(W,\mu-\nu)$ is {\it assumed} to have 
{\it the weak property of predictable representation} (See Chapter XIII in \cite{He-Wang-Yan}.).
For the general topics regarding stochastic calculus with random measures, see also \cite{Jacod}.
\end{remark}
%%%%%%%%%%%%%%%%%%%%%%%%%%%%%%%%%%%%
\subsection{Notation}
%%%%%%%%%%%%%%%%%%%%%%%%%%%%%%%%%%%%
We denote a generic constant by $C$, which may change line by line, is sometimes associated 
with several subscripts (such as $C_{K,T}$) showing its dependence 
when necessary. $\calt^T_0$ denotes the set of $\mbb{F}$-stopping times $\tau\in[0,T]$.  

Let us introduce a sup-norm for a $\mbb{R}^r$-valued function $x:[0,T]\rightarrow \mbb{R}^r$ as
\be
||x||_{[a,b]}:=\sup\{|x_t|,t\in[a,b]\}~ \nn
\ee
and write $||x||_t:=||x||_{[0,t]}$.
We use the following spaces for stochastic processes for $p\geq 2$:\\
\bull $\mbb{S}^p_r[s,t]$ is the set of $\mbb{R}^r$-valued adapted \cadlag
processes $X$ such that
\be
||X||_{\mbb{S}^p_r[s,t]}:=\mbb{E}\left[||X||_{[s,t]}^p\right]^{1/p}<\infty~. \nn
\ee
%%%%%%%%%%%%%%%%%%%%%%
\bull  $\mbb{S}^{\infty}_r$ is the set of $\mbb{R}^r$-valued essentially bounded \cadlag processes $X$ such that
\be
||X||_{\mbb{S}^{\infty}_r}:=\bigl|\bigl|\sup_{t\in[0,T]}|X_t|\bigr|\bigr|_{\infty}<\infty. \nn
\ee
%%%%%%%%%%%%%%%%%%%%
%%%%%%%%
\bull $\mbb{H}^p[s,t]$ is the set of progressively measurable $\mbb{R}^d$-valued processes $Z$ 
such that
\be
||Z||_{\mbb{H}^p_r[s,t]}:=\mbb{E}\Bigl[ \Bigl(\int_s^t |Z_u|^2 du\Bigr)^{\frac{p}{2}}\Bigr]^{\frac{1}{p}}<\infty. \nn 
\ee
%%%%%%%%%%%%
%%%%%%%%%%%%
\bull $\mbb{J}^p[s,t]$ is the set of $k$-dimensional functions $\psi=\{\psi^i,1\leq i\leq k\}$,
$\psi^i: \Omega\times [0,T]\times \mbb{R}_0 \rightarrow \mbb{R}$
which are $\calp\times \calb(\mbb{R}_0)$-measurable and satisfy
\bea
||\psi||_{\mbb{J}^p[s,t]}:=\mbb{E}\Bigl[\Bigl(\sum_{i=1}^k\int_s^t \int_{\mbb{R}_0}|\psi_u^{i}(x)|^2\nu^i(dx)du
\Bigr)^{\frac{p}{2}}\Bigr]^{\frac{1}{p}}<\infty. \nn
\eea
%%%%%%%%%%
\bull $\mbb{J}^\infty$ is the space of functions which are $d\mbb{P}\otimes \nu(dz)$ essentially bounded i.e.,
\be
||\psi||_{\mbb{J}^\infty}:=\bigl|\bigl| \sup_{t\in[0,T]}||\psi_t||_{\mbb{L}^\infty(\nu)}\bigr|\bigr|_{\infty}~<\infty,\nn
\ee
where $\mbb{L}^\infty(\nu)$ is the space of $\mbb{R}^k$-valued measurable functions
$\nu(dz)$-a.e. bounded endowed with the usual essential sup-norm.\\
%%%%%%%%%%%%%
\bull $\calk^p[s,t]$ is the set of functions $(Y,Z,\psi)$ in the space $\mbb{S}^p[s,t]\times \mbb{H}^p[s,t]\times \mbb{J}^p[s,t]$
with the norm defined by 
\be
||(Y,Z,\psi)||_{\calk^p[s,t]}:=\bigl(||Y||_{\mbb{S}^p[s,t]}^p+||Z||_{\mbb{H}^p[s,t]}^p+||\psi||_{\mbb{J}^p[s,t]}^p\bigr)^{\frac{1}{p}}\nn .
\ee

For notational simplicity, we use $(E,\cale)=(\mbb{R}_0^k,\calb(\mbb{R}_0)^k)$ and denote 
the maps $\{\psi^i,1\leq i\leq k\}$ defined above
as $\psi:\Omega\times [0,T]\times E\rightarrow \mbb{R}^{k}$ and say $\psi$ is $\calp\otimes \cale$-measurable 
without referring to each component.
We also use the notation such that
\bea
\int_s^t\int_E \psi_u(x)\wt{\mu}(du,dx):=\sum_{i=1}^k \int_s^t \int_{\mbb{R}_0}\psi_u^i(x)\wt{\mu}^i(du,dx)~ \nn
\eea
for simplicity. The similar abbreviation is used also for the integrals with respect to $\mu$ and $\nu$.
When we use $E$ and $\cale$, one should always interpret it in this way so that the integral
with the $k$-dimensional Poisson measure does make sense. On the other hand, when we use the range $\mbb{R}_0$
with the integrators $(\wt{\mu},\mu, \nu)$, for example, 
\bea
\int_{\mbb{R}_0}\psi_u(x)\nu(dx):=\Bigl(\int_{\mbb{R}_0}\psi_u^i(x)\nu^i(dx)\Bigr)_{1\leq i\leq k} \nn
\eea
we interpret it as a $k$-dimensional vector.

We frequently omit the subscripts specifying the dimension $r$ and the time interval $[s,t]$
when they are unnecessary or obvious in the context.  
We use $\bigl(\Theta_s,s\in[0,T]\bigr)$ as a collective argument
$\Theta_s=\bigl(Y_s,Z_s,\psi_s\bigr)$
to lighten the notation.
We use the notation of partial derivatives such that for $x\in\mbb{R}^d$
\bea
&&\part_x=(\part_{x_1},\cdots, \part_{x_d})=\Bigl(\frac{\part}{\part x_1},\cdots,\frac{\part}{\part x_d}\Bigr) \nn
\eea
and for $\Theta$, $\part_\Theta=\bigl(\part_y,\part_z,\part_\psi\bigr)$.
We use the similar notations for every higher order derivative without a detailed indexing. We suppress the obvious summation of indexes 
throughout the paper for notational simplicity.

%%%%%%%%%%%%%%%%%%%%%%%%%%%%%%%%%%%%%%%%%%%%%%%%%%%%%%
\subsection{BMO-martingale and its properties}
%%%%%%%%%%%%%%%%%%%%%%%%%%%%%%%%%%%%%%%%%%%%%%%%%%%%%%
The properties of the BMO-martingales play a crucial role throughout this work. 
This section summarizes the necessary facts used in the following discussions.
\begin{definition}
Let $M$ be a square integrable martingale. When it satisfies
\be
||M||^2_{BMO}:=\sup_{\tau\in\calt^T_0}\Bigl|\Bigl| \mbb{E}\Bigl[(M_T-M_{\tau-}\bold{1}_{\tau>0})^2|\calf_\tau\Bigr]\Bigr|\Bigr|_{\infty}<\infty \nn
\ee
then $M$ is called a BMO-martingale and denoted by $M\in BMO$.
\end{definition}

\begin{lemma}
\label{lemma-BMO-1}
Suppose $M$ is a square integrable martingale with initial value $M_0=0$.
If $M$ is a BMO-martingale, then its jump component is essentially bounded $\Del M\in\mbb{S}^\infty$.
On the other hand, if $\Del M\in\mbb{S}^\infty$ and
$\sup_{\tau\in\calt^T_0}\Bigl|\Bigl|\mbb{E}\Bigl[\langle M\rangle_T-\langle M \rangle_\tau|\calf_\tau\Bigr]\Bigr|\Bigr|_{\infty}<\infty$,
then $M$ is a BMO-martingale.
\begin{proof}
From Lemma 10.7 in \cite{He-Wang-Yan}, we have
\bea
||M||^2_{BMO}&=&\sup_{\tau \in \calt^T_0}\Bigl|\Bigl|\mbb{E}\Bigl[[M]_T-[M]_\tau |\calf_\tau\Bigr]+M_0^2\bold{1}_{\tau=0}+(\Del M_\tau)^2\Bigr|\Bigr|_{\infty}\nn \\
&=&\sup_{\tau \in \calt^T_0}\Bigl|\Bigl|\mbb{E}\Bigl[\langle M\rangle_T-\langle M\rangle_\tau |\calf_\tau\Bigr]+(\Del M_\tau)^2\Bigr|\Bigr|_{\infty}~.\nn
\eea
Thus,
\bea
&&\sup_{\tau \in \calt^T_0}\Bigl|\Bigl|\mbb{E}\Bigl[\langle M\rangle_T-\langle M\rangle_\tau |\calf_\tau\Bigr]\Bigr|\Bigr|_{\infty}
\vee ||\Del M||^2_{\mbb{S}^\infty}\leq ||M||^2_{BMO}\nn \\
&&\hspace{30mm} \leq \sup_{\tau \in \calt^T_0}\Bigl|\Bigl|\mbb{E}\Bigl[\langle M\rangle_T-\langle M\rangle_\tau |\calf_\tau\Bigr]\Bigr|\Bigr|_{\infty}+||\Del M||^2_{\mbb{S}^\infty} \nn
\eea
and hence the claim is proved.
\end{proof}
\end{lemma}

Let us introduce the following spaces.
$\mbb{H}^2_{BMO}$ is the set of progressively measurable $\mbb{R}^d$-valued functions $Z$ satisfying~\footnote{
We sometimes include a scalar function satisfying the rightmost inequality also in $\mbb{H}^2_{BMO}$. By 
multiplying a $d$-dimensional unit vector, one can always connect to it the BMO norm if necessary.}
\bea
||Z||^2_{\mbb{H}^2_{BMO}}&:=&\Bigl|\Bigl|\int_0^\cdot Z_s dW_s\Bigr|\Bigr|_{BMO}^2
=\sup_{\tau\in\calt^T_0} \Bigl|\Bigl|\mbb{E}\Bigl[\int_\tau^T|Z_s|^2 ds |\calf_\tau\Bigr]\Bigr|\Bigr|_{\infty}<\infty. \nn 
\eea
$\mbb{J}^2_{BMO}$ and $\mbb{J}^2_B$ are the sets of $\calp\otimes \cale$-measurable functions $\psi:\Omega\times [0,T]\times \mbb{E}\rightarrow \mbb{R}^k$ satisfying
\bea
||\psi||^2_{\mbb{J}^2_{BMO}}:=\Bigl|\Bigl|\int_0^\cdot \int_E \psi_s(x)\wt{\mu}(ds,dx)\Bigr|\Bigr|^2_{BMO}=\sup_{\tau\in\calt^T_0} \Bigl|\Bigl|\mbb{E}\Bigl[\int_{\tau}^T \int_E |\psi_s(x)|^2\mu(ds,dx) |\calf_\tau\Bigr]+(\Del M_\tau)^2\Bigr|\Bigr|_{\infty}<\infty~, \nn
\eea
where $\Del M_\tau$ is a jump of $M=\int_0^\cdot\int_E \psi_s(x) \wt{\mu}(ds,dx)$ at time $\tau$.
\bea
||\psi||^2_{\mbb{J}^2_B}&:=&\sup_{\tau\in\calt^T_0} \Bigl|\Bigl|\mbb{E}\Bigl[\int_\tau^T \int_E |\psi_s(x)|^2\nu(dx)ds |\calf_\tau\Bigr]\Bigr|\Bigr|_{\infty}<\infty, \nn
\eea
respectively. Note that
$(||\psi||^2_{\mbb{J}^2_B}\vee ||\psi||^2_{\mbb{J}^\infty})\leq ||\psi||^2_{\mbb{J}^2_{BMO}}\leq ||\psi||^2_{\mbb{J}^2_B}+||\psi||^2_{\mbb{J}^\infty}~$
from the proof of Lemma~\ref{lemma-BMO-1}.

\begin{lemma} [energy inequality]
\label{lemma-energy}
Let $Z\in \mbb{H}^2_{BMO}$ and $\psi\in \mbb{J}^2_{BMO}$. Then, for any $n\in\mbb{N}$,
\bea
&&\mbb{E}\Bigl[\Bigl(\int_0^T|Z_s|^2 ds\Bigr)^n\Bigr]\leq  n!\bigl(||Z||^2_{\mbb{H}^2_{BMO}}\bigr)^n,\nn \\
&&\mbb{E}\Bigl[\Bigl(\int_0^T\int_E |\psi_s(x)|^2\mu(ds,dx)\Bigr)^n\Bigr]\leq n!\bigl(||\psi||^2_{\mbb{J}^2_{BMO}}\bigr)^n\nn, \\
&&\mbb{E}\Bigl[\Bigl(\int_0^T\int_E |\psi_s(x)|^2\nu(dx)ds\Bigr)^n\Bigr]\leq n!\bigl(||\psi||^2_{\mbb{J}^2_{B}}\bigr)^n
\leq n!\bigl(||\psi||^2_{\mbb{J}^2_{BMO}}\bigr)^n. \nn
\eea
\begin{proof}
See proof of Lemma 9.6.5 in \cite{Cvitanic}.
\end{proof}
\end{lemma}

Let $\cale(M)$ be a Dol\'ean-Dade exponential of $M$.
\begin{lemma} [reverse H\"older inequality]
\label{lemma-R-Holder}
Let $\del>0$ be a positive constant and $M$ be a BMO-martingale
satisfying $\Del M_t\geq -1+\del$ $\mbb{P}$-a.s. for all $t\in[0,T]$.
Then,  $\bigl(\cale_t(M),t\in[0,T]\bigr)$ is a uniformly integrable martingale,  and 
for every stopping time $\tau\in\calt^T_0$, there exists some $p>1$
such that $
\mbb{E}\left[\cale_T(M)^p|\calf_\tau\right]\leq C_{p,M} \cale_\tau (M)^p$
with some positive constant $C_{p,M}$ depending only on $p$ and $||M||_{BMO}$. 
\begin{proof}
See Kazamaki (1979)~\cite{Kazamaki}, and also Remark 3.1 of Kazamaki (1994)~\cite{Kazamaki-note}.
\end{proof}
\end{lemma}
Note here that the condition $\Del M_t\geq -1+\del$ is the very reason why one needs
a stronger assumption than the Lipschitz continuity 
for the comparison principle to hold for the BSDEs with jumps (See Proposition 2.6 in Barles et.al. (1997)~\cite{Barles}.).
The following properties of the {\it continuous} BMO martingales by Kazamaki~\cite{Kazamaki-note}
are very useful.
\begin{lemma}
\label{lemma-BMO-PQ}
Let $M$ be a square integrable continuous martingale and $\hat{M}:=\langle M \rangle -M$.
Then, $M\in BMO(\mbb{P})$ if and only if $\hat{M}\in BMO(\mbb{Q})$
with $d\mbb{Q}/d\mbb{P}=\cale_T(M)$. Furthermore, $||\hat{M}||_{BMO(\mbb{Q})}$ is 
determined by some function of $||M||_{BMO(\mbb{P})}$ and vice versa.
\begin{proof}
See Theorem 3.3 and Theorem 2.4 in \cite{Kazamaki-note}.
\end{proof}
\end{lemma}

\begin{remark}
For continuous martingales, Theorem 3.1~\cite{Kazamaki-note} also tells that
there exists some decreasing function $\Phi(p)$ with $\Phi(1+)=\infty$
and $\Phi(\infty)=0$ such that if $||M||_{BMO(\mbb{P})}$ satisfies
$ ||M||_{BMO(\mbb{P})}<\Phi(p)$
then $\cale(M)$ satisfies the reverse H\"older inequality with power $p$.
This implies together with Lemma~\ref{lemma-BMO-PQ}, one can take 
a common positive constant $\bar{r}$ satisfying $1<\bar{r}\leq r^*$ such that both of 
the $\cale(M)$ and $\cale(\hat{M})$ satisfy the reverse H\"older inequality
with power $\bar{r}$ under the respective probability measure $\mbb{P}$ and $\mbb{Q}$.
Furthermore,  the upper bound $r^*$ is determined only by $||M||_{BMO(\mbb{P})}$ (or equivalently by $||M||_{BMO(\mbb{Q})}$).
\end{remark}

%%%%%%%%%%%%%%%%%%%%%%%%%%%%%%%%%%%%%%%%%%%%%%%%%%%%%%%%%%
\section{$Q_{\exp}$-growth BSDEs with Jumps}
%%%%%%%%%%%%%%%%%%%%%%%%%%%%%%%%%%%%%%%%%%%%%%%%%%%%%%%%%%%
%%%%%%%%%%%%%%%%%%%%%%%%%
\subsection{Universal Bound}
%%%%%%%%%%%%%%%%%%%%%%%%%

We now introduce, for $t\in[0,T]$,  the quadratic-exponential ($Q_{\exp}$) growth BSDE;
\bea
Y_t=\xi+\int_t^T f(s,Y_s,Z_s,\psi_s)ds-\int_t^T Z_s dW_s-\int_t^T \int_E \psi_s(x)\wt{\mu}(ds,dx)~, 
\label{eq-Qexp-BSDE}
\eea
where $\xi:\Omega\rightarrow \mbb{R}$, $f:\Omega\times [0,T]\times \mbb{R}\times \mbb{R}^d\times \mbb{L}^2(E,\nu;\mbb{R}^k)
\rightarrow \mbb{R}$ and denote $Z$ and $\psi$ as row vectors for simplicity.

Let us introduce the quadratic-exponential structure condition proposed by 
Barrieu \& El Karoui (2013)~\cite{Barrieu-ElKaroui} and extended to a
jump diffusion case by Ngoupeyou (2010)~\cite{Ngoupeyou}. See also El Karoui et.al. (2016)~\cite{ElKaroui-Matoussi}.
\begin{assumption} 
\label{assumption-Qexp}
(i)The map $(\omega,t)\mapsto f(\omega,t,\cdot)$ is $\mbb{F}$-progressively measurable.
For every $(y,z,\psi)\in\mbb{R}\times\mbb{R}^d\times \mbb{L}^2(E,\nu;\mbb{R}^k)$, 
there exist two constants $\beta\geq 0$ and $\gamma>0$ and a positive $\mbb{F}$-progressively measurable process $(l_t,t\in[0,T])$
such that
\bea
&&-l_t-\beta|y|-\frac{\gamma}{2}|z|^2-\int_Ej_\gamma (-\psi(x))\nu(dx) \nn \\
&&\hspace{20mm}\leq ~f(t,y,z,\psi) ~\leq~ l_t+\beta|y|+\frac{\gamma}{2}|z|^2+\int_E j_\gamma(\psi(x))\nu(dx) \nn
\eea
$dt\otimes d\mbb{P}$-a.e. $(\omega,t)\in\Omega\times[0,T]$, where $\displaystyle j_\gamma(u):=\frac{1}{\gamma} \bigl(e^{\gamma u}-1-\gamma u\bigr)$. \\
(ii) $|\xi|, (l_t,t\in[0,T])$ are essentially bounded, i.e.,  $||\xi||_{\infty}, ||l||_{S^\infty}<\infty$. 
\end{assumption}

The Assumption~\ref{assumption-Qexp}  yields useful universal bounds as Lemmas~\ref{lemma-BMO-bound}
and \ref{lemma-universal-bound} for the possible solutions of (\ref{eq-Qexp-BSDE}).
\begin{lemma}
\label{lemma-BMO-bound} Under Assumption~\ref{assumption-Qexp}, 
if there exists a solution $(Y,Z,\psi)\in \mbb{S}^{\infty}\times \mbb{H}^2\times \mbb{J}^2$ to the BSDE (\ref{eq-Qexp-BSDE}), then
$Z\in\mbb{H}^2_{BMO}$ and $\psi\in\mbb{J}^2_{BMO}$ (and hence $\psi\in \mbb{J}^{\infty}$)
and $||Z||_{\mbb{H}^2_{BMO}}$, $||\psi||_{\mbb{J}^2_{BMO}}$ are
bounded by some constant depending only on $(\gamma,\beta, T, ||\xi||_{\infty}, ||l||_{\mbb{S}^\infty}, ||Y||_{\mbb{S}^\infty})$.
\begin{proof}
Since $||\psi||_{J^{\infty}}\leq 2||Y||_{\mbb{S}^\infty}$, it is clear that $\psi\in\mbb{J}^{\infty}$.
Applying It\^o formula to $e^{2\gamma Y_t}$ and using the equality $2\gamma j_{2\gamma}(x)=(e^{\gamma x}-1)^2+2\gamma j_\gamma (x)$,
one obtains
\bea
&&\int_\tau^{T} e^{2\gamma Y_s}2\gamma^2|Z_s|^2ds+\int_\tau^{T} \int_E e^{2\gamma Y_s}
\bigl(e^{\gamma \psi_s(x)}-1\bigr)^2\nu(dx)ds\nn \\
&&=e^{2\gamma Y_{T}}-e^{2\gamma Y_\tau}+2\gamma \int_\tau^{T}e^{2\gamma Y_s}\Bigl(
f(s,Y_s,Z_s,\psi_s)-\int_E j_\gamma(\psi_s(x))\nu(dx)\Bigr)ds\nn \\
&&-\int_\tau^{T}e^{2\gamma Y_s} 2\gamma Z_s dW_s-\int_\tau^{T} \int_E 
e^{2\gamma Y_{s-}}\bigl(e^{2\gamma \psi_s(x)}-1\bigr)\wt{\mu}(ds,dx)~,\nn
\eea
where $\tau\in\calt^T_0$. Taking a conditional expectation and using Assumption~\ref{assumption-Qexp}, one obtains
\bea
&&\mbb{E}\left[\int_\tau^{T} e^{2\gamma Y_s}\gamma^2 |Z_s|^2 ds+\int_\tau^{T}\int_E e^{2\gamma Y_{s}}
\bigl(e^{\gamma \psi_s(x)}-1\bigr)^2\nu(dx)ds\Bigr|\calf_\tau\right]\nn \\
&&\qquad \leq \mbb{E}\left[e^{2\gamma Y_{T}}+2\gamma \int_\tau^{T} e^{2\gamma Y_s}\bigl(l_s+\beta |Y_s|\bigr)ds\Bigr|
\calf_\tau\right]\nn \\
&&\qquad \leq e^{2\gamma ||Y||_{\mbb{S}^\infty}}+2\gamma e^{2\gamma ||Y||_{\mbb{S}^\infty}}T\Bigl(\beta ||Y||_{\mbb{S}^\infty}+||l||_{\mbb{S}^\infty}\Bigr)~.\nn
\eea
Thus
\bea
&&\mbb{E}\left[\int_\tau^{T} \gamma^2 |Z_s|^2 ds+\int_\tau^{T}\int_E
\bigl(e^{\gamma \psi_s(x)}-1\bigr)^2\nu(dx)ds\Bigr|\calf_\tau\right]\nn \\
&&\quad \leq e^{4\gamma ||Y||_{\mbb{S}^\infty}}+2\gamma e^{4\gamma ||Y||_{\mbb{S}^\infty}}T\Bigl(\beta ||Y||_{\mbb{S}^\infty}+||l||_{\mbb{S}^\infty}\Bigr)~.
\label{eq-plus}
\eea
Similar calculation for $e^{-2\gamma Y_t}$ yields
\bea
&&\mbb{E}\left[\int_\tau^{T} \gamma^2 |Z_s|^2 ds+\int_\tau^{T}\int_E
\bigl(e^{-\gamma \psi_s(x)}-1\bigr)^2\nu(dx)ds\Bigr|\calf_\tau\right]\nn \\
&&\quad \leq e^{4\gamma ||Y||_{\mbb{S}^\infty}}+2\gamma e^{4\gamma ||Y||_{\mbb{S}^\infty}}T\Bigl(\beta ||Y||_{\mbb{S}^\infty}+||l||_{\mbb{S}^\infty}\Bigr)~. 
\label{eq-minus}
\eea
Let us mention the fact that $(e^{x}-1)^2+(e^{- x}-1)^2\geq x^2,~\forall x\in\mbb{R}~$.
Indeed, for $g(x):=(e^{x}-1)^2+(e^{-x}-1)^2-x^2$, we have 
$g^\prime(x)=2(e^{x}-1)e^x+2 (1-e^{-x})e^{-x} -2x$
which is an odd function. It is easy to see that  $g^\prime(x)\geq 0$ for $x\geq 0$ and $g^\prime(0)=0$. Thus $g(x)\geq g(0)=0$.
With the help of this relation, adding (\ref{eq-plus}) and (\ref{eq-minus}), and then taking $\sup_{\tau}||~||_{\infty}$
separately for $Z$ and $\psi$ terms yields
\bea
||Z||_{\mbb{H}^2_{BMO}}^2+||\psi||^2_{\mbb{J}^2_B}\leq \frac{e^{4\gamma ||Y||_{\mbb{S}^\infty}}}{\gamma^2}
\Bigl(3+6\gamma T\bigl(\beta ||Y||_{\mbb{S}^\infty}+||l||_{\mbb{S}^\infty}\bigr)\Bigr)<\infty. \nn
\eea
Since $||\psi||_{\mbb{J}^\infty}\leq 2 ||Y||_{\mbb{S}^\infty}$, one also sees $||\psi||_{\mbb{J}^2_{BMO}}\leq 
||\psi||_{\mbb{J}^2_B}+||\psi||_{\mbb{J}^\infty}<\infty$.
\end{proof}
\end{lemma}

The following result is an adaptation of Proposition 3.2 in \cite{Barrieu-ElKaroui} and 
Proposition 16 in \cite{Ngoupeyou} to our setting. 
Similar results can be fond in \cite{Briand-Hu-06} for a diffusion setup 
and in \cite{Morlais, Antonelli} with jumps. 
\begin{lemma}
\label{lemma-universal-bound}
Under  Assumption~\ref{assumption-Qexp}, 
if there exists a solution $(Y,Z,\psi)\in\mbb{S}^\infty\times \mbb{H}^2\times \mbb{J}^2$ to the BSDE (\ref{eq-Qexp-BSDE}), it satisfies
\bea
|Y_t|\leq \frac{1}{\gamma}\ln \mbb{E}\Bigl[\exp\Bigl(\gamma e^{\beta (T-t)}|\xi|+
\gamma \int_t^T e^{\beta(t-s)}l_s ds\Bigr)\Bigr|\calf_t\Bigr]~, \nn
\eea
and in particular, 
\be
||Y||_{\mbb{S}^\infty}\leq e^{\beta T}\bigl(||\xi||_{\infty}+T||l||_{\mbb{S}^\infty}\bigr)~. \nn
\ee
\begin{proof}
An application of Meyer-It\^o formula (Theorem 70 in \cite{Protter}) yields
\bea
&&d\bigl(e^{\beta s} |Y_s|\bigr)=e^{\beta s}\bigl(\beta|Y_s|ds+d|Y_s|\bigr)\nn \\
&&=e^{\beta s}\Bigl\{ \beta|Y_s|ds+{\rm sign}(Y_{s-})\Bigl(-f(s,\Theta_s)ds+
Z_s dW_s+\int_E \psi_s (x)\wt{\mu}(ds,dx)\Bigr)+dL_s^Y\Bigr\}\nn 
\eea
where $L^Y$ is a non-decreasing process including a local time of $Y$ at the origin.
Let us define the process $(B_s,s\in[0,T])$ with $B_0=0$ by
\bea
dB_s=-{\rm sign}(Y_s)f(s,\Theta_s)ds+\Bigl(l_s+\beta|Y_s|+\frac{\gamma}{2}|Z_s|^2+
\int_E j_\gamma ({\rm sign}(Y_s)\psi_s(x))\nu(dx)\Bigr)ds\nn 
\eea
which is also a non-decreasing process by Assumption~\ref{assumption-Qexp}.
Using this process, 
\bea
&&d(e^{\beta s}|Y_s|)=e^{\beta s} (dB_s+dL_s^Y)+e^{\beta s}{\rm sign}(Y_{s-})\Bigl(Z_s dW_s+
\int_E \psi_s(x){\wt{\mu}}(ds,dx)\Bigr) \nn \\
&&\quad -e^{\beta s}\Bigl(l_s+\frac{\gamma}{2}|Z_s|^2+\int_E j_\gamma ({\rm sign}(Y_s)\psi_s(x))\nu(dx)\Bigr)ds~,\nn
\eea
which is further transformed as
\bea
&&d(e^{\beta s}|Y_s|)=e^{\beta s}{\rm sign}(Y_{s-})\Bigl(Z_s dW_s+\int_E \psi_s(x)\wt{\mu}(ds,dx)\Bigr)-\frac{\gamma}{2}\bigl| e^{\beta s}{\rm sign}(Y_s)Z_s\bigr|^2ds\nn \\
&&\quad-\int_E j_\gamma(e^{\beta s}{\rm sign}(Y_s)\psi_s(x))\nu(dx)ds
-e^{\beta s}l_s ds+\frac{\gamma}{2}\Bigl(e^{2\beta s}|Z_s|^2-e^{\beta s}|Z_s|^2\Bigr)ds \nn \\
&&\quad +\int_E \Bigl(j_\gamma(e^{\beta s}{\rm sign}(Y_s)\psi_s(x))-e^{\beta s} j_\gamma({\rm sign}(Y_s)\psi_s(x))\Bigr)\nu(dx)ds+e^{\beta s}(dB_s+dL_s^Y)~.\nn
\eea
It is easy to confirm that for $k\geq 1$,
\be
j_{\gamma} (kx)-k j_\gamma(x)=\frac{1}{\gamma}(e^{k \gamma x}-ke^{\gamma x}-1+k)\geq 0~. \nn
\ee
Thus we obtain
\bea
&&d(e^{\beta s} |Y_s|)=e^{\beta s}{\rm sign}(Y_{s-})\Bigl(Z_s dW_s+\int_E \psi_s(x)\wt{\mu}(ds,dx)\Bigr)\nn \\
&&\quad -\frac{\gamma}{2}|e^{\beta s}{\rm sign}(Y_s)Z_s|^2 ds-\int_E 
j_{\gamma}(e^{\beta s}{\rm sign}(Y_s)\psi_s(x))\nu(dx)ds-e^{\beta s}l_s ds+dC_s,\nn
\eea
where $C$ is a non-decreasing process. 

Define the process $P$ by  $P_t:=\exp\Bigl(\gamma e^{\beta t}|Y_t|+\gamma \int_0^t e^{\beta s}l_s ds\Bigr)$.
Using another non-decreasing process $C^\prime$, one has
\bea
dP_t=P_{t-}\Bigl(\gamma e^{\beta t}{\rm sign}(Y_{t})Z_t dW_t+\int_E \Bigl(\exp\bigl(\gamma e^{\beta t}
{\rm sign}(Y_{t-})\psi_t(x)\bigr)-1\Bigr)\wt{\mu}(dt,dx)+\gamma dC_t^\prime \Bigr).~
\label{eq-dP}
\eea
The boundedness of $P$ and Lemma~\ref{lemma-BMO-bound} imply that the first two terms of (\ref{eq-dP}) are
true martingale and that the last term is an integrable increasing process.
Therefore $P$ is a submartingale and it follows that
\bea
\exp\Bigl(\gamma e^{\beta t}|Y_t|+\gamma \int_0^t e^{\beta s}l_s ds\Bigr)
\leq \mbb{E}\left[\exp\Bigl(\gamma e^{\beta T}|\xi|+\gamma \int_0^T e^{\beta s}l_s ds\Bigr)\Bigr|\calf_t\right]~,\nn
\eea
for $\forall t\in[0,T]$, and the claim is proved.
\end{proof}
\end{lemma}

%%%%%%%%%%%%%%%%%%%%%%%%%%%%%%%%%%%%%%%%
\subsection{Stability and Uniqueness}
%%%%%%%%%%%%%%%%%%%%%%%%%%%%%%%%%%%%%%%%
We now introduce local Lipschitz conditions to derive the stability and uniqueness result
for a bounded solution.
\begin{assumption} 
\label{assumption-LLC}
For each $M>0$, and for every $(y,z,\psi), (y^\prime,z^\prime,\psi^\prime)\in\mbb{R}\times \mbb{R}^d\times \mbb{L}^2(E,\nu;\mbb{R}^k)$
satisfying 
\be
|y|, |y^\prime|, ||\psi||_{\mbb{L}^\infty(\nu)}, ||\psi^\prime||_{\mbb{L}^\infty(\nu)}\leq M\nn
\ee
there exists some positive constant $K_M$ possibly depending on $M$ such that
\bea
&&\bigl|f(t,y,z,\psi)-f(t,y^\prime,z^\prime,\psi^\prime)\bigr|\leq K_M\bigl(|y-y^\prime|+||\psi-\psi^\prime||_{\mbb{L}^2(\nu)}\bigr)\nn \\
&&\qquad+K_M\bigl(1+|z|+|z^\prime|+||\psi||_{\mbb{L}^2(\nu)}+||\psi^\prime||_{\mbb{L}^2(\nu)}\bigr)|z-z^\prime|\nn
\eea
$dt\otimes d\mbb{P}$-a.e. $(\omega,t)\in \Omega\times [0,T]$.
\end{assumption}

Consider the two BSDEs with $i\in\{1,2\}$ satisfying Assumptions~\ref{assumption-Qexp} and \ref{assumption-LLC};
\bea
Y^i_t=\xi^i+\int_t^T f^i(s,Y^i_s,Z_s^i,\psi_s^i)ds-\int_t^T Z_s^i dW_s-\int_t^T \int_E \psi_s^i(x) \wt{\mu}(ds,dx), 
\label{eq-two-BSDEs}
\eea
for $t\in[0,T]$ and let us denote
\bea
&&\del Y:=Y^1-Y^2, \quad \del Z:=Z^1-Z^2, \quad \del \psi:=\psi^1-\psi^2, \nn \\
&&\del f(s):=(f^1-f^2)(s,Y_s^1,Z_s^1,\psi_s^1)~. \nn
\eea

\begin{lemma}
\label{lemma-uniqueness-pre}
Suppose Assumptions~\ref{assumption-Qexp} and \ref{assumption-LLC} hold
for the two BSDEs (\ref{eq-two-BSDEs}) with $i\in\{1,2\}$. Then,
if there exists a solution $(Y^i,Z^i,\psi^i)\in\mbb{S}^\infty\times \mbb{H}^2\times \mbb{J}^2, i\in\{1,2\}$ to the BSDEs, the following inequalities are  
satisfied;
\bea
&&(a)~||\del Z||_{\mbb{H}^2_{BMO}}+||\del \psi||_{\mbb{J}^2_{BMO}}\leq 
C\Bigl(||\del Y||_{\mbb{S}^\infty}+||\del \xi||_{\infty}+\sup_{\tau\in\calt^T_0} \Bigl|\Bigl|
\mbb{E}\left[\int_\tau^T |\del f(s)|ds\Bigr|\calf_\tau\right]\Bigr|\Bigr|_{\infty}\Bigr)  \nn \\
&&(b)~\bigl|\bigl|(\del Y, \del Z,\del \psi)\bigr|\bigr|^p_{\calk^p[0,T]}
\leq C^\prime \Bigl(\mbb{E}\left[|\del \xi|^{p\bar{q}^2}+\Bigl(\int_0^T |\del f(s)|ds\Bigr)^{p\bar{q}^2}
\right]\Bigr)^{\frac{1}{\bar{q}^2}}, \forall p\geq 2, ~\forall \bar{q}\geq q_* \nn
\eea
Here, $C$ and $q_*~(>1)$ are positive constants depending only on $(K_M,\gamma,\beta,T,||\xi||_{\infty}, ||l||_{\mbb{S}^\infty})$
and the constant $M$ is chosen such that $||Y^i||_{\mbb{S}^\infty}$, $||\psi^i||_{\mbb{J}^\infty}\leq M$ for both $i\in\{1,2\}$.
$C^\prime$ is a positive constant depending only on $(p,\bar{q},K_M, \gamma,\beta,T,||\xi||_{\infty},||l||_{\mbb{S}^\infty})$.

\begin{proof}
{\it Proof for (a)}\\
Firstly, due to the universal bounds, it is obvious that one can choose $M$ such that $||Y^i||_{\mbb{S}^\infty}\leq M$
and $||\psi^i||_{\mbb{J}^{\infty}}\leq M$ for both $i\in\{1,2\}$.
For $\forall \tau\in\calt^T_0$, one has
\bea
&&|\del Y_\tau|^2+\int_\tau^{T}|\del Z_s|^2 ds+\int_\tau^{T}\int_E |\del \psi_s(x)|^2\mu(ds,dx)\nn \\
&&=|\del \xi|^2+\int_\tau^{T}2\del Y_s\Bigl(\del f(s)+f^2(s,\Theta_s^1)-f^2(s,\Theta_s^2)\Bigr)ds\nn \\
&&\quad-\int_\tau^{T}2 \del Y_s \del Z_s dW_s-\int_\tau^{T}\int_E 2\del Y_{s-}\del \psi_s(x)\wt{\mu}(ds,dx)~.\nn
\eea
Taking the conditional expectation, one obtains
\bea
&&|\del Y_{\tau}|^2+\mbb{E}\Bigl[\int_\tau^T |\del Z_s|^2 ds|\calf_\tau\Bigr]+\mbb{E}\Bigl[\int_\tau^T \int_E |\del \psi_s(x)|^2\mu(ds,dx)\Bigr|\calf_\tau\Bigr]\nn \\
&&=\mbb{E}\left[|\del \xi|^2+\int_\tau^T 2\del Y_s\Bigl(\del f(s)+f^2(s,\Theta_s^1)-f^2(s,\Theta_s^2)\Bigr)ds\Bigr|\calf_\tau\right]~.\nn
\eea
Taking $\sup_{\tau\in\calt^T_0}$ for each term in the left gives 
\bea
&&||\del Z||^2_{\mbb{H}^2_{BMO}}+||\del\psi||^2_{\mbb{J}^2_{B}} \leq 2||\del \xi||^2_{\infty}\nn \\
&&~+4||\del Y||_{\mbb{S}^\infty}
\sup_{\tau\in\calt^T_0}\Bigl|\Bigl|\mbb{E}\left[\int_\tau^T \Bigl(|\del f(s)|+K_M\bigl(|\del Y|_s+||\del \psi_s||_{\L2nu}+H_s|\del Z_s|\bigr)
\Bigr)ds\Bigr|\calf_\tau\right]\Bigr|\Bigr|_{\infty}~,\nn
\eea
where the process $H$ is defined by $H_s:=1+\sum_{i=1}^2\bigl(|Z_s^i|+||\psi_s^i||_{\L2nu}\bigr)$. It is clear that $H\in \mbb{H}^2_{BMO}$
whose norm is dominated by the universal bounds given in Lemma~\ref{lemma-BMO-bound}.
One can see
\bea
\sup_{\tau\in\calt^T_0}\Bigl|\Bigl|\mbb{E}\left[\int_\tau^T H_s|\del Z_s|ds\Bigr|\calf_\tau\right]\Bigr|\Bigr|_{\infty} 
&\leq& \sup_{\tau\in\calt^T_0}\Bigl|\Bigl|\mbb{E}\left[\int_\tau^T |H_s|^2 ds\Bigr|\calf_\tau\right]^{\frac{1}{2}}\Bigr|\Bigr|_{\infty}
~ \sup_{\tau\in\calt^T_0}\Bigl|\Bigl|\mbb{E}\left[\int_\tau^T |\del Z_s|^2ds\Bigr|\calf_\tau\right]^{\frac{1}{2}}\Bigr|\Bigr|_{\infty}\nn \\
&\leq& ||H||_{\mbb{H}^2_{BMO}}||\del Z||_{\mbb{H}^2_{BMO}}~.\nn
\eea
Thus, with an arbitrary positive constant $\ep>0$, 
\bea
&&||\del Z||^2_{\mbb{H}^2_{BMO}}+||\del \psi||^2_{\mbb{J}^2_{B}}
\leq 2||\del \xi||^2_\infty+2\sup_{\tau\in\calt^T_0}\Bigl|\Bigl| \mbb{E}\left[\int_\tau^T |\del f(s)|ds\Bigr|\calf_\tau\right]
\Bigr|\Bigr|_{\infty}^2\nn\\
&&+||\del Y||^2_{\mbb{S}^\infty}\Bigl(2+4K_M T+\frac{4K_M^2}{\ep}+\frac{4K_M^2}{\ep}||H||^2_{\mbb{H}^2_{BMO}}\Bigr)+
\ep\Bigl(||\del Z||^2_{\mbb{H}^2_{BMO}}+||\del\psi||^2_{\mbb{J}^2_B}\Bigr)\nn.
\eea
Choosing $\ep<1$ and noticing the fact that $||\del\psi||_{\mbb{J}^2_{BMO}}\leq ||\del \psi||_{\mbb{J}^2_{B}}+2||\del Y||_{\mbb{S}^{\infty}}$,
one obtains the desired result.
\\\\
{\it Proof for (b)}\\
Define a $d$-dimensional $\mbb{F}$-progressively measurable process $(b_s,s\in[0,T])$ by
\bea
b_s:=\frac{f^{2}(s,Y_s^1,Z_s^1,\psi_s^1)-f^{2}(s,Y_s^1,Z_s^2,\psi_s^1)}{|\del Z_s|^2}\bold{1}_{\del Z_s\neq 0}\del Z_s\nn
\eea 
and also the map $\wt{f}:\Omega\times [0,T]\times \mbb{R}\times \mbb{L}^2(E,\nu;\mbb{R}^k)
\rightarrow \mbb{R}$ by
\bea
\wt{f}(\omega, s,\wt{y},\wt{\psi}):=\del f(\omega,s)-f^2(\omega, s,\Theta_s^2)+f^2\bigl(\omega,s,\wt{y}+Y_s^2,Z_s^2,\wt{\psi}+\psi_s^2\bigr)~.\nn
\eea
Then, $(\del Y, \del Z, \del \psi)$ can be interpreted as the solution to the BSDE
\bea
\del Y_t=\del \xi+\int_t^T \Bigl(\wt{f}(s,\del Y_s,\del \psi_s)+b_s\cdot \del Z_s \Bigr)ds-\int_t^T \del Z_s dW_s
-\int_t^T \int_E \del \psi_s(x)\wt{\mu}(ds,dx).
\label{eq-delY-bmo-like}
\eea
Since $|b_s|\leq K_M(1+|Z_s^1|+|Z_s^2|+2||\psi_s^1||_{\L2nu})$, the process $b$ belongs to $\mbb{H}^2_{BMO}$.
Furthermore, $\wt{f}$ satisfies the linear growth property $|\wt{f}(s,\wt{y},\wt{\psi})|\leq |\del f(s)|+K_M(|\wt{y}|+||\wt{\psi}||_{\L2nu})$. 
Thus, the BSDE (\ref{eq-delY-bmo-like}) satisfies Assumption~\ref{assumption-apriori-bmolike}
with $g=|\del f|$.
One obtains the desired result by applying Lemma~\ref{lemma-bmolike-apriori}.
The dependency of the constants $C^\prime, q_*$ is obtained from the universal bounds in Lemmas~\ref{lemma-universal-bound}
and \ref{lemma-BMO-bound}, as well as the properties of the reverse H\"older inequality in Lemma~\ref{lemma-R-Holder}
and the remarks that follow.
\end{proof}
\end{lemma}

We now gives the uniqueness result:
\begin{proposition}
\label{prop-uniqueness}
Suppose the BSDE (\ref{eq-Qexp-BSDE}) satisfies Assumption~\ref{assumption-Qexp} and \ref{assumption-LLC}.
Then, if there exists a solution $(Y,Z,\psi)\in \mbb{S}^\infty\times \mbb{H}^2\times \mbb{J}^2$ 
to (\ref{eq-Qexp-BSDE}), it is unique in the space $\mbb{S}^\infty\times \mbb{H}^2_{BMO}
\times \mbb{J}^2_{BMO}$.
\begin{proof}
By Lemmas~\ref{lemma-universal-bound} and \ref{lemma-BMO-bound}, if there exists such a solution
it satisfies $(Y,Z,\psi)\in\mbb{S}^{\infty}\times \mbb{H}^2_{BMO} \times \mbb{J}^2_{BMO}$. 
Firstly, by Lemma~\ref{lemma-uniqueness-pre} (b), the solution is unique 
in the space $\calk^p[0,T]$ for $\forall p\geq 2$.
Since $Y\in\mbb{S}^\infty$, the uniqueness of $Y$ in $\mbb{S}^p$ gives 
the uniqueness of $Y$ also in the space $\mbb{S}^\infty$.  
This can be easily shown from an argument of contradiction
by assuming $||Y^1-Y^2||^p_{\mbb{S}^p}=0$
but not equal in $\mbb{S}^\infty$.
%Suppose that there exist two solution $Y^1, Y^2 \in \mbb{S}^\infty$ which are equal 
%in the space of $\mbb{S}^p$ i.e., 
%This implies that there exists some constant $a>0$ such that
%\bea
%\Bigl|\Bigl| \sup_{t\in[0,T]}|Y_t^1-Y_t^2|\Big|\Bigr|_{\infty}=a~. \nn 
%\eea
%Then, for any $0<b<a$, there exists some positive constant $0<c\leq1$ such that
%\bea
%\mbb{P}\Bigl(\sup_{t\in[0,T]}|Y_t^1-Y_t^2|\geq b\Bigr)=c~. \nn
%\eea
%This gives $||Y^1-Y^2||^p_{\mbb{S}^p}\geq b^p~c>0$ and hence yields a contradiction.
%Combined with Lemma~\ref{lemma-uniqueness-pre} (a), the solution $(Y,Z,\psi)$
%is unique in the space $\mbb{S}^\infty\times \mbb{H}^2_{BMO}\times \mbb{J}^2_{BMO}$.
\end{proof}
\end{proposition}

%%%%%%%%%%%%%%%%%%%%%%%%%%%%%%%%%%%%%%%%%%%%%%%%%%%%%%%%%%%%%%%%%%%%%%%%%%%
\section{Existence of solution to a $Q_{\exp}$-growth BSDE}
\label{sec-existence-QexpBSDE}
%%%%%%%%%%%%%%%%%%%%%%%%%%%%%%%%%%%%%%%%%%%%%%%%%%%%%%%%%%%%%%%%%%%%%%%%%%
%\subsection{Some preparations}
In this section, we prove the existence of the solution to the BSDE (\ref{eq-Qexp-BSDE}). 
Although one may use the stability of quadratic semimartingales as \cite{ElKaroui-Matoussi},
we provide a concrete, less abstract strategy similar to that of Kobylanski~\cite{Kobylanski}.  
We need another assumption so that we can apply the comparison principle.
\begin{assumption} ($A_\Gamma$-condition)\\
\label{assumption-AGamma}
For all $t\in[0,T]$, $M>0$, $y\in\mbb{R}, z\in \mbb{R}^d, \psi,\psi^\prime \in \mbb{L}^2(E,\nu;\mbb{R}^k)$ with
$|y|, ||\psi||_{\mbb{L}^{\infty}(\nu)}, ||\psi^\prime||_{\mbb{L}^{\infty}(\nu)}\leq M$, there exists
a $\calp\otimes \cale$-measurable process $\Gamma^{y,z,\psi,\psi^\prime}$ satisfying $dt\otimes d\mbb{P}$-a.e.
\bea
f(t,y,z,\psi)-f(t,y,z,\psi^\prime)\leq \int_E \Gamma_t^{y,z,\psi,\psi^\prime}(x)\bigl[\psi(x)-\psi^\prime(x)\bigr]\nu(dx) 
\label{eq-Agamma}
\eea
and
$C^1_M(1\wedge |x|)\leq \Gamma_t^{y,z,\psi,\psi^\prime}(x)\leq C^2_M(1\wedge |x|)$ with two constants $C^1_M, C^2_M$.
Here, $C^1_M> -1$ and $C^2_M>0$ depend on $M$. (Hereafter, we frequently omit the superscripts $y,z$
to lighten the notation.)~\footnote{ $A_\Gamma$-condition implies $M$-dependent {\it local Lipschitz} continuity with respect to $\psi$,
which is known to be satisfied in the case of the exponential utility optimization~\cite{Morlais}.} 
\end{assumption}
 
Let us introduce a sequence of smooth truncation functions $\varphi_m:\mbb{R}\rightarrow\mbb{R}$ with $m\in\mbb{N}$
with the following properties:
\bea 
\varphi_m(x)=\begin{cases} 
-(m+1) & {\text{for}~~}x\leq -(m+2) \\
x & {\text{for}~~} |x|\leq m\\
m+1 & {\text{for}~~} x\geq m+2
\end{cases}  
\label{eq-varphi}
\eea
and $|\part_x \varphi_m(x)|\leq 1$ uniformly in $x\in\mbb{R}$.\footnote{The smoothness is introduced 
just for convenience so that one can use the same function later when proving Malliavin differentiability.} 
We denote $\ol{f}:=f\vee 0$, $\ul{f}:=f\wedge 0$ and introduce the following regularization of the driver:
\bea
&&\ol{f}^n(t,y,z,\psi):=\inf_{w\in\mbb{R}^d}\{\ol{f}(t,y,w,\psi)+n|z-w|\} \nn \\
&&\ul{f}^m(t,y,z,\psi):=\sup_{w\in\mbb{R}^d}\{\ul{f}(t,y,w,\psi)-m|z-w|\} \nn  \\
&&\ol{f}^{n,k}(t,y,z,\psi):=\ol{f}^n(t,\varphi_k(y),z,\varphi_k(\psi))\nn \\
&&\ul{f}^{m,k}(t,y,z,\psi):=\ul{f}^m(t,\varphi_k(y),z,\varphi_k(\psi))\nn
\eea
and $f^{n,m}:=\ol{f}^{n}+\ul{f}^m$, $f^{n,m,k}:=\ol{f}^{n,k}+\ul{f}^{m,k}$.
For $\psi$, the mollifier $\varphi_k$ should be applied component-wise.
\begin{lemma}
\label{lemma-nmk}
For a driver $f$ satisfying Assumptions~\ref{assumption-Qexp}, \ref{assumption-LLC} and \ref{assumption-AGamma}, we have \\
(i) $\ol{f}^n,\ul{f}^m,\ol{f}^{n,k},\ul{f}^{m,k}, f^{n,m}, f^{n,m,k}$ satisfy the structure condition of Assumption~\ref{assumption-Qexp}
uniformly in $n,m,k\in\mbb{N}$.\\
(ii) $\ol{f}^n,\ul{f}^m$ and $f^{n,m}$ satisfy $A_\Gamma$-condition (\ref{eq-Agamma}) uniformly in $n,m\in\mbb{N}$.\\
(iii) $\ol{f}^{n,k}, \ul{f}^{m,k}, f^{n,m,k}$ are globally Lipschitz continuous for each $n,m,k\in\mbb{N}$.
\begin{proof}
(i) One can easily confirm the assertion from the fact that $0\leq \ol{f}^n\leq \ol{f}^{n+1}\leq \ol{f}$, $\ul{f}\leq \ul{f}^{m+1}\leq \ul{f}^m\leq 0$ and that 
$j_\gamma(\cdot)$ is convex.
(ii) Firstly, let us check the condition for $\ol{f}^n$. Since
\bea
&&\ol{f}^n(t,y,z,\psi)-\ol{f}^n(t,y,z,\psi^\prime)\nn \\
&&=\inf_{w\in\mbb{R}^d}\{\ol{f}(t,y,w,\psi)+n|z-w|\}
-\inf_{w\in\mbb{R}^d}\{\ol{f}(t,y,w,\psi^\prime)+n|z-w|\}\nn\\
&&\leq \sup_{w\in \mbb{R}^d}\{\ol{f}(t,y,w,\psi)-\ol{f}(t,y,w,\psi^\prime)\}\nn
\eea
One sees the desired result by considering the four cases of signs $(f(t,y,w, \psi),f(t,y,w, \psi^\prime))=(+,+),(+,-),(-,+),(-,-)$.
The first two cases are bounded by $f(\cdot,\psi)-f(\cdot,\psi^\prime)$.  The last two cases are bounded by 0 and hence the condition is trivially satisfied. Similar analysis yields the same conclusion for $\ul{f}^m$.
Finally, let us consider $f^{n,m}$. Based on the same categorization of signs $(f(\cdot, \psi),f(\cdot, \psi^\prime))$, we have
\bea
f^{n,m}(\cdot,\psi)-f^{n,m}(\cdot,\psi^\prime)=\begin{cases}
\ol{f}^n(\cdot,\psi)-\ol{f}^n(\cdot,\psi^\prime)\quad ~{\rm if} \quad (+,+) \\
\ul{f}^m(\cdot,\psi)-\ul{f}^m(\cdot,\psi^\prime)\quad {\rm if} \quad (-,-)\\
\ul{f}^m(\cdot,\psi)-\ol{f}^n(\cdot,\psi^\prime)\quad ~{\rm if} \quad (-,+)\\
\ol{f}^n(\cdot,\psi)-\ul{f}^m(\cdot,\psi^\prime)\quad ~{\rm if} \quad (+,-)
\end{cases} \nn
\eea
The first two cases satisfies $A_\Gamma$-condition by the previous discussion. The third case is trivial since it is bounded by 0.
As  for the last case, one sees $\ol{f}^n(\cdot,\psi)-\ul{f}^m(\cdot,\psi^\prime)\leq \ol{f}(\cdot,\psi)-\ul{f}(\cdot,\psi^\prime)
=f(\cdot,\psi)-f(\cdot,\psi^\prime)$
and hence the conclusion follows.
(iii) Lipschitz continuity with respect to $y,\psi$ arguments can be shown similarly as (ii) above.
Consider now the following obvious inequality
$\ol{f}(t,\varphi_k(y),w, \varphi_k(\psi))+n|z-w|\leq \ol{f}(t,\varphi_k(y),w, \varphi_k(\psi))+n|z^\prime-w|+n|z-z^\prime|$.
By taking $\inf_{w}$ in the both hands, we get $\ol{f}^{n,k}(t,y,z,\psi)\leq \ol{f}^{n,k}(t,y,z^\prime,\psi)+n|z-z^\prime|$.
The desired result follows by flipping the role of $z,z^\prime$. The same conclusion follows similarly for $\ul{f}^{m,k}$
and hence also $f^{n,m,k}$.
\end{proof}
\end{lemma}
The above regularization is inspired by \cite{Lepeltier-Martin, ElKaroui-Matoussi, Cvitanic} as an application to quadratic BSDEs.
However, notice the differences from the one used in \cite{ElKaroui-Matoussi} regarding the arguments of $y,\psi$. 
The following result is an extension of Lemma 9.6.6 in \cite{Cvitanic} for our setting.
\begin{proposition}
\label{prop-monotone}
Suppose $\xi\in\calf_T$ is bounded and the sequence $\{f^n,n\geq 1\}$ and $f$ of the drivers are such that that
(i) They are continuous mappings and satisfy Assumptions~\ref{assumption-Qexp} and \ref{assumption-AGamma}  uniformly. 
(ii) $f^n\downarrow f$ (resp. $f^n\uparrow f$).
(iii) If $y^n\rightarrow y$ in $\mbb{R}$, $z^n\rightarrow z$ in $\mbb{R}^d$  and
$\psi^n\rightarrow \psi$ in $\mbb{L}^2(\nu)$,  then $f^n(\cdot,y^n,z^n,\psi^n)\rightarrow f(\cdot,y,z,\psi)$ in $\mbb{R}$.
(iv) There exists a solution $(Y^n,Z^n,\psi^n)\in \mbb{S}^\infty\times \mbb{H}^2\times \mbb{J}^2$
 to the BSDE for each $n$
\bea
Y^n_t=\xi+\int_t^T f^n(s,Y_s^n,Z_s^n,\psi_s^n)ds-\int_t^T Z^n_s dW_s-\int_t^T \int_E \psi_s^n(x)\wt{\mu}(ds,dx),~t\in[0,T], \nn
\eea 
for which the comparison principle holds i.e. $Y^{n+1}_t\leq Y^n_t$ (resp. $Y^{n}_t\leq Y^{n+1}_t$) for $\forall t\in[0,T]$ a.s.
Then, there exists $(Y,Z,\psi)\in\mbb{S}^\infty\times \mbb{H}^2_{BMO}\times \mbb{J}^2_{BMO}$ such that 
$Y^{n}\rightarrow Y$ in $\mbb{S}^\infty$, $Z^{n}\rightarrow Z$ in $\mbb{H}^2$ and $\psi^{n}\rightarrow \psi$ in $\mbb{J}^2$
and solves the BSDE
\bea
Y_t=\xi+\int_t^T f(s,Y_s,Z_s,\psi_s)ds-\int_t^T Z_s dW_s-\int_t^T \int_E \psi_s(x)\wt{\mu}(ds,dx),~t\in[0,T].
\label{eq-limit-BSDE}
\eea
\begin{proof}
It suffices to consider the case $f^n\downarrow f$ with monotonically decreasing sequence of $Y^n$. By condition (i), the solution $(Y^n,Z^n,\psi^n)$ satisfies the universal bounds given in Lemmas~\ref{lemma-BMO-bound}
and \ref{lemma-universal-bound} uniformly in $n$. By monotonicity, $(Y^n)$ converges, for all $t\in[0,T]$,
$Y^n_t \downarrow Y_t$ $\mbb{P}$-a.s. to its limit process $Y:=lim_n Y^n$.
Furthermore, there exists $(Z,\psi)$ satisfying the universal bounds,
such that $Z^n\rightharpoonup Z$ weakly in $\mbb{H}^2$ as well as
$\psi^n\rightharpoonup \psi$ weakly in $\mbb{J}^2$ under an appropriate subsequence (still denoted by the same $n$).
By condition (i), each driver $f^n$ satisfies,  $dt\otimes d\mbb{P}$-a.e.,
$\displaystyle f^n(t,Y^n_t,Z_t^n,\psi_t^n) \leq  f^n(t,Y^n_t,Z^n_t,0)+\int_E\Gamma_t^{\psi^n,0}(x)\psi^n_t(x)\nu(dx)
\leq  l_t+\beta|Y^n_t|+\frac{\gamma}{2}|Z_t^n|^2+C_M||\psi^n_t||_{\mbb{L}^2(\nu)} $
and similarly $-f^n(t,Y^n_t,Z^n_t,\psi^n_t)\leq l_t+\beta|Y^n_t|+\frac{\gamma}{2}|Z_t^n|^2+C_M||\psi^n_t||_{\mbb{L}^2(\nu)}$,
where $C_M$ is a constant depending only on the universal bounds.

Let $\phi:\mbb{R}\rightarrow \mbb{R}$ is a smooth {\it convex} function such that $\phi(0)=0$, $\phi^\prime(0)=0$, which will be specified later.
We put $\del Y^{n,m}:=Y^n-Y^m$, $\del Z^{n,m}:=Z^n-Z^m$, $\del \psi^{n,m}:=\psi^n-\psi^m$, and assume $m\geq n$. Note that $\del Y^{n,m}_T=0$ and
$\del Y^{n,m}\geq 0$ for $m\geq n$.  It\^o formula gives 
\bea
&&\phi(\del Y_t^{n,m})+\int_t^T \frac{1}{2}\phi^{\prime\prime}(\del Y_s^{n,m})|\del Z^{n,m}_s|^2ds+\int_t^T\int_E \bigl[\phi(\del Y_{s-}^{n,m}+\del\psi_s^{n,m}(x))-\phi(\del Y^{n,m}_{s-})\nn \\
&&-\phi^\prime(\del Y^{n,m}_{s-})\del\psi_s^{n,m}(x)\bigr]\mu(ds,dx)=\int_t^T \phi^{\prime}(\del Y_s^{n,m})\bigl[f^n(s,\Theta_s^n)-f^m(s,\Theta_s^m)\bigr]ds\nn \\
&&-\int_t^T \phi^\prime(\del Y^{n,m}_s)\del Z_s^{n,m}dW_s-\int_t^T \int_E \phi^\prime(\del Y_{s-}^{n,m})\wt{\mu}(ds,dx)~.\nn
\eea
Using the previous driver's bound and noticing that $\phi^\prime(y)\geq 0$ for $y\geq 0$, there exist constants $C_M,C_0$ independent of $n,m$ satisfying 
\bea
&&\mbb{E}\int_0^T \frac{1}{2}\phi^{\prime\prime}(\del Y_s^{n,m})|\del Z_s^{n,m}|^2 ds
+\mbb{E}\int_0^T\int_E \bigl[\phi(\del Y_{s-}^{n,m}+\del\psi_s^{n,m}(x))-\phi(\del Y^{n,m}_{s-})\nn \\
&&-\phi^\prime(\del Y^{n,m}_{s-})\del\psi_s^{n,m}(x)\bigr]\mu(ds,dx)
\leq \mbb{E}\int_0^T C_M \phi^\prime(\del Y_s^{n,m})\Bigl(\frac{1}{\ep}+|Y^n_t|+|Y^{m}_t|\nn \\
&&+|Z_t^n|^2+|Z_t^m|^2+\ep ||\psi^n_t||^2_{\mbb{L}^2(\nu)}+\ep ||\psi^m_t||^2_{\mbb{L}^2(\nu)}\Bigr)ds \nn \\
&&\leq \mbb{E}\int_0^TC_0\phi^\prime(\del Y_s^{n,m})\Bigl(\frac{1}{\ep}+|\del Z^{n,m}_s|^2+|Z_s^n-Z_s|^2+|Z_s|^2\nn \\
&&\hspace{30mm}+\ep||\del \psi^{n,m}_s||^2_{\mbb{L}^2(\nu)}+\ep||\psi_s^n-\psi||^2_{\mbb{L}^2(\nu)}+\ep||\psi_s||^2_{\mbb{L}^2(\nu)}\Bigr)ds
\label{eq-phi-ito}
\eea
for any constant $\ep>0$. We now choose $\phi$ as
\be
\phi(y):=\frac{1}{8C_0^2}\bigl[e^{4C_0 y}-4C_0 y-1\bigr],\quad \phi^\prime(y)=\frac{1}{2C_0}\bigl[e^{4C_0 y}-1\bigr], \quad \phi^{\prime\prime}(y)=2 e^{4C_0 y}.\nn
\ee
By the mean-value theorem and the universal bound of Lemma~\ref{lemma-universal-bound}  for  $\del Y^{n,m}_{s},\del Y^{n,m}_{s-}$,
\bea
 c_M|\del \psi^{n,m}_s(x)|^2 \leq \phi(\del Y_{s-}^{n,m}+\del \psi^{n,m}_s(x))-\phi(\del Y_{s-}^{n,m})-\phi^\prime(\del Y_{s-}^{n,m})\del \psi^{n,m}_s(x)~ \nn 
\eea
holds uniformly in $(n,m)$ by choosing $c_M:=\exp\bigl(-8C_0 e^{\beta T}(||\xi||_{\infty}+T||l||_{\mbb{S}^\infty})\bigr)$.
Similarly, one can choose the constant $\ep$ such that
$C_0\phi^\prime\bigl(2e^{\beta T}(||\xi||_{\infty}+T||l||_{\mbb{S}^\infty})\bigr)\ep= c_M/4$~.
Then (\ref{eq-phi-ito}) implies (note that $\phi^{\prime\prime}(y)=4C_0\phi^\prime(y)+2$),
\bea
&&\mbb{E}\int_0^T \bigl[C_0\phi^\prime(\del Y^{n,m}_s)+1\bigr]|\del Z_s^{n,m}|^2 ds+
\mbb{E}\int_0^T \int_E \frac{3}{4}c_M |\del \psi_s^{n,m}(x)|^2 \nu(dx)ds\nn \\
&&\leq \mbb{E}\int_0^T C_0\phi^\prime(\del Y_s^{n,m})\Bigl(\frac{1}{\ep}+|Z_s^n-Z_s|^2+|Z_s|^2+
\ep||\psi^n_s-\psi_s||^2_{\mbb{L}^2(\nu)}+\ep ||\psi_s||^2_{\mbb{L}^2(\nu)}\Bigr)ds \nn.
\eea
Let fix $n$. $\del\psi^{n,m}\rightharpoonup \psi^n-\psi$ weakly in $\mbb{J}^2$.
Since $\del Y^{n,m}$ is bounded and strongly converges $\forall t\in[0,T]$ $\del Y^{n,m}_t\rightarrow Y^n_t-Y_t$ a.s., 
it is easy to see that
$\sqrt{C_0\phi^\prime(\del Y^{n,m})+1} [\del Z^{n,m}]$ converges weakly to $\sqrt{C_0\phi^\prime(Y^n-Y)+1} [Z^n-Z]$
in $\mbb{H}^2$. From Proposition 3.5 (iii)~\cite{Brezis}, by passing to the limit $m\rightarrow \infty$,
\bea
&&\mbb{E}\int_0^T \bigl[C_0\phi^\prime(Y_s^n-Y_s)+1\bigr]|Z^n_s-Z_s|^2ds+
\mbb{E}\int_0^T \frac{3}{4}c_M||\psi^n_s-\psi_s||^2_{\mbb{L}^2(\nu)}ds\nn \\
&&\leq \lim\inf_{m\rightarrow \infty}\mbb{E}\int_0^T \bigl[C_0\phi^\prime(\del Y^{n,m}_s)+1\bigr]|\del Z_s^{n,m}|^2 ds+
\mbb{E}\int_0^T \int_E  \frac{3}{4}c_M |\del \psi_s^{n,m}(x)|^2 \nu(dx)ds\nn \\
&&\leq \mbb{E}\int_0^T C_0\phi^\prime(Y^n_s-Y_s)\Bigl(\frac{1}{\ep}+|Z_s^n-Z_s|^2+|Z_s|^2+
\ep||\psi^n_s-\psi_s||^2_{\mbb{L}^2(\nu)}+\ep ||\psi_s||^2_{\mbb{L}^2(\nu)}\Bigr)ds \nn,
\eea
which then yields
\bea
&&\mbb{E}\int_0^T|Z_s^n-Z_s|^2ds+\mbb{E}\int_0^T \frac{c_M}{2} ||\psi_s^n-\psi_s||^2_{\mbb{L}^2(\nu)}ds\nn \\
&&\qquad  \leq \mbb{E}\int_0^T C_0\phi^\prime(Y^n_s-Y_s)\Bigl(\frac{1}{\ep}+|Z_s|^2+\ep ||\psi_s||^2_{\mbb{L}^2(\nu)}\Bigr)ds~.
\label{eq-H2J2}
\eea
Since $\phi^\prime(Y^n_s-Y_s)\rightarrow 0$ a.s. as $n\rightarrow \infty$, one concludes $Z^n\rightarrow Z$ in $\mbb{H}^2$ 
and $\psi^n\rightarrow \psi$ in $\mbb{J}^2$ by the dominated convergence theorem.

Therefore, one can extract a subsequence such that $Z^n\rightarrow Z$ $dt\otimes d\mbb{P}$-a.s. and 
$\psi^n\rightarrow \psi$ $\nu(dx)dt\otimes d\mbb{P}$-a.s.
Thus condition (iii) implies $f^n(t,Y^n_t,Z^n_t,\psi^n_t)\rightarrow f(t,Y_t,Z_t,\psi_t)$ $dt\otimes d\mbb{P}$-a.s.
Moreover, by extracting further subsequence if necessary, one sees from Lemma 2.5 of \cite{Kobylanski} that
$G_z:=\sup_n |Z^n|^2$, $G_\psi:=\sup_n ||\psi^n||^2_{\mbb{L}^2(\nu)}$ are in $\mbb{L}^1([0,T]\times \Omega)$.
By assumption (i), for almost all $\omega$,  $|f^n(\cdot, Y^n, Z^n, \psi^n)|$ is dominated by $C_M\bigl(1+G_z+G_\psi)\in \mbb{L}^1([0,T])$ with some constant $C_M$ depending only on the universal bounds. Note also that 
$f(\cdot,Y,Z,\psi)\in \mbb{L}^1([0,T])$ a.s.
Thus one obtains, for almost all $\omega$,
$\int_0^T|f^n(s,Y^n_s,Z^n_s,\psi_s^n)-f(s,Y_s,Z_s,\psi_s)|ds\rightarrow 0$
by Lebesgue's dominated convergence theorem. 
From (\ref{eq-H2J2}) and the Burkholder-Davis-Gundy inequality\footnote{See, for example, Theorem 48 in IV.4. of \cite{Protter}.}, one can also extract a subsequence in which $\sup_{t\in[0,T]}\Bigl|\int_t^T (Z^n_s-Z_s)dW_s\Bigr|\rightarrow 0$, $\sup_{t\in[0,T]}\Bigl|\int_t^T\int_E (\psi^n_s(x)-\psi_s(x))\wt{\mu}(ds,dx)\Bigr|\rightarrow 0$ a.s.
By passing to the limit $m\rightarrow \infty$ and taking supremum over $t$ in
\bea
&&|Y^n_t-Y^m_t|\leq \int_t^T |f^n(s, \Theta_s^n)-f^m(s,\Theta_s^m)|ds+\Bigl|\int_t^T (Z^n_s-Z_s^m) dW_s\Bigr|\nn \\
&&\qquad +\Bigl|\int_t^T \int_E (\psi^n_s(x)-\psi^m_s(x))\wt{\mu}(ds,dx)\Bigr|~,\nn
\eea
one obtains 
\bea
&&\sup_{t\in[0,T]}|Y^n_t-Y_t|\leq \int_0^T |f^n(s,\Theta^n_s)-f(s,\Theta_s)|ds+\sup_{t\in[0,T]}
\Bigl|\int_t^T (Z_s^n-Z_s)dW_s\Bigr|\nn \\
&& \qquad +\sup_{t\in[0,T]}\Bigl|\int_t^T \int_E (\psi^n_s(x)-\psi_s(x))\wt{\mu}(ds,dx)\Bigr|,\nn
\eea
from which one concludes the uniform convergence $\sup_{t\in[0,T]}|Y^n_t-Y_t|\rightarrow 0$ a.s.
(hence $||Y^n-Y||_{\mbb{S}^\infty}\rightarrow 0$)
under an appropriate subsequence and $(Y,Z,\psi)$ solves (\ref{eq-limit-BSDE}).
One can check that $\mbb{S}^\infty$ convergence actually occurs in the entire sequence.
If this is not the case, 
there exists a subsequence $(n_j)\subset(n)$ such that $||Y^{n_j}-Y||_{\mbb{S}^\infty}>c$
with some $c>0$ for all $n_j$, where $Y=\lim_n Y^n$ is independent of the choice of subsequence due to the monotonicity.
However, one can extract 
a further subsequence $(n_{jk})\subset (n_j)$
such that $\sup_{t\in[0,T]}|Y^{n_{jk}}_t-Y_t|\rightarrow 0$ a.s. by repeating the same discussion given above
and hence $||Y^{n_{jk}}-Y||_{\mbb{S}^\infty}\rightarrow 0$, which is a contradiction. 
\end{proof}
\end{proposition}

\begin{remark}
\label{remark-strong-conv}
By applying It\^o-formula to $|Y^n-Y|^2$,
\bea
&&|Y^n_\tau-Y_\tau|^2 +\mbb{E}\Bigl[\int_\tau^T|Z^n_s-Z_s|^2 ds\Bigr|\calf_\tau\Bigr]+
\mbb{E}\Bigl[\int_\tau^T \int_E|\psi^n_s(x)-\psi_s(x)|^2\mu(ds,dx) \Bigr|\calf_\tau\Bigr] \nn \\
&&\leq 2||Y^n-Y||_{\mbb{S}^\infty}\mbb{E}\Bigl[\int_\tau^T|f^n(s,Y^n_s,Z_s^n,\psi^n_s)-f(s,Y_s,Z_s,\psi_s)|ds\Bigr|\calf_\tau\Bigr]\nn
\eea
for any $\tau\in\calt^T_0$. It follows that the uniform convergence of $Y^n\rightarrow Y$
implies $Z^n\rightarrow Z$ and $\psi^n\rightarrow \psi$ in $\mbb{H}^2_{BMO}$
and $\mbb{J}^2_{BMO}$ respectively, because
$\sup_{\tau\in\calt^T_0}\Bigl|\Bigl|\mbb{E}\Bigl[\int_\tau^T |f^n(s,\Theta^n_s)-f(s,\Theta_s)|ds\Bigr|\calf_\tau\Bigr]\Bigr|\Bigr|_{\infty}
\leq C\bigl(1+||Z^n||^2_{\mbb{H}^2_{BMO}}+||\psi^n||^2_{\mbb{J}^2_{BMO}}\bigr)\leq C $
with some constant $C$ depending only on the universal bounds.~\footnote{ 
Convergence in the norm of $\calk^p=\mbb{S}^p\times \mbb{H}^p\times \mbb{J}^p$ with $\forall p\geq 2$ is actually 
enough for the discussions on Malliavin's differentiability.}
\end{remark}

\begin{theorem}
\label{theorem-existence}
Under Assumptions~\ref{assumption-Qexp}, \ref{assumption-LLC} and \ref{assumption-AGamma},
there exists a unique bounded solution $(Y,Z,\psi)\in\mbb{S}^\infty\times \mbb{H}^2_{BMO}\times \mbb{J}^2_{BMO}$
of the BSDE (\ref{eq-Qexp-BSDE}).
\begin{proof}
From Proposition~\ref{prop-uniqueness}, it suffices to prove the existence.
Firstly, consider the BSDE with data $(\xi,f^{n,m,k})$. Since $f^{n,m,k}$ is globally Lipschitz,
there exists a unique solution $(Y^{n,m,k},Z^{n,m,k},\psi^{n,m,k})$ for each $n,m,k$.
One also sees $Y^{n,m,k}\in\mbb{S^\infty}$ by Lemma~\ref{lemma-ynmk-bound}.
Since the driver $f^{n,m,k}$ satisfies the $Q_{\exp}$-structure condition by Lemma~\ref{lemma-nmk},
$(Y^{n,m,k},Z^{n,m,k},\psi^{n,m,k})$ satisfies the universal bounds of Lemmas~\ref{lemma-BMO-bound} and \ref{lemma-universal-bound}
uniformly in $n,m,k$. In particular, since $||Y^{n,m,k}||_{\mbb{S}^\infty}, ||\psi^{n,m,k}||_{\mbb{J}^\infty}$ are bounded uniformly,
$(Y^{n,m,k},Z^{n,m,k},\psi^{n,m,k})$ also consists of a solution of the BSDE
\bea
Y^{n,m}_t=\xi+\int_t^T f^{n,m}(s,Y^{n,m}_s,Z^{n,m}_s,\psi^{n,m}_s)ds-\int_t^T Z_s^{n,m}dW_s
-\int_t^T \int_E \psi_s^{n,m}(x)\wt{\mu}(ds,dx)~
\label{eq-fnm-BSDE}
\eea
for each $n,m$ provided $k$ is large enough. By Lemma~\ref{lemma-fnm-BSDE}, this is actually the unique solution of (\ref{eq-fnm-BSDE})
and satisfies the comparison principle $Y^{n,m+1}\leq Y^{n,m}\leq Y^{n+1,m}$ for every $n,m\in\mbb{N}$.
Thus, from Lemma~\ref{lemma-nmk}, we can apply Proposition~\ref{prop-monotone} with a fixed $n$. In particular, the condition (iii) 
follows from the continuity of the driver and the property of inf(sup)-convolution (see, Lemma~1 of \cite{Lepeltier-Martin}).
We then obtain $Y^{n,m}\rightarrow \wt{Y}^n$ in $\mbb{S}^\infty$, $Z^{n,m}\rightarrow \wt{Z}^n$ in $\mbb{H}^2$ and
$\psi^{n,m}\rightarrow \wt{\psi}^n$ in $\mbb{J}^2$, which solves  
\bea
\wt{Y}^n_t=\xi+\int_t^T \wt{f}^n(s,\wt{Y}^n_s,\wt{Z}^n_s,\wt{\psi}^n_s)ds-\int_t^T \wt{Z}_s^n dW_s
-\int_t^T \int_E  \wt{\psi}_s^n(x)\wt{\mu}(ds,dx)~, 
\eea
for each $n\in \mbb{N}$, where $\wt{f}^n:=\ol{f}^n+\ul{f}$. $\wt{f}^n$ satisfies the structure as well as $A_\Gamma$-conditions
uniformly in $n$. By Lemma~\ref{lemma-fn-tilde}, one can once again apply Proposition~\ref{prop-monotone}
to the monotone sequence $\wt{f}^n\uparrow f$.
Then there exists $(Y,Z,\psi)\in \mbb{S}^\infty\times \mbb{H}^2_{BMO}\times \mbb{J}^2_{BMO}$ 
with the convergence $\wt{Y}^n\rightarrow Y$ in $\mbb{S}^\infty$, 
$\wt{Z}^n\rightarrow Z$ in $\mbb{H}^2$, $\wt{\psi}^n\rightarrow \psi$ in $\mbb{J}^2$,
which solves the BSDE (\ref{eq-Qexp-BSDE}). By the Remark~\ref{remark-strong-conv},
one also obtains the convergence in the stronger norms.
\end{proof}
\end{theorem}

Although we have used a specific regularization to obtain a monotone sequence of drivers, 
we can actually weaken the condition of monotonicity. The following result is 
the adaptation of Theorem 2.8 of \cite{Kobylanski} to our setting.
\begin{proposition}
\label{prop-conv-general}
Suppose $\xi\in\calf_T$ is bounded and the sequence $\{f^n,n\geq 1\}$ and $f$ of the drivers are such that
(i) They are continuous mappings and satisfy Assumptions~\ref{assumption-Qexp}, \ref{assumption-LLC}, \ref{assumption-AGamma}  uniformly
in $n$.  (ii) If $y^n\rightarrow y$ in $\mbb{R}$, $z^n\rightarrow z$ in $\mbb{R}^d$  and
$\psi^n\rightarrow \psi$ in $\mbb{L}^2(\nu)$,  then $f^n(\cdot,y^n,z^n,\psi^n)\rightarrow f(\cdot,y,z,\psi)$ in $\mbb{R}$.
(iii) Let $(Y^n,Z^n,\psi^n)\in \mbb{S}^\infty\times \mbb{H}^2_{BMO}\times \mbb{J}^2_{BMO}$ be
 the unique solution of the BSDE (which is guaranteed by Theorem~\ref{theorem-existence})
\bea
Y^n_t=\xi+\int_t^T f^n(s,Y_s^n,Z_s^n,\psi_s^n)ds-\int_t^T Z^n_s dW_s-\int_t^T \int_E \psi_s^n(x)\wt{\mu}(ds,dx),\quad  t\in[0,T] \nn
\eea 
for each $n$. Then $Y^n\rightarrow Y$ in $\mbb{S}^\infty$, $Z^n\rightarrow Z$ in $\mbb{H}^2_{BMO}$
and $\psi^n\rightarrow \psi$ in $\mbb{J}^2_{BMO}$ where $(Y,Z,\psi)$ is a unique solution of (\ref{eq-Qexp-BSDE})
with data $(\xi,f)$.
\begin{proof}
Let us define two drivers such that $G^n:=\sup_{m\geq n} f^m$, $H^n:=\inf_{m\geq n} f^m$.
Then we have $G^n \downarrow f$, $H^n\uparrow f$ as $n\rightarrow \infty$. By condition (i), both $G^n$
and $H^n$ satisfy Assumptions~\ref{assumption-Qexp} and \ref{assumption-LLC} uniformly in $n$.
Moreover the relations $G^n(\cdot,\psi)-G^n(\cdot,\psi^\prime)\leq \sup_{m\geq n}\bigl[f^m(\cdot,\psi)-f^m(\cdot,\psi^\prime)]$
and $H^n(\cdot,\psi)-H^n(\cdot,\psi^\prime)\leq \sup_{m\geq n}\bigl[f^m(\cdot,\psi)-f^m(\cdot,\psi^\prime)\bigr]$
imply $A_\Gamma$-condition of Assumption~\ref{assumption-AGamma} holds uniformly.
Thus, by Theorem~\ref{theorem-existence}, there exists a unique solution $(Y^{n*},Z^{n*},\psi^{n*})$
(resp. $(Y^n_*,Z^n_*,\psi^n_*)$) in $S^\infty\times \mbb{H}^2_{BMO}\times \mbb{J}^2_{BMO}$
to the BSDEs with data $(\xi,G^n)$ (resp. $(\xi,H^n)$) for each $n$. 
By the local Lipschitz continuity, $A_\Gamma$-condition, and the universal bounds of the solutions
make the measure change used in the comparison principle well defined.
Hence,  by similar arguments of Lemma~\ref{lemma-fn-tilde}, it is straightforward to confirm that the 
comparison principle holds among $(Y^{n*},Y_*^n,Y^n)$. One has
$Y_*^n \leq Y^n\leq Y^{n*}$
for every $n\in \mbb{N}$. Furthermore, Proposition~\ref{prop-monotone} 
also imply the convergence $Y^{n*}\downarrow Y$
and $Y^n_*\uparrow Y$ in $\mbb{S}^\infty$. Thus we have $Y^n\rightarrow Y$ in $\mbb{S}^\infty$.
Remark~\ref{remark-strong-conv} gives the convergence of $Z^n,\psi^n$ in the desired norms.
\end{proof}
\end{proposition}

%%%%%%%%%%%%%%%%%%%%%%%%%%%%%%%%%%%%%%%%%%%%%%%%%%%%%%%%%%%%%%%%%%%%%
\section{Malliavin Differentiability}
%%%%%%%%%%%%%%%%%%%%%%%%%%%%%%%%%%%%%%%%%%%%%%%%%%%%%%%%%%%%%%%%%%%%%
In the reminder of the paper, we study the Malliavin differentiability of
the quadratic-exponential growth BSDEs.
Among the various ways to develop Malliavin's calculus, we follow 
the conventions based on the chaos expansion used in Delong \& Imkeller (2010)~\cite{Delong-Imkeller}
and Delong (2013)~\cite{Delong}, which were adopted from the work of Sol\'e et.al. (2007)~\cite{Sole}.
See also Di Nunno et.al. (2009)~\cite{Nunno} for an extension to a multi-dimensional setup and
other applications (with only a slight adjustment of conventions).
For the detailed conventions, see Section 3 of \cite{Delong-Imkeller}.
Following the extension given in Section 17 of \cite{Nunno}, 
we denote $(D^i_{t,0}, i\in\{1,\cdots,d\})$ and $(D^i_{t,z}, i\in\{1,\cdots,k\})$
as the Malliavin derivatives with respect to $(W_i(t), i\in\{1,\cdots,d\})$
and $(\wt{\mu}^i(dt,dz), i\in\{1,\cdots,k\})$, respectively.

Note that  a random variable $F$ is Malliavin differentiable if and only if 
$F\in\mbb{D}^{1,2}$. Here, the space $\mbb{D}^{1,2}\subset \mbb{L}^2(\mbb{P})$ is defined by
the completion with respect to the norm $||\cdot||_{1,2}$ which is given by
\bea
||F||_{1,2}^2:=\mbb{E}\Bigl[|F|^2\Bigr]+\sum_{i=1}^d\mbb{E}\left[
\int_0^T |D_{s,0}^i F|^2 ds\right]+\sum_{i=1}^k\mbb{E}
\left[\int_0^T\int_{\mbb{R}_0}|D_{s,z}^i F|^2 z^2\nu^i(dz)ds\right]~.\nn
\eea
For notational convenience, let us introduce two types of finite measures $m^i(dz)=\bold{1}_{z\neq 0} z^2\nu^i(dz)$ with $i\in\{1,\cdots,k\}$ defined on whole $\mbb{R}$, and $q$ defined on $\wt{E}:=[0,T]\times \mbb{R}^k$ by
\bea
&q(dt,dz):=\bold{1}_{z=0}dt+\sum_{i=1}^k m^i(dz)dt~.\nn 
\eea
We also introduce a space $\mbb{L}^{1,2}(\mbb{R}^n)$
of product measurable and $\mbb{F}$-adapted processes $\chi:\Omega\times[0,T]\times \mbb{R}^k\rightarrow \mbb{R}^n$
satisfying
\bea
&&\mbb{E}\Bigl[\int_{\wt{E}}|\chi(s,y)|^2q(ds,dy)\Bigr]<\infty,\nn \\
&&\chi(s,y)\in\mbb{D}^{1,2}(\mbb{R}^n), {\text{~~for $q$-a.e.~}} (s,y)\in\wt{E}, \nn \\
&&\mbb{E}\Bigl[\int_{\wt{E}}\int_{\wt{E}}|D_{t,z}\chi(s,y)|^2q(ds,dy)q(dt,dz)\Bigr]<\infty.\nn
\eea
Note that the space $\mbb{L}^{1,2}$ is a Hilbert space endowed with the norm
\bea
||\chi||^2_{\mbb{L}^{1,2}}:=\mbb{E}\Bigl[\int_{\wt{E}}|\chi(s,y)|^2q(ds,dy)\Bigr]+\mbb{E}\Bigl[
\int_{\wt{E}}\int_{\wt{E}}|D_{t,z}\chi(s,y)|^2q(ds,dy)q(dt,dz)\Bigr]~.\nn
\eea
The fact that the Malliavin derivative is a closed operator in $\mbb{L}^{1,2}$
(See, Theorem 12.6 in \cite{Nunno}) plays a crucial role later.

Suppose that $(t,z)$ is a jump of size $z$ at time $t$ in a random measure $\mu^i$.
We denote by $\omega_{\mu^i}^{t,z}$ a transformed family of $\omega_{\mu^i}=((t_1,z_1),(t_2,z_2),\cdots)\in\Omega_{\mu^i}$
into a new family with additional jump at $(t,z)$; $\omega_{\mu^i}^{t,z}=((t,z),(t_1,z_1),(t_2,z_2),\cdots)\in\Omega_{\mu^i}$~.
As for an element $\omega=(\omega_W,\omega_{\mu^1},\omega_{\mu^2},\cdots,\omega_{\mu^k})\in \Omega$ in the full canonical product space, 
we denote $\omega^{t,z}\in\Omega$ as the above transformation only in the corresponding element, such as $\omega^{t,z}=(\omega_W,\omega_{\mu^1},\cdots,\omega_{\mu^i}^{t,z},\cdots,\omega_{\mu^k})\in\Omega$
without specifying the relevant coordinate for notational simplicity.
By the same reason, we also frequently omit $i$ denoting the direction of derivative $D_{s,z}^i$
by assuming that we consider each Wiener $(z=0, i\in\{1,\cdots,d\})$ and jump $(z\neq 0, i\in\{1,\cdots,k\})$) direction separately (and summing them up whenever necessary, such as when considering 
integration on $\wt{E}$).

In this section, we consider Malliavin's differentiability of the following BSDE;
\bea
Y_t=\xi+\int_t^T f\Bigl(s,Y_s,Z_s,\int_{\mbb{R}_0}\rho(x)G(s,\psi_s(x))\nu(dx)\Bigr)ds-\int_t^T Z_s dW_s -\int_t^T \int_E \psi_x(x)\wt{\mu}(ds,dx),  \nn \\
\label{eq-Qexp-BSDE-M}
\eea
for $t\in[0,T]$ where $\xi:\Omega\rightarrow \mbb{R}$, $f:\Omega\times[0,T]\times \mbb{R}\times \mbb{R}^d\times \mbb{R}^k\rightarrow \mbb{R}$,
and $\rho^i:\mbb{R}\rightarrow \mbb{R}$, $G^i:[0,T]\times \mbb{R}\rightarrow \mbb{R}$ for each $i\in\{1,\cdots,k\}$.
The last arguments of the driver denotes a $k$-dimensional vector whose $i$-th element is given by
$\int_{\mbb{R}_0}\rho^i(x)G^i(s,\psi_s^i(x))\nu^i(dx)$. With slight abuse of notation, we adopt
$ \Theta_r:=\Bigl(Y_r,Z_r,\int_{\mbb{R}_0}\rho(z)G(r,\psi_r(z))\nu(dz)\Bigr)$, $r\in[0,T] $
as a collective argument in this section.

\begin{remark}
In Sol\'e et.al.~\cite{Sole} and Delong \& Imkeller~\cite{Delong-Imkeller},  the conventions
\bea
\psi(x) \rightarrow \psi(x)/x, \quad \wt{\mu}(dt,dx)\rightarrow x \wt{\mu}(dt,dx) \quad x\in\mbb{R}_0 \nn
\eea
are used. For the convenience when discussing the $\mbb{L}^{1,2}$-norm, we introduce the notation
$\overline{\phi}(x):=\phi(x)/x, x\in\mbb{R}_0$
for the control variables of the random measure, $\phi=\psi,\psi^m$ etc. See, in particular, Section~3.5 of \cite{Delong}.
\end{remark}

\begin{assumption}
\label{assumption-rho-G}
(i) For every $i\in\{1,\cdots,k\}$, $\rho^i$ is a continuous function satisfying $\int_{\mbb{R}_0} |\rho^i(x)|^2\nu^i(dx)<\infty$.
(ii) For every $i\in\{1,\cdots,k\}$, $G^i(s,v)$ is a continuous function in the both arguments and
one-time continuously differentiable with respect to $v$ with continuous derivative. Moreover, for every $R>0$, 
\bea
&&G_{R}:=\sup_{(s,v)\in[0,T]\times (|v|\leq R)}\sum_{i=1}^k |G^i(s,v)|<\infty, \quad G_{R}^\prime:=\sup_{(s,v)\in[0,T]\times (|v|\leq R)}\sum_{i=1}^k|\part_v G^i(s,v)|<\infty~.\nn
\eea
We put without loss of generality that $G^i(\cdot,0)=0$ for every $i\in\{1,\cdots,k\}$.
\end{assumption}

\begin{assumption} 
\label{assumption-Qexp-M}
The driver $F$ defined by $F(s,y,z,\psi):=f(s,y,z,\int_{\mbb{R}_0}\rho(x)G(s,\psi(x))\nu(dx))$
for $s\in[0,T],y\in \mbb{R},z\in\mbb{R}^d, \psi\in \mbb{L}^2(E,\nu;\mbb{R}^k)$ and the data $(\xi,l)$ satisfies both
Assumption~\ref{assumption-Qexp} and \ref{assumption-AGamma}.
\end{assumption}

\begin{assumption} 
\label{assumption-LLC-M}
For each $M>0$, and for every $(y,z,\psi), (y^\prime,z^\prime,\psi^\prime)\in\mbb{R}\times \mbb{R}^d\times \mbb{L}^2(E,\nu;\mbb{R}^k)$
satisfying 
$|y|, |y^\prime|, ||\psi||_{\mbb{L}^\infty(\nu)}, ||\psi^\prime||_{\mbb{L}^\infty(\nu)}\leq M$, 
there exists some positive constant $K_M$ possibly depending on $M$ such that
\bea
&&\bigl|f\bigl(t,y,z,u_t\bigr)-f\bigl(t,y^\prime,z^\prime,u_t^\prime \bigr)\bigr| \leq K_M\bigl(|y-y^\prime|+|u_t-u_t^\prime|\bigr)\nn \\
&&\qquad+K_M\bigl(1+|z|+|z^\prime|+|u_t|+|u_t^\prime|\bigr)|z-z^\prime|\nn
\eea
$d\mbb{P}\otimes dt$-a.e. $(\omega,t)\in \Omega\times [0,T]$, where we have used
$u_t:=\int_{\mbb{R}_0}\rho(x)G(t,\psi(x))\nu(dx)$ and $u_t^\prime:=\int_{\mbb{R}_0}\rho(x)G(t,\psi^\prime(x))\nu(dx)$
for notational simplicity.
\end{assumption}

%%%%%%%%%%%%%%%%%%%%%%%%%%%%
\begin{remark}
\label{remark-u}
%%%%%%%%%%%%%%%%%%%%%%%%%%%%
In the above assumption, using the fact that
\bea
&&|u_t|\leq ||\rho||_{\L2nu}G_M^\prime||\psi||_{\L2nu}, \quad|u_t-u_t^\prime|\leq ||\rho||_{\L2nu}G^\prime_{M}||\psi-\psi^\prime||_{\L2nu}~, \nn
\eea
one can see the consistency with Assumption~\ref{assumption-LLC}.
Therefore, under Assumptions~\ref{assumption-rho-G}, \ref{assumption-Qexp-M} and \ref{assumption-LLC-M}, there exists a
unique solution $(Y,Z,\psi)\in\mbb{S}^\infty\times \mbb{H}^2_{BMO}\times \mbb{J}^2_{BMO}$ to
the BSDE (\ref{eq-Qexp-BSDE-M}) by Theorem~\ref{theorem-existence}.
\end{remark}
For Malliavin differentiability, we need the following additional assumptions:
\begin{assumption}
\label{assumption-Qexp-MD}
With the notation $u_t=\int_{\mbb{R}_0}\rho(x)G(t,\psi(x))\nu(dx)$, $u^\prime_t=\int_{\mbb{R}_0}\rho(x)G(t,\psi^\prime(x))\nu(dx)$, \\
(i) The terminal value is Malliavin differentiable; $\xi\in\mbb{D}^{1,2}$. \\
(ii) For each $M>0$, and for every $(y,z,\psi)\in\mbb{R}\times \mbb{R}^d\times \mbb{L}^2(E,\nu;\mbb{R}^k)$
satisfying $
|y|, ||\psi||_{\mbb{L}^\infty(\nu)} \leq M$, the driver $\bigl(f(t,y,z,u_t),t\in[0,T]\bigr)$ belongs to $\mbb{L}^{1,2}(\mbb{R})$
and its Malliavin derivative is denoted by $(D_{s,z}f)(t,y,z,u_t)$. Furthermore, the driver $f$ is 
one-time  continuously differentiable with respect to its spacial variables with continuous derivatives.
 \\
(iii) For every Wiener as well as jump direction, for every $M>0$ and $d\mbb{P}\otimes dt$-a.e. $(\omega,t)\in\Omega\times [0,T]$, and 
for every $(y,z,\psi), (y^\prime,z^\prime,\psi^\prime)\in\mbb{R}\times \mbb{R}^d\times \mbb{L}^2(E,\nu;\mbb{R}^k)$
satisfying $|y|, |y^\prime|, ||\psi||_{\mbb{L}^\infty(\nu)}, ||\psi^\prime||_{\mbb{L}^\infty(\nu)}\leq M$,
the Malliavin derivative of the driver satisfies the following local Lipschitz conditions;
\bea
&&\bigl|(D_{s,0}^if)(t,y,z,u_t)-(D_{s,0}^if)(t,y^\prime,z^\prime,u^\prime)\bigr|\nn \\
&&\leq K_{s,0}^{M,i}(t)\bigl(|y-y^\prime|+|u_t-u_t^\prime|+(1+|z|+|z^\prime|+|u_t|+|u_t^\prime|)|z-z^\prime|\bigr)\nn
\eea
for $ds$-a.e. $s\in[0,T]$ with $i\in\{1,\cdots,d\}$, and 
\bea
&&\bigl|(D_{s,z}^if)(t,y,z,u_t)-(D_{s,z}^i f)(t,y^\prime,z^\prime,u_t^\prime)\bigr|\nn \\
&&\leq K_{s,z}^{M,i}(t)\bigl(|y-y^\prime|+|u_t-u_t^\prime|+(1+|z|+|z^\prime|+|u_t|+|u_t^\prime|)|z-z^\prime|\bigr)\nn
\eea
for $m^i(dz)ds$-a.e. $(s,z)\in[0,T]\times \mbb{R}_0$ with $i\in\{1,\cdots,k\}$.
For every $M>0$ and $(s,z)$,  $\bigl(K_{s,0}^{M,i}(t),t\in[0,T]\bigr)_{i\in\{1,\cdots,d\}}$ and $\bigl(K_{s,z}^{M,i}(t), 
t\in[0,T]\bigr)_{i\in\{1,\cdots,k\}}$ are $\mbb{R}_+$-valued $\mbb{F}$-progressively measurable processes.\\
(iv) There exists  some positive constant $p\geq 2$ such that
\bea
\int_{\wt{E}}\Bigl(\mbb{E}\Bigl[|D_{s,z}\xi|^{pq}+\Bigl(\int_0^T |(D_{s,z}f)(r,0)|dr\Bigr)^{pq}+||K_{s,z}^M||^{2pq}_T\Bigr]\Bigr)^{\frac{1}{q}}
q(ds,dz)<\infty \nn 
\eea
hold for $\forall q\geq 1$ and $\forall M>0$.
\end{assumption}
%%%%%%%%%%%%%%%%%%%%%%%%%%%%%%
\begin{remark}
\label{remark-dominated}
%%%%%%%%%%%%%%%%%%%%%%%%%%%%%%
Assumption~\ref{assumption-Qexp-MD} (iv) implies, for each $(s,z)$ in $\wt{E}$ $q(ds,dz)$-a.e., 
\bea
&\mbb{E}\Bigl[ |D_{s,z}\xi|^{p^\prime}+\Bigl(\int_0^T |(D_{s,z}f)(r,0)|dr\Bigr)^{p\prime}+
||K_{s,z}^M||_T^{2p^\prime}\Bigr]<\infty \nn
\eea
for $\forall p^\prime\geq 2$.  In particular, $K^M_{s,0}\in \mbb{S}^{p^\prime}$ for $ds$-a.e. $s\in[0,T]$
and $K^M_{s,z}\in \mbb{S}^{p^\prime}$ 
for $z^2\nu(dz)ds$-a.e. $(s,z)\in[0,T]\times \mbb{R}_0$ for $\forall p^\prime \geq 2$.
%It also implies that
%\bea
%&\int_0^T\int_{\mbb{R}}\Bigl(\mbb{E}\Bigl[|D_{s,z}\xi|^{pq}+\Bigl(\int_0^T |(D_{s,z}f)(r,0)|dr\Bigr)^{pq}+||K_{s,z}^M||^{2pq}_T\Bigr]\Bigr)^{\frac{1}{p}}\bold{1}_{\{|z|\leq \ep\}}m^i(dz)ds\nn \\
%&=\int_0^T\int_{|z|\leq \ep}\Bigl(\mbb{E}\Bigl[|D_{s,z}\xi|^{pq}+\Bigl(\int_0^T |(D_{s,z}f)(r,0)|dr\Bigr)^{pq}+||K_{s,z}^M||^{2pq}_T\Bigr]\Bigr)^{\frac{1}{p}}m^i(dz)ds \rightarrow 0 \nn 
%\eea
%as $\ep\downarrow 0$ by Lebesgue's dominated convergence for  $\forall i\in\{1,\cdots,k\}$, $\forall q\geq 1$ and $\forall M>0$.
\end{remark}

We now give the main result of this section.
\begin{theorem}
\label{theorem-Qexp-MD}
Suppose that Assumptions~\ref{assumption-rho-G}, \ref{assumption-Qexp-M}, \ref{assumption-LLC-M}
and \ref{assumption-Qexp-MD} hold true and denote the solution to the BSDE (\ref{eq-Qexp-BSDE-M}) 
as $(Y,Z,\psi)\in\mbb{S}^\infty\times \mbb{H}^2_{BMO}\times \mbb{J}^2_{BMO}$.
Then, the following statements hold:
(a) For each Wiener direction $i\in\{1,\cdots,d\}$ and $ds$-a.e. $s\in[0,T]$, there exists a 
unique solution $(Y^{s,0,i},Z^{s,0,i},\psi^{s,0,i})\in\calk^{p^\prime}[0,T]$ with $\forall p^\prime \geq 2$ to the BSDE
\bea
Y_t^{s,0,i}=D_{s,0}^i\xi+\int_t^T f^{s,0,i}(r)dr-\int_t^T Z_r^{s,0,i}dW_r
-\int_t^T \int_E \psi_r^{s,0,i}(x)\wt{\mu}(dr,dx)
\label{eq-Qexp-S0}
\eea
for $0\leq s\leq t\leq T$, where
\bea
f^{s,0,i}(r)&:=&(D_{s,0}^if)(r,\Theta_r)+\part_\Theta f(r,\Theta_r)\Theta_r^{s,0,i}\nn \\
&:=&(D_{s,0}^i f)(r,\Theta_r)+\part_y f(r,\Theta_r)Y_r^{s,0,i}+\part_z f(r,\Theta_r)Z_r^{s,0,i}\nn \\
&&+\part_u f(r,\Theta_r)\int_{E}\rho(x)\part_vG(r,\psi_r(x))\psi_r^{s,0,i}(x)\nu(dx)~.\nn
\eea
The solution also satisfies $\int_0^T ||(Y^{s,0,i},Z^{s,0,i},\psi^{s,0,i})||^p_{\calk^p[0,T]}ds<\infty~$.\\
(b) For each jump direction $i\in\{1,\cdots,k\}$ and $m^i(dz)ds$-a.e $(s,z)\in[0,T]\times \mbb{R}_0$,
there exists a unique solution $(Y^{s,z,i},Z^{s,z,i},\psi^{s,z,i})\in\mbb{S}^\infty\times \mbb{H}^2_{BMO}\times \mbb{J}^2_{BMO}$ 
to the BSDE
\bea
Y_t^{s,z,i}=D_{s,z}^i\xi+\int_t^T f^{s,z,i}(r)dr-\int_t^T Z_r^{s,z,i}dW_r-\int_t^T \int_E
\psi_r^{s,z,i}(x)\wt{\mu}(dr,dx)
\label{eq-Qexp-SZ}
\eea
for $0\leq s\leq t\leq T$ and $z\neq 0$, where
\bea
f^{s,z,i}(r)&:=&\frac{1}{z}\Bigl(f(\omega^{s,z},r,\Theta_r+z\Theta_r^{s,z,i})-f(\omega,r,\Theta_r)\Bigr):=\frac{1}{z}\Bigl\{f\Bigl(\omega^{s,z}, r,Y_r+zY_r^{s,z,i}\nn \\
&& ,Z_r+zZ_r^{s,z,i},\int_{\mbb{R}_0}\rho(x)G\bigl(r,\psi_r(x)+z\psi_r^{s,z,i}(x)\bigr)\nu(dx)\Bigr)-f(\omega,r,\Theta_r)\Bigr\}~.\nn
\eea
The solution also satisfies $ \int_0^T\int_{\mbb{R}} ||(Y^{s,z,i},Z^{s,z,i},\psi^{s,z,i})||^p_{\calk^p[0,T]}
m^i(dz)ds<\infty~$.\\
(c) The solution of the BSDE (\ref{eq-Qexp-BSDE-M}) is Malliavin differentiable $(Y,Z,\overline{\psi})\in\mbb{L}^{1,2}\times \mbb{L}^{1,2}
\times \mbb{L}^{1,2}$. Put, for every $i$, $Y_t^{s,\cdot,i}=Z_t^{s,\cdot,i}=\psi_t^{s,\cdot,i}(\cdot)\equiv 0$ for 
$t<s\leq T$, then $\bigl((Y_t^{s,z,i},Z_t^{s,z,i},\psi_t^{s,z,i}(x)), 0\leq s,t\leq T, x\in\mbb{R}_0, z\in\mbb{R}\bigr)$
is a version of the Malliavin derivative $\bigl((D_{s,z}^iY_t, D_{s,z}^iZ_t, D_{s,z}^i \psi_t(x)),0\leq s,t\leq T,
x\in\mbb{R}_0, z\in\mbb{R}\bigr)$ for every Wiener and jump direction.
\begin{proof}
Firstly, from Assumptions~\ref{assumption-rho-G}, \ref{assumption-Qexp-M} and \ref{assumption-LLC-M},
Theorem~\ref{theorem-existence} tells us that there exists a unique solution $(Y,Z,\psi)\in\mbb{S}^\infty\times 
\mbb{H}^2_{BMO}\times \mbb{J}^2_{BMO}$ to the BSDE (\ref{eq-Qexp-BSDE-M}). Since $||Y||_{\mbb{S}^\infty}, ||\psi||_{\mbb{J}^{\infty}}$
are bounded by the universal bounds,
one can choose a  constant $M>0$ big enough so that the local Lipschitz conditions hold true
for the whole relevant range. We choose one such $M$ and fix it throughout the proof.
We also omit the superscript $i$ denoting the direction of derivative
by assuming that we always discuss each direction separately.
\\ \\
%%%%%%%%%%%%%%%%%%%%%%%%%%%%%%%%%%%%%%%%
{\it Proof for (a):}
Firstly, the continuous differentiability of $f$ and 
the local Lipschitz conditions imply that, for the relevant range of variables,
\bea
&|\part_y f(t,y,z,u_t)|\leq K_M, \quad |\part_u f(t,y,z,u_t)|\leq K_M, \nn \\
&|\part_z f(t,y,z,u_t)|\leq K_M(1+2|z|+2|u_t|)~. 
\label{eq-deriv-estimate}
\eea
It is easy to check that the BSDE (\ref{eq-Qexp-S0}) satisfies Assumption~\ref{assumption-BMO-BSDE}. 
Indeed, its second condition follows from the relation
\bea
|(D_{s,0}f)(r,\Theta_r)|&\leq& |(D_{s,0}f)(r,0)|+ K_{s,0}^M(|Y_r|+||\rho||_{\L2nu}G^\prime_M||\psi_r||_{\L2nu})\nn \\
&&+K_{s,0}^M(1+|Z_r|+||\rho||_{\L2nu}G^\prime_M||\psi||_{\L2nu})|Z_r|~, \nn
\eea
Lemma~\ref{lemma-energy}  and Remark~\ref{remark-dominated}.
Thus, Theorem~\ref{th-BMO-existence} implies that there exists a unique solution
$(Y^{s,0},Z^{s,0},\psi^{s,0})\in\calk^{p^\prime}_{[0,T]}$ to the BSDE (\ref{eq-Qexp-S0}) satisfying 
\bea
&&||(Y^{s,0},Z^{s,0},\psi^{s,0})||^{p^\prime}_{\calk^{p^\prime}}
\leq C_{p^\prime}\Bigl(1+\mbb{E}\Bigl[ |D_{s,0}\xi|^{p^\prime \bar{q}^2}+\Bigl(\int_0^T |(D_{s,0}f)(r,0)|dr\Bigr)^{p^\prime \bar{q}^2}+||K_{s,0}^M||
^{2p^\prime \bar{q}^2}_T \nn \\
&&\quad+||Y||^{2p^\prime \bar{q}^2}_T+\Bigl(\int_0^T |Z_r|^2dr\Bigr)^{2p^\prime \bar{q}^2}+\Bigl(\int_0^T ||\psi_r||^2_{\L2nu}dr\Bigr)^{2p^\prime \bar{q}^2}
\Bigr]\Bigr)^{\frac{1}{\bar{q}^2}}<\infty, \nn
\eea
%where $C$ is a positive constant depending only on $(p^\prime,\bar{q},\beta,\gamma,T,||\xi||_{\infty},||l||_{\mbb{S}^\infty},K_M)$
%and $\bar{q}$ is a positive constant satisfying $1<q_*\leq \bar{q}<\infty$ where the lower bound $q_*$
%is an increasing function of the $\mbb{H}^2_{BMO}$-norm of $\part_z f(\cdot, \Theta_\cdot)$, which is 
%also controlled by the universal bound given by $(\beta,\gamma, T,||\xi||_{\infty},||l||_{\mbb{S}^\infty})$.
for $\forall p^\prime\geq 2$, where $C_{p^\prime}$ and $\bar{q}>1$ are positive constants.
Assumption~\ref{assumption-Qexp-MD} (iv) also gives the
2nd claim $\int_0^T ||(Y^{s,0},Z^{s,0},\psi^{s,0})||^p_{\calk^p[0,T]}ds<\infty~$.
\\ \\
%%%%%%%%%%%%%%%%%%%%%%%%%%%%%%%%%%%%%%%%%%%%%%
{\it Proof for (b):} Let us first consider the BSDE
\bea
&&\caly^{s,z}_t=\xi(\omega^{s,z})+\int_t^T f\Bigl(\omega^{s,z},r,\caly^{s,z}_r,\calz^{s,z}_r,
\int_{\mbb{R}_0}\rho(x)G(r,\Psi^{s,z}_r(x))\nu(dx)\Bigr)dr\nn \\
&&-\int_t^T \calz_r^{s,z}dW_r-\int_t^T \int_E\Psi_r^{s,z}(x)\wt{\mu}(dr,dx)~.
\label{eq-aggregate}
\eea
For every $(s,z)\in[0,T]\times \mbb{R}_0$, $m(dz)ds$-a.e,  Assumptions \ref{assumption-rho-G}, \ref{assumption-Qexp-M},
\ref{assumption-LLC-M} are all satisfied. 
Thus, by Theorem~\ref{theorem-existence}, there exists a unique solution $(\caly^{s,z},\calz^{s,z},\Psi^{s,z})
\in\mbb{S}^\infty\times \mbb{H}^2_{BMO}\times \mbb{J}^2_{BMO}$ to the BSDE (\ref{eq-aggregate}) satisfying the 
universal bounds.
Now, let us define for $z\in\mbb{R}_0$,
\bea
Y^{s,z}:=\frac{\caly^{s,z}-Y}{z},\quad Z^{s,z}:=\frac{\calz^{s,z}-Z}{z}, \quad \psi^{s,z}:=
\frac{\Psi^{s,z}-\psi}{z}~, \nn
\eea
and then $(Y^{s,z},Z^{s,z},\psi^{s,z})\in\mbb{S}^\infty\times \mbb{H}^2_{BMO}\times \mbb{J}^2_{BMO}$ 
is the unique solution to the BSDE (\ref{eq-Qexp-SZ}).
Note that $D_{s,z}\xi:=\frac{1}{z}(\xi(\omega^{s,z})-\xi(\omega))$.
%The uniqueness easily follows from that of $(Y,Z,\psi)$ and $(\caly^{s,z},\calz^{s,z},\Psi^{s,z})$.

We use a new collective argument 
$\Xi^{s,z}:=\bigl(\caly^{s,z},\calz^{s,z},\int_{\mbb{R}_0}\rho(x)G(r,\Psi^{s,z}_r(x))\nu(dx)\bigr)$.
Let us introduce
\bea
f^{s,z}(r)&:=&\frac{1}{z}\bigl(f(\omega^{s,z},r,\Xi^{s,z})-f(\omega,r,\Theta_r)\bigr)\nn \\
&=&(D_{s,z}f)(r,\Theta_r)+\frac{f(\omega^{s,z},r,\Xi^{s,z}_r)-f(\omega^{s,z},r,\Theta_r)}{z}~, \nn
\eea
a $d$-dimensional $\mbb{F}$-progressively measurable process $(b^{s,z}_r,r\in[0,T])$,
\bea
b^{s,z}_r(\omega)&:=&\frac{1}{|\calz_r^{s,z}-Z_r|^2}
\Bigl\{ f\Bigl(\omega^{s,z},r,Y_r,\calz_r^{s,z},\int_{\mbb{R}_0}\rho(x)G(r,\psi_r(x))\nu(dx)\Bigr) \nn \\
&&- f\Bigl(\omega^{s,z},r,Y_r,Z_r,\int_{\mbb{R}_0}\rho(x)G(r,\psi_r(x))\nu(dx)\Bigr)\Bigr\}\bold{1}_{\calz_r^{s,z}-Z_r\neq 0}
(\calz_r^{s,z}-Z_r) \nn
\eea
and also the map $\wt{f}^{s,z}:\Omega\times [0,T]\times \mbb{R}\times \mbb{L}^2(E,\nu;\mbb{R}_k) \rightarrow \mbb{R}$, 
\bea
\wt{f}^{s,z}(\omega,r,\wt{y},\wt{\psi})&:=&(D_{s,z}f)(r,\Theta_r)+\frac{1}{z}\Bigl\{f\Bigl(\omega^{s,z},r,z\wt{y}+Y_r,\calz_r^{s,z},\int_{\mbb{R}_0}
\rho(x)G(r,z\wt{\psi}(x)+\psi_r(x))\nu(dx)\Bigr)\nn \\
&&-f\Bigl(\omega^{s,z},r,Y_r,\calz_r^{s,z},\int_{\mbb{R}_0}\rho(x)G(r,\psi_r(x))\nu(dx)\Bigr)\Bigr\}\nn~.
\eea
Then, $(Y^{s,z},Z^{s,z},\psi^{s,z})$ can also be expressed as a solution to the BSDE
\bea
Y_t^{s,z}=D_{s,z}\xi+\int_t^T\Bigl(\wt{f}^{s,z}(r,Y_r^{s,z},\psi_r^{s,z})+b_r^{s,z}\cdot Z_r^{s,z}\Bigr)dr
-\int_t^T Z_r^{s,z}dW_r-\int_t^T \int_E \psi_r^{s,z}(x)\wt{\mu}(dr,dx)~.\nn
\eea
%Note that $|b_r|\leq H_r$, $r\in[0,T]$ where $H_r:=K_M\Bigl(1+|\calz_r|^{s,z}+|Z_r|+2||\rho||_{\L2nu}G_M^\prime ||\psi_r||_{\L2nu}\Bigr)$
%and that $H\in\mbb{H}^2_{BMO}$. Furthermore, the new driver satisfies  
%a linear growth property $|\wt{f}(r,\bar{y},\bar{\psi})|\leq |(D_{s,z}f)(r,\Theta_r)|+
%K_M\Bigl(|\bar{y}|+||\rho||_{\L2nu}G_M^\prime ||\bar{\psi}||_{\L2nu}\Bigr)$.
It is straightforward to check that Assumption~\ref{assumption-apriori-bmolike} is satisfied. Thus, Lemma~\ref{lemma-bmolike-apriori} 
gives
\bea
&&||(Y^{s,z},Z^{s,z},\psi^{s,z})||^{p^\prime}_{\calk^{p^\prime}}
%\leq C\left(
%\mbb{E}\left[|D_{s,z}\xi|^{p^\prime \bar{q}^2}+\Bigl(\int_0^T |(D_{s,z}f)(r,\Theta_r)|dr\Bigr)^{p^\prime \bar{q}^2}\right]
%\right)^{\frac{1}{\bar{q}^2}}\nn \\
\leq C_{p^\prime} \Bigl(1+\mbb{E}\Bigl[ |D_{s,z}\xi|^{p^\prime \bar{q}^2}+\Bigl(\int_0^T |(D_{s,z}f)(r,0)|dr\Bigr)^{p^\prime \bar{q}^2}
+||K_{s,z}^M||^{2p^\prime\bar{q}^2}_T \nn \\
&&\qquad +||Y||^{2p^\prime \bar{q}^2}_T+\Bigl(\int_0^T |Z_r|^2dr\Bigr)^{2p^\prime \bar{q}^2}+
\Bigl(\int_0^T ||\psi_r||^2_{\L2nu}dr\Bigr)^{2p^\prime \bar{q}^2}\Bigr]\Bigr)^{\frac{1}{\bar{q}^2}}<\infty~\nn 
\eea
for $\forall p^\prime \geq 2$, 
%where a positive constant $C$ depending only on $(p^\prime,\bar{q},\beta,\gamma,T,||\xi||_{\infty},||l||_{\mbb{S}^\infty},K_M)$
%and $\bar{q}$ is a positive constant satisfying $q_*\leq \bar{q}<\infty$ where $q_*>1$ is determined by $||b^{s,z}||_{\mbb{H}^2_{BMO}}$.
where $C_{p^\prime}$ and $\bar{q}>1$ are the positive constants.
Choosing $p^\prime=p$, one can show 
$\int_0^T\int_{\mbb{R}}||(Y^{s,z},Z^{s,z},\psi^{s,z})||^p_{\calk^p}m(dz)ds<\infty$
from Assumption~\ref{assumption-Qexp-MD} (iv), which proves the second claim of (b).
Note that, we also have
$\int_{\wt{E}}||(Y^{s,z},Z^{s,z},\psi^{s,z})||^p_{\calk^p}q(ds,dz)<\infty$
by combining the results (a) and (b).
\\ \\
%%%%%%%%%%%%%%%%%%%%%%%%%%%%%%%%%%%%%%%%%%%%%%%%%%%%
{\it Proof for (c): First step (Approximating sequence of globally Lipschitz BSDEs)}\\
%%%%%%%%%%%%%%%%%%%%%%%%%%%%%%%%%%%%%%
We finally proceed to the proof for (c). Firstly, let us define for each $m\in \mbb{N}$
\bea
G_m(s,\psi(x)):=G(s,\varphi_m(\psi \circ \zeta_m(x)), \quad f_m(s,y,z,u):=f(s,\varphi_m(y),\varphi_m(z),u)\nn
\eea
where $\varphi_m$ is the smooth truncation function defined in (\ref{eq-varphi}),
and  $\psi\circ\zeta_m(x):=\psi(x)\bold{1}_{|x|\geq 1/m}$, which are applied component-wise for $z$ and $\psi$.
Let us now define a sequence of regularized drivers $(F_m,m\in\mbb{N})$ by
$\displaystyle F_m(s,y,z,\psi):=f_m\bigl(s,y,z,\int_{\mbb{R}_0}\rho(x)G_m(s,\psi(x))\nu(dx)\bigr)$ for $s\in[0,T], y\in \mbb{R},z\in \mbb{R}^d, \psi\in \mbb{L}^2(E,\nu;\mbb{R}^k)$.
Note that
\be
||\varphi_m(\psi\circ\zeta_m)||_{\mbb{L}^2(\nu)}^2=\int_E |\varphi_m(\psi\circ\zeta_m(x))|^2\nu(dx)\leq (m+1)^2 C_m \nn
\ee
where $C_m:=k \max_{1\leq i\leq k}\int_{\mbb{R}_0}\bold{1}_{|x|\geq 1/m}\nu^i(dx)$.
Combined with Assumption~\ref{assumption-LLC-M} and Remark~\ref{remark-u},
one sees $F_m$ is globally Lipschitz for each $m\in\mbb{N}$.
One can also check $|F_m|$ is bounded.
Thus, for each $m\in \mbb{N}$, there exists a unique solution $(Y^m,Z^m,\psi^m)$
of the BSDE 
\bea
Y^m=\xi+\int_t^T F_m(s,Y^m_s,Z^m_s,\psi^m_s) ds-\int_t^T Z_s^m dW_s-\int_t^T \int_E \psi_s^m(x)\wt{\mu}(ds,dx), 
\label{eq-app-BSDE-M}
\eea
with $Y^m\in\mbb{S}^\infty$. Moreover, the convexity of positive function $j_\gamma(\cdot)$ and Assumption~\ref{assumption-Qexp-M}
imply that $F_m$ satisfy the $Q_{\exp}$-structure condition {\it uniformly} in $m$. 
Therefore, $(Y^m,Z^m,\psi^m)$ satisfies 
the universal bounds of Lemmas~\ref{lemma-BMO-bound} and \ref{lemma-universal-bound}. Since $||Y^m||_{\mbb{S}^\infty}$
and $||\psi^m||_{\mbb{J}^\infty}$ are bounded uniformly in $m$, the truncation $\varphi_m$ for $(y,\psi)$ becomes
irrelevant provided $m$ is large enough.
Thus, for large $m$, $(Y^m,Z^m,\psi^m)$
also consists of a unique bounded solution\footnote{ Using the universal bounds, uniqueness is checked similarly 
as in the standard Lipschitz BSDE.} to the BSDE with data $(\xi,\wt{F}_m)$
where 
\be 
\wt{F}_m(s,y,z,\psi):=f\Bigl(s,y,\varphi_m(z),\int_{\mbb{R}_0}\rho(x)G(s,\psi\circ\zeta_m(x))\nu(dx)\Bigr)~.\nn
\ee
Since $(\wt{F}_m)$ satisfies $A_\Gamma$-condition uniformly in $m$,
and also $\wt{F}_m \rightarrow F$ locally uniformly in the spacial variables, 
Proposition~\ref{prop-conv-general} implies $Y^m\rightarrow Y$ in $\mbb{S}^\infty$, $Z^m\rightarrow Z$ in $\mbb{H}^2_{BMO}$
and $\psi^m\rightarrow \mbb{\psi}$ in $\mbb{J}^2_{BMO}$ where $(Y,Z,\psi)$ is a unique solution of the BSDE (\ref{eq-Qexp-BSDE-M}).
One can also check that, for each $m\in\mbb{N}$, the BSDE (\ref{eq-app-BSDE-M}) 
satisfies Assumptions~\ref{assumption-Lipschitz} as well as \ref{assumption-M-Lipschitz}.
Therefore Theorem~\ref{theorem-Malliavin-Lipschitz} implies that 
the approximating BSDEs are Malliavin differentiable and
$(Y^m,Z^m,\overline{\psi}^m)\in (\mbb{L}^{1,2})^3$ for $\forall m\in\mbb{N}$.
\\
\\
%%%%%%%%%%%%%%%%%%%%%%%%%%%%%%%%%%%%%%%%%%%%%%%%%%%%%%%%%%%%%%%%%%%%%%%
{\it Second step (Uniform boundedness of $\mbb{L}^{1,2}$-norm of the approximating BSDEs)}\\
From the {\it first step}, one can define the Malliavin derivatives of $(Y^m, Z^m,\psi^m)$ for every $m\in\mbb{N}$
as the solution to the following BSDEs:
For every Wiener direction $i\in\{1,\cdots,d\}$, $ds$-a.e. $s\in[0,T]$ and $s\leq t\leq T$,
\bea
&&D_{s,0}^i Y_t^m=D_{s,0}^i\xi+\int_t^T D_{s,0}^i f_m(r)dr-\int_t^T D_{s,0}^i Z_r^m dW_r-
\int_t^T \int_E D_{s,0}^i \psi^m_r(x)\wt{\mu}(dr,dx), \nn \\
&&D_{s,0}^i f_m(r):=(D_{s,0}f_m)(r,\Theta_r^m)+\part_\Theta f_m(r,\Theta^m_r)D_{s,0}^i \Theta_r^m,
\label{eq-DS0-Ym}
\eea
and for jump direction $i\in\{1,\cdots,k\}$, $m^i(dz)ds$-a.e. $(s,z)\in[0,T]\times \mbb{R}_0$ and $s\leq t\leq T$,
\bea
&& D_{s,z}^i Y_t^m=D_{s,z}^i\xi+\int_t^T D_{s,z}^i f_m(r)dr-\int_t^T D^i_{s,z} Z_r^m dW_r
-\int_t^T \int_E \psi^m_r(x)\wt{\mu}(dr,dx),\nn \\
&& D_{s,z}^i f_m(r):=\frac{1}{z}\bigl(f_m(\omega^{s,z},r,\Theta_r^m+z D_{s,z}^i \Theta^m_r)-f_m(\omega,r,\Theta^m_r)\bigr)\nn \\
&&\hspace{5mm}=(D_{s,z}^if_m)(r,\Theta^m_r)+\frac{1}{z}\bigl(f_m(\omega^{s,z},r,\Theta^m_r+z D_{s,z}^i\Theta^m_r)-f_m(\omega^{s,z},r,\Theta^m_r)\bigr)~.
\label{eq-DSZ-Ym}
\eea
Here, we have defined $
\Theta^m_r:=\bigl(Y^m_r,Z^m_r,\int_{\mbb{R}_0}\rho(x)G_m(r,\psi_r^m(x))\nu(dx)\bigr)$ for $r\in[0,T]$
and slightly abused its notation in such a way that
$\displaystyle f_m(\omega^{s,z},r,\Theta_r^m+zD_{s,z}^i\Theta_r^m)
:=f_m\bigl(\omega^{s,z},r,Y_r^m+zD_{s,z}^i Y_r^m,~
Z_r^m+zD_{s,z}^i Z_r^m ,\int_{\mbb{R}_0}\rho(x)G_m\bigl(r,\psi_r^m(x)+z D_{s,z}^i \psi_r^m(x)\bigr)\nu(dx)\bigr)$
to save the space. For $0\leq t<s$, one has $D_{s,z}\Theta^m_t\equiv 0$.

One can check that the unique solution of (\ref{eq-DS0-Ym})
satisfies $(D_{s,0}Y^m,D_{s,0}Z^m,D_{s,0}\psi^m)\in\calk^{p^\prime}[0,T]$ for $\forall p^\prime\geq 2$
by Theorem~\ref{th-BMO-existence}.
Let us also define (for each direction $i\in\{1,\cdots,k\}$)
\bea
\caly^m_{s,z}(t):=Y^m_t+z D_{s,z}Y^m_t, \quad \calz^m_{s,z}(t):=Z^m_t+z D_{s,z}Z^m_t,\quad \Psi^m_{s,z}(t,\cdot):=\psi^m_t(\cdot)+z D_{s,z}\psi^m_t(\cdot)~,\nn
\eea
for $(s,z)\in[0,T]\times \mbb{R}_0$ and $t\in[0,T]$, and denote its collective argument as
$\Xi^m_{s,z}(t):=\bigl(\caly^m_{s,z}(t),\calz^m_{s,z}(t),\int_{\mbb{R}_0}\rho(x)G_m(t,\Psi^m_{s,z}(t,x))\nu(dx)\bigr)$. 
Note that $(\caly^m_{s,z},\calz^m_{s,z}, \Psi^m_{s,z})$ is a solution 
to a Lipschitz BSDE (\ref{eq-aggregate}) with $f, G$ replaced by $f_m, G_m$.
Since it satisfies the structure condition uniformly in $m$, 
$(\caly^m_{s,z},\calz^m_{s,z},\Psi^m_{s,z})$ satisfies the  universal bounds.
It then shows $(D_{s,z}Y^m,D_{s,z}Z^m,D_{s,z}\psi^m)\in \mbb{S}^\infty\times \mbb{H}^2_{BMO}\times \mbb{J}^2_{BMO}$ for $z\neq 0$.
Moreover, by the same analysis given in the {\it first step},  one observes the convergence
$(\caly^m_{s,z},\calz^m_{s,z},\Psi^m_{s,z})\rightarrow (\caly^{s,z},\calz^{s,z},\Psi^{s,z})$
in the space $\mbb{S}^\infty\times \mbb{H}^2_{BMO}\times \mbb{J}^2_{BMO}$.

By the same arguments used in the proofs for (a) and (b),
one can apply Theorem~\ref{th-BMO-existence} to the BSDE (\ref{eq-DS0-Ym}) and
Lemma~\ref{lemma-bmolike-apriori} to the BSDE (\ref{eq-DSZ-Ym})
to obtain 
\bea
&&\hspace{-5mm}\bigl|\bigl|(D_{s,z}Y^m,D_{s,z}Z^m,D_{s,z}\psi^m)\bigr|\bigr|^{p^\prime}_{\calk^{p^\prime}[0,T]} \nn \\
% \leq C\left(\mbb{E}\left[ |D_{s,z}\xi|^{p^\prime \bar{q}^2}+\Bigl(\int_0^T |(D_{s,z}f_m)(r,\Theta_r^m)|dr\Bigr)^{p^\prime \bar{q}^2}
%\right]\right)^{\frac{1}{\bar{q}^2}}\nn \\
&&\quad \leq C_{p^\prime}\Bigl(1+\mbb{E}\Bigl[ |D_{s,z}\xi|^{p^\prime \bar{q}^2}+\Bigl(\int_0^T |(D_{s,z}f)(r,0)|dr\Bigr)^{p^\prime \bar{q}^2}+
||K_{s,z}||^{2p^\prime \bar{q}^2}_T \nn \\
&&\qquad \qquad +||Y^m||^{2p^\prime \bar{q}^2}_T+\Bigl(\int_0^T |Z_r^m|^2dr\Bigr)^{2p^\prime \bar{q}^2}+\Bigl(\int_0^T
||\psi^m_r||^2_{\L2nu}dr\Bigr)^{2p^\prime \bar{q}^2}\Bigr]\Bigr)^{\frac{1}{\bar{q}^2}} \nn
\eea
with $\forall p^\prime\geq 2$,  for the Wiener ($z=0$) as well as the jump ($z\neq 0$) directions. 
Here, $C_{p^\prime}$ and $\bar{q}>1$ are positive constants independent of $m$. 
Assumption~\ref{assumption-Qexp-MD} (iv), the universal bounds for $\Theta^m$ and the energy inequality give
\bea
\int_{\wt{E}}~\sup_{m\in\mbb{N}}\bigl|\bigl|(D_{s,z}Y^m,D_{s,z}Z^m,D_{s,z}\psi^m)\bigr|\bigr|^p_{\calk^p[0,T]}q(ds,dz)<\infty~.
\label{eq-Dsz-Ym-L12-bound}
\eea
It then easily follows that $\mbb{L}^{1,2}$-norm of $(Y^m,Z^m,\overline{\psi}^m)$ is bounded uniformly in $m$.
%Remark~\ref{remark-dominated} and the fact that $(\Theta^m)_{m\in\mbb{N}}$ satisfy the universal bound
%also imply that the convergence
The estimate (\ref{eq-Dsz-Ym-L12-bound}) also gives
\bea
&&\sum_{i=1}^k \int_0^T\int_{|z|>\ep}\bigl|\bigl|(D_{s,z}^iY^m,D_{s,z}^iZ^m,D_{s,z}^i\psi^m)\bigr|\bigr|^p_{\calk^p[0,T]}
m^i(dz)ds \nn \\
&&\rightarrow \sum_{i=1}^k \int_0^T\int_{\mbb{R}_0}\bigl|\bigl|(D_{s,z}^iY^m,D_{s,z}^iZ^m,D_{s,z}^i\psi^m)\bigr|\bigr|^p_{\calk^p[0,T]}
m^i(dz)ds 
\label{eq-Dsz-ep-conv}
\eea
as $\ep\downarrow 0$ uniformly in $m\in\mbb{N}$ by the Lebesgue's dominated convergence theorem. 
%(See the discussion given just below (\ref{eq-uniform-ep}). Thanks to the 
%universal bound, the arguments are much simpler here.)
%Here, $C$ is a positive constant depending only on $(p^\prime,\bar{q},\beta,\gamma,T,||\xi||_{\infty},||l||_{\mbb{S}^\infty},K_M)$
%and $\bar{q}$ is a positive constant satisfying $1<q_*\leq \bar{q}<\infty$ where the lower bound $q_*$
%is an increasing function of the $\mbb{H}^2_{BMO}$-norm of $D_{s,z}Z^m$'s coefficients, which are 
%also controlled by the universal bound given by $(\beta,\gamma, T,||\xi||_{\infty},||l||_{\mbb{S}^\infty})$.
%In particular, they are independent of $m$.
%For $\forall p^\prime\geq 2$, due to the energy inequality in Lemma~\ref{lemma-energy}, together with Lemmas~\ref{lemma-universal-bound} %and \ref{lemma-BMO-bound}
%imply that the right-hand side of (\ref{eq-apriori-DszYm}) is bounded by some positive constant independent of $m$
%for $q(ds,dz)$-a.e. $(s,z)\in\wt{E}$. 
%In particular, $D_{s,z}\Theta^m \in \calk^{p^\prime}[0,T]$ for $\forall p^\prime\geq 2$.
\\
%%%%%%%%%%%%%%%%%%%%%%%%%%%%%%%%%%%%%%%%%%%%%%%%%%%%%%%%%%%%%%%%%%%%%%%%%%%%%%
\\
{\it Third step (Convergence of $D_{s,0}\Theta^m\rightarrow \Theta^{s,0}$)}\\
For $ds$-a.e. $s\in[0,T]$ and $m\in\mbb{N}$, set
\bea
\Del^{s,0}Y^m:=Y^{s,0}-D_{s,0}Y^m, \quad \Del^{s,0}Z^m:=Z^{s,0}-D_{s,0}Z^m, \quad \Del^{s,0} \psi^m:=\psi^{s,0}-D_{s,0}\psi^m\nn
\eea
and then $(\Del^{s,0} Y^m, \Del^{s,0}Z^m, \Del^{s,0}\psi^m)\in\calk^{p^\prime}[0,T]$ with 
$\forall p^\prime\geq 2$ is the unique solution to the BSDE
\bea
\Del^{s,0}Y^m_t=\int_t^T \Bigl(f^{s,0}(r)-D_{s,0}f_m(r)\Bigr)dr-\int_t^T \Del^{s,0}Z_r^m dW_r-
\int_t^T\int_E \Del^{s,0}\psi^m_r(x)\wt{\mu}(dr,dx)~.\nn
\eea
We claim
\be
\lim_{m\rightarrow \infty}\int_0^T \bigl|\bigl|(\Del^{s,0}Y^m, \Del^{s,0}Z^m, \Del^{s,0}\psi^m)\bigr|\bigr|^p_{\calk^p[0,T]}ds
=0~.
\label{eq-Ds0Theta-conv}
\ee
The proof is straightforward and we give the details in Appendix~\ref{app-QMD}.
\\
\\
%%%%%%%%%%%%%%%%%%%%%%%%%%%%%%%%%%%%%%%%%%%%%%%%%%%%%%%%%%%%%%%%
{\it Fourth step (Convergence of $D_{s,z}\Theta^m\rightarrow \Theta^{s,z}~(z\neq 0)$)}\\
For each direction of jump, let us put 
\bea
\Del^{s,z} Y^m:=Y^{s,z}-D_{s,z}Y^m,\quad \Del^{s,z} Z^m=Z^{s,z}-D_{s,z} Z^m, \quad \Del^{s,z} \psi^m=\psi^{s,z}-D_{s,z}\psi^m~.\nn
\eea
Then, $(\Del^{s,z} Y^m, \Del^{s,z} Z^m, \Del^{s,z} \psi^m)\in\mbb{S}^\infty\times \mbb{H}^2_{BMO}\times \mbb{J}^2_{BMO}$
is the unique solution to
\bea
\Del^{s,z}Y^m_t=\int_t^T \Bigl(f^{s,z}(r)-D_{s,z}f_m(r)\Bigr)dr-\int_t^T \Del^{s,z}Z^m_r dW_r
-\int_t^T\int_E \Del^{s,z}\psi^m_r(x)\wt{\mu}(dr,dx)~,\nn
\eea
with $t\in[0,T]$. 
As in the third step, we claim
\bea
\lim_{m\rightarrow 0}\int_0^T\int_{\mbb{R}_0} \bigl|\bigl| (\Del^{s,z} Y^m,\Del^{s,z} Z^m, \Del^{s,z} \psi^m)\bigr|\bigr|^p_{\calk^p[0,T]}m(dz)ds=0
\label{eq-DszTheta-conv}.
\eea
The proof is tedious but straightforward and we give the details in Appendix~\ref{app-QMD2}.\\ \\
%%%%%%%%%%%%%%%%%%%%%%%%%%%%%%
{\it Final step}\\
From the previous steps, one sees $(Y^m, Z^m,\overline{\psi}^m)$
converges to $\bigl((Y,Z,\overline{\psi}),(Y^{s,z}, Z^{s,z},\overline{\psi}^{s,z})\bigr)$
in $\mbb{L}^2(0,T;\mbb{D}^{1,2})=\mbb{L}^{1,2}$.
The closability of the Malliavin derivatives in $\mbb{L}^{1,2}$
(See Theorem 12.6 in \cite{Nunno}.), one concludes $(Y,Z,\overline{\psi})\in\mbb{L}^{1,2}$
and that $(Y^{s,z},Z^{s,z},\psi^{s,z})$ is a version of $(D_{s,z}Y,D_{s,z}Z,D_{s,z}\psi)$.
\end{proof}
\end{theorem}

\begin{corollary}
\label{corollary-version}
Under the assumptions of Theorem~\ref{theorem-Qexp-MD}, we have \\
(i) $\Bigl((D_{t,0}^iY_t)^{\calp},t\in[0,T]\Bigr)$ is a version of $\Bigl(Z^i_t,t\in[0,T]\Bigr)$ for $i\in\{1,\cdots,d\}$,\\
(ii) $\Bigl( (zD_{t,z}^iY_t)^{\calp},(t,z)\in[0,T]\times \mbb{R}_0\Bigr)$ is a version of $\Bigl(\psi^i_t(z), (t,z)\in[0,T]\times \mbb{R}_0\Bigr)$ for $i\in\{1,\cdots,k\}$, \\
where $(\cdot)^{\calp}$ denotes the predictable projection of a process.
\begin{proof}
See Corollory 4.1 in \cite{Delong-Imkeller}.
\end{proof}  
\end{corollary}

%%%%%%%%%%%%%%%%%%%%%%%%%%%%%%%%%%%%%%%%%%%%%%
\section{An application: Markovian forward-backward system}
%%%%%%%%%%%%%%%%%%%%%%%%%%%%%%%%%%%%%%%%%%%%%%
\subsection{Forward SDE}
As an important application, we consider a $Q_{\exp}$-growth BSDE driven by 
an $n$-dimensional Markovian process $\bigl(X^{t,x}_s,s\in[0,T]\bigr)$ defined by the next SDE:
\bea
X_s^{t,x}=x+\int_t^s b(r,X_r^{t,x})dr+\int_t^s \sigma(r,X_r^{t,x})dW_r+\int_t^s \int_E \gamma(r,X_{r-}^{t,x},e)\wt{\mu}(dr,de)
\eea
for $s\in[t,T]$ and put $X_s^{t,x}\equiv x$ for $s<t$.
Here, $x\in\mbb{R}^n$, $b:[0,T]\times \mbb{R}^n\rightarrow \mbb{R}^n$, $\sigma:[0,T]\times \mbb{R}^n\rightarrow \mbb{R}^{n\times d}$
and $\gamma:[0,T]\times \mbb{R}^n\times E\rightarrow \mbb{R}^{n\times k}$.
Let us  introduce $\eta:\mbb{R}\rightarrow \mbb{R}_+$ by $\eta(e)=1\wedge |e|$.

\begin{assumption}
\label{assumption-X}
The functions $b(t,x)$, $\sigma(t,x)$ and $\gamma(t,x,e)$ are continuous in all their arguments and one-time
continuously differentiable with respect to $x$ with continuous derivatives. Furthermore, there exists some positive constant $K$ such that \\
(i) $|b(t,0)|+|\sigma(t,0)|\leq K$ uniformly in $t\in[0,T]$. \\
(ii) $|\part_x b(t,x)|+|\part_x \sigma(t,x)|\leq K$ uniformly in $(t,x)\in[0,T]\times \mbb{R}^n$. \\
(iii) For each column vector $i\in\{1,\cdots,k\}$,  $|\gamma^i(t,0,e)|\leq  K\eta(e)$
uniformly in $(t,e)\in[0,T]\times \mbb{R}_0$. \\
(iv) For each column vector $i\in\{1,\cdots,k\}$, $|\part_x \gamma^i(t,x,e)| \leq  K\eta(e)$
uniformly in $(t,x,e)\in[0,T]\times \mbb{R}^n\times \mbb{R}_0$.
\end{assumption}

We have the following result:
\begin{proposition}
\label{prop-X}
Under Assumption~\ref{assumption-X}, there exists a unique solution $X^{t,x}\in\mbb{S}^p[0,T]$
with $\forall p\geq 2$ for every initial data $(t,x)\in[0,T]\times \mbb{R}^n$.
Furthermore, the process $X^{t,x}$ is Malliavin differentiable $X^{t,x}\in\mbb{L}^{1,2}$ and satisfies,
for $\forall p\geq 2$, 
\bea
\int_{\wt{E}}\mbb{E}\Bigl[||D_{u,z}X^{t,x}||_T^p\Bigr]q(du,dz)\leq C(1+|x|^p)\nn
\eea 
with some positive constant $C$ depending only on $(p,T,K)$.
\begin{proof}
The fact that $X^{t,x}\in\mbb{S}^p[0,T]$ with $\forall p\geq 2$ is rather standard. See, for example,
Lemma A.3 in \cite{FT-BSDEJ}. The existence of Malliavin derivative follows from Theorem 3 
of Petrou (2008)~\cite{Petrou}. This implies, for $u\in[t,s]$ and $i\in\{1,\cdots,d\}$, 
\bea
D_{u,0}^i X_s^{t,x}&=&\sigma^i(u,X_u^{t,x})+\int_u^s\part_x b(r,X_r^{t,x})D_{u,0}^iX_r^{t,x}
+\int_u^s \part_x \sigma(r,X_r^{t,x})D_{u,0}^iX_r^{t,x} dW_r\nn \\
&&+\int_u^s\int_E \part_x \gamma(r,X_{r-}^{t,x},e)D_{u,0}^i X_r^{t,x} \wt{\mu}(dr,de)~,\nn
\eea
and for $(u,z)\in[t,s]\times \mbb{R}_0$ and $i\in\{1,\cdots,k\}$,
\bea
D_{u,z}^i X_s^{t,x}&=&\frac{\gamma^i(u,X_{u-}^{t,x},z)}{z}+
\int_u^s D_{u,z}^i b(r,X_r^{t,x})dr+\int_u^s D_{u,z}^i \sigma(r,X_r^{t,x})dW_r\nn \\
&&+\int_u^s\int_E D_{u,z}^i \gamma(r,X_{r-}^{t,x},e)\wt{\mu}(dr,de)~,\nn
\eea
where both $\sigma^i$ and $\gamma^i$ denote the $i$-th column vectors of dimension $n$, and for $\varphi=b,\sigma,\gamma$,
\bea
&&D_{u,z}^i \varphi(r,X_r^{t,x}):=\frac{\varphi(r,X_r^{t,x}+z D_{u,z}^i X_r^{t,x})-\varphi(r,X_r^{t,x})}{z}. \nn
\eea

By Lemma A.3~\cite{FT-BSDEJ}, the above SDEs satisfy the a priori estimates
\bea
&&\mbb{E}\Bigl[||D_{u,0} X^{t,x}||^p_T\Bigr]\leq C_{p,T,K}\mbb{E}\Bigl[|\sigma(u,X_u^{t,x})|^p\Bigr]\nn \\
&&\qquad \leq C_{p,T,K}\mbb{E}\Bigl[|\sigma(u,0)|^p+||X^{t,x}||_T^p\Bigr]\leq C_{p,T,K}(1+|x|^p)\nn
\eea
and 
\bea
&&\mbb{E}\Bigl[||D_{u,z}X^{t,x}||^p_T\Bigr]\leq C_{p,T,K}\mbb{E}\Bigl[\Bigl|\frac{\gamma(u,X_{u-}^{t,x},z)}{z}\Bigr|^p\Bigr]\nn \\
&&\qquad \leq C_{p,T,K}\mbb{E}\Bigl[\Bigl|\frac{\gamma(u,0,z)}{z}\Bigr|^p+||X^{t,x}||^p_T \Bigr]
\leq C_{p,T,K}(1+|x|^p)~.\nn
\eea
Since $q(du,dz)$ on $\wt{E}$ is a finite measure, the claim is proved.
\end{proof}
\end{proposition}

%%%%%%%%%%%%%%%%%%%%%%%%%%%%%%%%%%%%%%%%%%%%%%%%%%%%
\subsection{$Q_{\exp}$-growth BSDE driven by $X^{t,x}$}
%%%%%%%%%%%%%%%%%%%%%%%%%%%%%%%%%%%%%%%%%%%%%%%%%%%%
In many applications, there appears a BSDE driven by a Markovian forward process.
Let us consider a $Q_{\exp}$-BSDE driven by the process $\bigl(X^{t,x}_s,s\in[0,T]\bigr)$ introduced in 
the last section;
\bea
&&Y_s^{t,x}=\xi(X_T^{t,x})+\int_s^T f\Bigl(r,X_r^{t,x},Y_r^{t,x},Z_r^{t,x},\int_{\mbb{R}_0}\rho(e)G(r,\psi_r(e))\nu(de)\Bigr)dr\nn \\
&&\qquad -\int_s^t Z_r^{t,x} dW_r-\int_s^T \int_E \psi_r^{t,x}(e)\wt{\mu}(dr,de)~
\label{eq-Qexp-Y-X}
\eea
for $s\in[t,T]$ and put $(Y_s^{t,x},Z_s^{t,x},\psi_s^{t,x})\equiv (Y_t^{t,x},0,0)$ for $s<t$.
Here, $\xi:\mbb{R}^n\rightarrow \mbb{R}$, $f:[0,T]\times \mbb{R}^{n}\times \mbb{R}\times \mbb{R}^d\times \mbb{R}^k\rightarrow \mbb{R}$
are measurable functions.
We treat $Z$ and $\psi$ as row vectors for notational simplicity.
In this setup, the driver $f$ is deterministic without explicit dependence on $\omega$,
which is now provided by the dependence on $X^{t,x}$.

\begin{assumption}
\label{assumption-Qexp-X}
(i)For every $(x,y,z,\psi)\in \mbb{R}^n\times \mbb{R}\times \mbb{R}^d \times 
\mbb{L}^2(E,\nu;\mbb{R}^k)$, there exist two positive constants $\beta\geq 0$, $\gamma>0$
and the non-negative measurable function $l:[0,T]\rightarrow \mbb{R}_+$ such that
the measurable function $f$ satisfies
\bea
&&-l_t-\beta|y|-\frac{\gamma}{2}|z|^2-\int_E j_\gamma\bigl(-\psi(e)\bigr)\nu(de)\leq f\Bigl(t,x,y,z,\int_{\mbb{R}_0}\rho(e)G(t,\psi(e))
\nu(de)\Bigr)\nn \\
&&\qquad \qquad \leq l_t+\beta|y|+\frac{\gamma}{2}|z|^2+\int_E j_\gamma\bigl(\psi(e)\bigr)\nu(de)\nn
\eea
$dt$-a.e. $t\in[0,T]$, where $ j_\gamma(u):=\frac{1}{\gamma} \bigl(e^{\gamma u}-1-\gamma u\bigr)$. 
(ii) $|\xi(x)|+l_t$ is bounded uniformly in $(t,x)\in[0,T]\times \mbb{R}^n$. 
(iii) $F(t,x,y,z,\psi):=f\bigl(t,x,y,z,\int_{\mbb{R}_0}\rho(e)G(t,\psi(e))\nu(de)\bigr)$
satisfies the $A_\Gamma$-condition (Assumption~\ref{assumption-AGamma}).
\end{assumption}

\begin{assumption}
\label{assumption-LLC-X}
For each $M>0$, for every $x\in\mbb{R}^n$ and $(y,z,\psi), (y^\prime,z^\prime,\psi^\prime)\in \mbb{R}\times 
\mbb{R}^d\times \mbb{L}^2(E,\nu;\mbb{R}^k)$ satisfying
\be
|y|,|y^\prime|,||\psi||_{\mbb{L}^\infty(\nu)}, ||\psi^\prime||_{\mbb{L}^\infty(\nu)}\leq M, \nn
\ee
there exists some positive constant $K_M$ (possibly dependent on $M$) such that
\bea
&&\bigl| f(t,x,y,z,u_t)-f(t,x,y^\prime,z^\prime,u^\prime_t) \bigr|\nn \\
&&\leq K_M\bigl(|y-y^\prime|+|u_t-u_t^\prime|\bigr)+K_M\bigl(1+|z|+|z^\prime|+|u_t|+|u_t^\prime|\bigr)|z-z^\prime|\nn
\eea
with the short-hand notation $u_t:=\int_{\mbb{R}_0}\rho(e)G(t,\psi(e))\nu(de)$
and $u_t^\prime:=\int_{\mbb{R}_0}\rho(e)G(t,\psi^\prime(e))\nu(de)$~.
\end{assumption}

\begin{lemma}
Under Assumptions~\ref{assumption-rho-G}, \ref{assumption-X}, \ref{assumption-Qexp-X} and \ref{assumption-LLC-X},
there exists a unique solution $(Y^{t,x},Z^{t,x},\psi^{t,x})\in\mbb{S}^{\infty}_{[0,T]}\times \mbb{H}^2_{BMO[0,T]}
\times \mbb{J}^2_{BMO[0,T]}$ to the BSDE (\ref{eq-Qexp-Y-X}) for every $(t,x)\in[0,T]\times \mbb{R}^n$.
\begin{proof}
This is a special case of Theorem~\ref{theorem-existence}.
\end{proof}
\end{lemma}
We denote $\Theta_r^{t,x}:=\bigl(Y^{t,x},Z^{t,x},\int_{\mbb{R}_0}\rho(e)G(r,\psi_r^{t,x}(e))\nu(de)\bigr)$
as a collective argument of the solution indexed by the initial data $(t,x)$.

\begin{assumption}
\label{assumption-MD-Y-X}
(i) $\xi$ and the driver $f$ are one-time continuously differentiable with respect to the spacial variables
with continuous derivatives.\\
(ii) There exists some positive constant $K$ such that $|\part_x \xi(x)|\leq K$
as well as $|\part_x f(t,x,0,0,0)|\leq K$ uniformly in $(t,x)\in[0,T]\times \mbb{R}^n$.\\
(iii) For each $M>0$, for every $x\in\mbb{R}^n$ and  $(y,z,\psi), (y^\prime,z^\prime,\psi^\prime)\in \mbb{R}\times 
\mbb{R}^d\times \mbb{L}^2(E,\nu;\mbb{R}^k)$ satisfying
\be
|y|,|y^\prime|,||\psi||_{\mbb{L}^\infty(\nu)}, ||\psi^\prime||_{\mbb{L}^\infty(\nu)}\leq M, \nn
\ee
there exists some positive constant $K_M$ (possibly dependent on M) such that
\bea
&&\bigl|\part_x f(t,x,y,z,u_t)-\part_x f(t,x,y^\prime,z^\prime,u_t^\prime)\bigr|\nn \\
&&\leq K_M\bigl(|y-y^\prime|+|u_t-u_t^\prime|\bigr)+K_M\bigl(1+|z|+|z^\prime|+|u_t|+|u_t^\prime|\bigr)|z-z^\prime|\nn
\eea
with the short-hand notation $u_t:=\int_{\mbb{R}_0}\rho(e)G(t,\psi(e))\nu(de)$
and $u_t^\prime:=\int_{\mbb{R}_0}\rho(e)G(t,\psi^\prime(e))\nu(de)$~.
\end{assumption}

%\subsubsection*{Remark}
One sees that Assumption~\ref{assumption-MD-Y-X}, together with Assumption~\ref{assumption-LLC-X}, implies 
\bea
&&|\part_x f(t,x,y,z,u_t)|\leq C K_M \bigl(1+|y|+|z|^2+|u_t|^2\bigr),\quad |\part_y f(t,x,y,z,u_t)|\leq K_M, \nn \\
&&|\part_z f(t,x,y,z,u_t)|\leq K_M\bigl(1+2|z|+2|u_t|\bigr), \quad |\part_u f(t,x,y,z,u_t)|\leq K_M~,\nn
\eea
where $C$ is some positive constant.

\begin{theorem}
\label{theorem-Qexp-YX-MD}
Under Assumptions~\ref{assumption-rho-G}, \ref{assumption-X}, \ref{assumption-Qexp-X}, \ref{assumption-LLC-X}
and \ref{assumption-MD-Y-X}, the solution of the BSDE (\ref{eq-Qexp-Y-X}) is 
Malliavin differentiable $(Y^{t,x},Z^{t,x},\psi^{t,x})\in\mbb{L}^{1,2}\times \mbb{L}^{1,2}\times \mbb{L}^{1,2}$
for every initial data $(t,x)\in[0,T]\times \mbb{R}^n$.
(i) A version of $\bigl((D_{s,0}^i Y_r^{t,x},D_{s,0}^i Z_r^{t,x}, D_{s,0}^i \psi_r^{t,x}(e)),
0\leq s,r\leq T, e\in\mbb{R}_0\bigr)_{i\in\{1,\cdots,d\}}$ is the unique solution to the BSDE
\bea
&&D_{s,0}^i Y_u^{t,x}=D_{s,0}^i Z_u^{t,x}=D_{s,0}^i \psi_u^{t,x}(\cdot)=0, \qquad 0\leq u<s\leq T, \nn \\
&& D_{s,0}^i Y_u^{t,x}= \part_x \xi(X_T^{t,x})D_{s,0}^i X_T^{t,x}+
\int_u^T f^{s,0,i}(r) dr  - \int_u^T D_{s,0}^i Z_r^{t,x} dW_r\nn \\
&&\qquad -\int_u^T \int_E D_{s,0}^i \psi_r^{t,x}\wt{\mu}(dr,de), \quad u\in[s,T]\nn 
\eea
where
$f^{s,0,i}(r):=\part_x f(r,X_r^{t,x},\Theta_r^{t,x})D_{s,0}X_r^{t,x}+\part_\Theta f(r,X_r^{t,x},\Theta_r^{t,x})D_{s,0}\Theta_r^{t,x}$. 
Moreover, for a given $ds$-a.e. $s\in[0,T]$, $(D_{s,0}^iY^{t,x}, D_{s,0}^iZ^{t,x},D_{s,0}^i\psi^{t,x})\in\calk^p[0,T]$ with 
$\forall p\geq 2$. \\
(ii) A version of $\bigl((D_{s,z}^i Y_r^{t,x},D_{s,z}^i Z_r^{t,x}, D_{s,z}^i \psi_r^{t,x}(e)),
0\leq s,r\leq T, e,z \in\mbb{R}_0\bigr)_{i\in\{1,\cdots,k\}}$ is the unique solution to the BSDE
\bea
&&D_{s,z}^i Y_u^{t,x}=D_{s,z}^i Z_u^{t,x}=D_{s,z}^i \psi_u^{t,x}(\cdot)=0, \qquad 0\leq u<s\leq T, \nn \\
&&D_{s,z}^i Y_u^{t,x}=\xi^{s,z,i}+\int_u^T f^{s,z,i}(r)dr-\int_u^T D_{s,z}^i Z_r^{t,x}dW_r -\int_u^T\int_E D_{s,z}^i \psi_r^{t,x}(e)\wt{\mu}(dr,de)~, \nn
\eea
for $u\in[s,T]$ where
\bea
&&\xi^{s,z,i}:=\frac{\xi(X_T^{t,x}+z D_{s,z}^i X_T^{t,x})-\xi(X_T^{t,x})}{z},\nn \\
&&f^{s,z,i}(r):=\frac{1}{z}\Bigl\{f\Bigr(r,X_r^{t,x}+z D_{s,z}^i X_r^{t,x},~Y_r^{t,x}+zD_{s,z}^i Y_r^{t,x},
~Z_r^{t,x}+zD_{s,z}^i Z_r^{t,x} \nn \\
&&\qquad\qquad  ,\int_{\mbb{R}_0}\rho(e)G(r,\psi_r^{t,x}(e)+z D_{s,z}^i \psi_r^{t,x}(e))\nu(e)de\Bigr)
-f(r,X_r^{t,x},\Theta_r^{t,x})
\Bigr\}~.\nn
\eea
Moreover, for a given $m^i(dz)ds$-a.e. $(s,z)\in[0,T]\times \mbb{R}_0$, 
$(D_{s,z}^iY^{t,x}, D_{s,z}^iZ^{t,x},D_{s,z}^i\psi^{t,x})\in \mbb{S}^\infty[0,T]\times \mbb{H}^2_{BMO}[0,T]\times \mbb{J}^2_{BMO}[0,T]$.
\begin{proof}
It suffices to check Assumption~\ref{assumption-Qexp-MD} to hold so that Theorem~\ref{theorem-Qexp-MD} can be applied.
$(i),(ii)$ are obviously satisfied due to the Malliavin's differential rule (Theorem 3.5 and Theorem 12.8 in \cite{Nunno}).
The local Lipschitz condition (iii) is satisfied if we replace
$K_{s,z}^M(r)$ by  $K_M|D_{s,z}X_r^{t,x}|$. This is easy to see for a Wiener direction $(z=0)$.
For a jump direction $(z\neq 0)$,  notice that
\bea
(D_{s,z}f)(r,y,z,u_r)&=&\frac{1}{z}\bigl[ f(r,X_r^{t,x}+zD_{s,z}X_r^{t,x},y,z,u_r)-f(r,X_r^{t,x},y,z,u_r)\bigr]\nn \\
&=&\left(\int_0^1 \part_x f\Bigl(r,X_r^{t,x}+\theta z D_{s,z}X_r^{t,x}, y,z,u_r\Bigr)d\theta\right)D_{s,z}X_r^{t,x}~, \nn
\eea
which implies
\bea
&&\bigl| (D_{s,z}f)(r,y,z,u_r)-(D_{s,z}f)(r,y^\prime,z^\prime,u_r^\prime)\bigr|\nn \\
&&\leq |D_{s,z}X_r^{t,x}|\int_0^1\Bigl| \part_x f(r,X_r^{t,x}+\theta z D_{s,z}X_r^{t,x},y,z,u_r)
-\part_x f(r,X_r^{t,x}+\theta z D_{s,z}X_r^{t,x},y^\prime,z^\prime,u_r^\prime)\Bigr|d\theta  \nn \\
&&\leq K_M |D_{s,z}X_r^{t,x}|\bigl( |y-y^\prime|+|u_r-u_r^\prime|+(1+|z|+|z^\prime|+|u_r|+|u_r^\prime|)|z-z^\prime|\bigr)~.
\nn
\eea
Since $|D_{s,z}\xi|\leq K|D_{s,z}X_T^{t,x}|$ and $|(D_{s,z}f)(r,0,0,0)|\leq K|D_{s,z}X_r^{t,x}|$,
one can confirm that the condition $(iv)$ are satisfied from an inequality 
\bea
&&\mbb{E}\left[ |D_{s,z}\xi|^p+\Bigl(\int_0^T |(D_{s,z}f)(r,0,0,0)|dr\Bigr)^p+K_M^{2p}|| D_{s,z}X^{t,x}||_T^{2p}\right]\nn \\
&&\leq C_{p,K,K_M,T}\mbb{E}\Bigl[ 1+||D_{s,z}X^{t,x}||_T^{2p}\Bigr]\leq C_{p,K,K_M,T}(1+|x|^{2p}) \nn
\eea
uniformly in $(s,z)\in[0,T]\times \mbb{R}$ for $\forall p\geq 2$ (See, proof of Proposition~\ref{prop-X}.).
\end{proof}
\end{theorem}

\begin{corollary}
Under the assumptions of Theorem~\ref{theorem-Qexp-YX-MD}, let us define the deterministic function $u:[0,T]\times \mbb{R}^n\rightarrow 
\mbb{R}$ by $u(t,x):=Y_t^{t,x}$. Then, $u(t,x)$ is continuous in $(t,x)$, one-time continuously differentiable
with respect to $x$ with continuous derivative. Moreover, 
\bea
&&\bigl(Z^{t,x}(s)\bigr)^i=\part_x u(s,X_{s-}^{t,x})\sigma^i(s,X_{s-}^{t,x}),\quad t\leq s\leq T~,i\in\{1,\cdots,d\} \nn \\
&&\bigl(\psi^{t,x}_s(z)\bigr)^i=u(s,X_{s-}^{t,x}+\gamma^i(s,X_{s-}^{t,x},z))-u(s,X_{s-}^{t,x}), 
\quad t\leq s\leq T~,i\in\{1,\cdots,k\} \nn 
\eea
where $\sigma^i$ and $\gamma^i$ denotes the $i$-th column vectors.
\begin{proof}
By replacing {\it a priori} estimates for the Lipschitz BSDEs of Lemma 5.1 in \cite{FT-BSDEJ} with the local Lipschitz ones
given in Theorem~\ref{th-BMO-existence} and Lemma~\ref{lemma-BMO-stability}, 
one can follow the same arguments in Theorem 3.1 in \cite{Ma-Zhang} to show that the function $u(t,x)$
is continuous in the both arguments and one-time continuously differentiable with respect to $x$ with continuous derivatives.
Then the fact that 
\be
D_{s,0}^iX_s^{t,x}=\sigma^i(s,X_s^{t,x}), \quad zD_{s,z}^iX_s^{t,x}=\gamma^i(s,X^{t,x}_s,z)~, \nn
\ee
Corollary~\ref{corollary-version}, and the Malliavin differential rule for a continuously differentiable function 
give the desired result.
\end{proof}
\end{corollary}

%%%%%%%%%%%%%%%%%%%%%%%%%%%%%%%%%%%%%%%%%
\begin{appendix}
%%%%%%%%%%%%%%%%%%%%%%%%%%%%%%%%%%%%%%%%%%%%%%%%%%%%%%%%
\section{An a priori estimate and BMO-Lipschitz BSDEs}
%%%%%%%%%%%%%%%%%%%%%%%%%%%%%%%%%%%%%%%%%%%%%%%%%%%%%%%%%
\subsection{An a priori estimate}
%%%%%%%%%%%%%%%%%%%%%%%%%%%%%%%%%%%%%%%%%%%%%%%%%%%%
Firstly, we establish a priori estimate which plays a crucial role 
throughout the paper. Although it is similar to that of 
BMO-Lipschitz BSDEs, which will be discussed in the next section,
it has a much wider range of applications.
See discussion in Section 3 of Ankirchner et.al.~\cite{Imkeller-Reis} for a diffusion setup. 
Let us consider the BSDE, for $t\in[0,T]$,
\bea
Y_t=\xi+\int_t^T f(s,Y_s,Z_s,\psi_s)ds-\int_t^T Z_s dW_s-\int_t^T \int_E \psi_s(x)\wt{\mu}(ds,dx)~,
\label{eq-BMO-BSDE}
\eea
where $\xi: \Omega\rightarrow \mbb{R}$, $f:\Omega\times [0,T]\times \mbb{R}\times \mbb{R}^d\times
\mbb{L}^2(E,\nu;\mbb{R}^k)\rightarrow \mbb{R}$. We treat $Z$, $\psi$ are row vectors for simplicity.
We introduce another driver $\wt{f}: \Omega\times [0,T]\times \mbb{R}\times \mbb{R}^d\times
\mbb{L}^2(E,\nu;\mbb{R}^k)\rightarrow \mbb{R}$.
The crucial point of the next assumption is that the process $(H_t)_{t\in[0,T]}$ is not forbidden
to be a function of $(Y_t,Z_t,\psi_t)_{t\in[0,T]}$.
\begin{assumption}
\label{assumption-apriori-bmolike}
(i) The maps $(\omega,t)\mapsto f(\omega,t,\cdot), \wt{f}(\omega,t,\cdot)$ are $\mbb{F}$-progressively 
measurable. $\xi$ is an $\calf_T$-measurable random variable.\\
(ii) There exists a solution $(Y,Z,\psi)$ to the BSDE (\ref{eq-BMO-BSDE}) satisfying $Y\in\mbb{S}^p$ for $\forall p\geq 2$. \\
(iii) For every $(y,z,\psi)\in \mbb{R}\times \mbb{R}^d\times \mbb{L}^2(E,\nu;\mbb{R}^k)$, the driver $\wt{f}$
satisfies with some positive constant $K$ such that~\footnote{This can be generalized to a monotone condition.}
\be
|\wt{f}(\omega,t,y,z,\psi)|\leq g_t+K\bigl(|y|+|z|+||\psi||_{\L2nu}\bigr) \nn
\ee 
$d\mbb{P}\otimes dt$-a.e. $(\omega,t)\in\Omega\times[0,T]$, where $(g_t,t\in[0,T])$
is an $\mbb{F}$-progressively measurable positive process. Moreover, $\xi$ and $g$ satisfy, for $\forall p\geq 2$,
$\mbb{E}\Bigl[|\xi|^p+\Bigl(\int_0^T g_s ds\Bigr)^{p}\Bigr]<\infty$. \\
(iv) 
With the solution $(Y,Z,\psi)$ to the BSDE (\ref{eq-BMO-BSDE}), there exists an $\mbb{F}$-progressively measurable positive process $(H_t,t\in[0,T])$, $H\in\mbb{H}^2_{BMO}$ such that
\be
|f(s,Y_s,Z_s,\psi_s)-\wt{f}(s,Y_s,Z_s,\psi_s)|\leq H_s |Z_s| \nn
\ee 
for $d\mbb{P}\otimes ds$-a.e. $(\omega,s)\in\Omega\times[0,T]$.
\end{assumption}

\begin{lemma}
\label{lemma-bmolike-apriori}
Suppose Assumption~\ref{assumption-apriori-bmolike} hold true. Then the solution $(Y,Z,\psi)$
to the BSDE (\ref{eq-BMO-BSDE}) satisfies, for $\forall p\geq 2$,
\bea
\bigl|\bigl|(Y,Z,\psi)\bigr|\bigr|^p_{\calk^p[0,T]}\leq
C\Bigl(\mbb{E}\Bigl[|\xi|^{p\bar{q}^2}+\Bigl(\int_0^T g_s ds\Bigr)^{p\bar{q}^2}\Bigr]\Bigr)^{\frac{1}{\bar{q}^2}}~\nn
\eea
with a positive constant $\bar{q}$ satisfying $q_{*}\leq \bar{q}<\infty$ whose lower bound $q_*>1$ is
controlled only by $||H||_{\mbb{H}^2_{BMO}}$, and some positive constant $C$ depending only on 
$(p,\bar{q},T,K,||H||_{\mbb{H}^2_{BMO}})$.
\begin{proof}
Define a $d$-dimensional progressively measurable process $(b_s,s\in[0,T])$ by
\bea
b_s:=\frac{f(s,Y_s,Z_s,\psi_s)-\wt{f}(s,Y_s,Z_s,\psi_s)}{|Z_s|^2}\bold{1}_{Z_s\neq 0} Z_s, \nn
\eea
which satisfies $|b_s|\leq H_s$ and hence $b\in\mbb{H}^2_{BMO}$ whose norm is bounded by $||H||_{\mbb{H}^2_{BMO}}$.
Using the process $b$, (\ref{eq-BMO-BSDE}) can be written as
\bea
Y_t=\xi+\int_t^T \Bigl(\wt{f}(s,Y_s,Z_s,\psi_s)+b_s\cdot Z_s\Bigr)ds-\int_t^T Z_s dW_s-\int_t^T \int_E 
\psi_s(x)\wt{\mu}(ds,dx)~\nn
\eea
and hence under the new measure $\mbb{Q}$ defined by $d\mbb{Q}/d\mbb{P}=\cale_T(b*W)$, one obtains
\bea
Y_t=\xi+\int_t^T \wt{f}(s,Y_s,Z_s,\psi_s)ds-\int_t^T Z_s dW_s^{\mbb{Q}}-\int_t^T \int_E \psi_s(x)\wt{\mu}^{\mbb{Q}}(ds,dx)
\label{eq-bmolike-Q}
\eea
where $W^{\mbb{Q}}:=W-\int_0^\cdot b_s ds$ and $\wt{\mu}^{\mbb{Q}}=\wt{\mu}$ due to the independence of $(W,\wt{\mu})$.
By the linear growth property of $\wt{f}$, one has
\bea
Y_s \wt{f}(s,Y_s,Z_s,\psi)\leq |Y_s|\bigl(g_s+K(|Y_s|+|Z_s|+||\psi_s||_{\L2nu})\bigr)~, \nn
\eea
and hence for $\forall \lambda>0$ 
\bea
Y_s \wt{f}(s,Y_s,Z_s,\psi)\leq |Y_s|^2\bigl(K+K^2/(2\lambda)\bigr)+|Y_s|g_s+\lambda (|Z_s|^2+||\psi_s||^2_{\L2nu})~. \nn
\eea
Thus by choosing $V_t^\lambda:=\bigl(K+\frac{K^2}{2\lambda}\bigr)t$ and $N_t^\lambda=\int_0^t g_s ds$, 
the BSDE (\ref{eq-bmolike-Q}) satisfies Assumption~B.1 in \cite{FT-BSDEJ}.
Then Lemma~B.1 in \cite{FT-BSDEJ} of an a prior estimate for the BSDEs with a monotone driver implies, for $\forall p\geq 2$, 
\bea
\bigl|\bigl|(Y,Z,\psi)\bigr|\bigr|^p_{\calk^p(\mbb{Q})[0,T]}\leq C \mbb{E}^{\mbb{Q}}\Bigl[
|\xi|^p+\Bigl(\int_0^T g_s ds\Bigr)^p\Bigr] \nn
\eea 
with some positive constant $C=C_{p,K,T}$ depending only on $(p,K,T)$.

By the properties of the BMO martingales, one can choose $\bar{r}>1$ with which both $\cale(b*W)$ and $\cale(-b*W^\mbb{Q})$
satisfy the reverse H\"older inequality (See Lemma~\ref{lemma-BMO-PQ} and the following remark.).
Define $\bar{q}=\frac{\bar{r}}{\bar{r}-1}$ as its dual.
Let us put $D:=\max\bigl(||\cale(b*W)||_{\mbb{L}^{\bar{r}}(\mbb{P})},||\cale(-b*W^{\mbb{Q}})||_{\mbb{L}^{\bar{r}}(\mbb{Q})}\bigr)$,
which is dominated by some constant depending only on $||H||_{\mbb{H}^2_{BMO}(\mbb{P})}$. 
Then one obtains
\bea
&&\bigl|\bigl|(Y,Z,\psi)\bigr|\bigr|^p_{\calk^p(\mbb{P})_{[0,T]}}=\mbb{E}^\mbb{Q}\Bigl[\cale_T(-b*W^{\mbb{Q}})\Bigl(||Y||^p_T+
\Bigl(\int_0^T |Z_s|^2 ds\Bigr)^{\frac{p}{2}}+\Bigl(\int_0^T ||\psi_s||^2_{\L2nu}ds\Bigr)^{\frac{p}{2}}\Bigr)\Bigr]\nn \\
&&\hspace{20mm}\leq D \bigl|\bigl|(Y,Z,\psi)\bigr|\bigr|^p_{\calk^{p\bar{q}}(\mbb{Q})[0,T]} 
\leq C_{p,\bar{q},K,T}D\Bigl(\mbb{E}^{\mbb{Q}}\Bigl[|\xi|^{p\bar{q}}+\Bigl(\int_0^T g_s ds\Bigr)^{p\bar{q}}\Bigr]
\Bigr)^{\frac{1}{\bar{q}}}\nn \\
&&\hspace{20mm}\leq C_{p,\bar{q},K,T}D^{1+\frac{1}{\bar{q}}}\Bigl(\mbb{E}\Bigl[|\xi|^{p\bar{q}^2}+\Bigl(\int_0^T g_s ds\Bigr)^{p\bar{q}^2}
\Bigr]\Bigr)^{\frac{1}{\bar{q}^2}}~,\nn
\eea
which proves the desired result.
\end{proof}
\end{lemma}

%%%%%%%%%%%%%%%%%%%%%%%%%%%%%%%%
\subsection{BMO-Lipschitz BSDE}
In this subsection, we study the properties of the BSDE with a locally Lipschitz driver where
the Lipschitz coefficient for the control variable belongs to $\mbb{H}^2_{BMO}$.
In the diffusion setup, the details have been
discussed by Briand \& Confortola (2008)~\cite{Briand-Confortola}.
As we have announced before, we keep the reverse H\"older property only to the 
continuous part and assume only the standard Lipschitz continuity for the 
jump coefficient.
\begin{assumption}
\label{assumption-BMO-BSDE}
The map $(\omega,t)\mapsto f(\omega,t,\cdot)$ is $\mbb{F}$-progressively measurable.\\
(i)~There exist a positive constant $K$ and a positive $\mbb{F}$-progressively measurable process $(H_t,t\in[0,T])\in \mbb{H}^2_{BMO}$
such that, for every 
$(y,z,\psi),(y^\prime,z^\prime,\psi^\prime) \in \mbb{R}\times \mbb{R}^d \times \mbb{L}^2(E,\nu;\mbb{R}^k)$,
\bea
|f(\omega,t,y,z,\psi)-f(\omega,t,y^\prime,z^\prime,\psi^\prime)|\leq K\bigl(|y-y^\prime|+||\psi-\psi^\prime||_{\L2nu}\bigr)+
H_t(\omega)|z-z^\prime|~ \nn
\eea
$d\mbb{P}\otimes dt$-a.e. $(\omega,t)\in\Omega\times [0,T]$.\\
(ii)~$\xi$ is $\calf_T$-measurable and, for $\forall p\geq 2$, 
\be
\mbb{E}\Bigl[|\xi|^p+\Bigl(\int_0^T |f(s,0,0,0)|ds\Bigr)^p\Bigr]<\infty~.\nn
\ee
\end{assumption}

\begin{theorem}
\label{th-BMO-existence}
Under Assumption~\ref{assumption-BMO-BSDE}, there exists a unique solution $(Y,Z,\psi)$ to 
the BSDE (\ref{eq-BMO-BSDE}) and it satisfies, for $\forall p\geq 2$,
\bea
\bigl|\bigl|(Y,Z,\psi)\bigr|\bigr|_{\calk^p[0,T]}^p\leq C 
\Bigl(\mbb{E}\Bigl[|\xi|^{p\bar{q}^2}+\Bigl(\int_0^T |f(s,0,0,0)|ds\Bigr)^{p\bar{q}^2}\Bigr]\Bigr)^{\frac{1}{\bar{q}^2}}\nn
\eea
with a positive constant $\bar{q}$ satisfying $q_*\leq \bar{q}<\infty$ whose lower bound $q_*>1$ is controlled only by $||H||_{\mbb{H}^2_{BMO}}$,
and some positive constant $C$ depending only on $(p,\bar{q},T,K,||H||_{\mbb{H}^2_{BMO}})$.
%%%%%%%%%%%%%%%%%%
\begin{proof}
Define a progressively measurable process $(b_s,s\in[0,T])$ taking values in $\mbb{R}^d$ by
\bea
b_s:=\frac{f(s,Y_s,Z_s,\psi_s)-f(s,Y_s,0,\psi_s)}{|Z_s|^2}\bold{1}_{Z_s\neq 0} Z_s\nn
\eea
then $|b_s|\leq H_s$ and hence $b\in\mbb{H}^2_{BMO}$ and its norm is dominated by $||H||_{\mbb{H}^2_{BMO}}$.
Under the measure $\mbb{Q}$ defined by $d\mbb{Q}/d\mbb{P}=\cale_T(b*W)$,
\bea
Y_t=\xi+\int_t^T f(s,Y_s,0,\psi_s)ds-\int_t^T Z_s dW_s^{\mbb{Q}}-\int_t^T \psi_s(x)\wt{\mu}^{\mbb{Q}}(ds,dx)
\label{eq-BMO-BSDE-Q}
\eea
where $W^\mbb{Q}=W-\int_0^\cdot b_s ds$ and $\wt{\mu}^{\mbb{Q}}=\wt{\mu}$.
As discussed in Lemma~\ref{lemma-bmolike-apriori},
one can choose $\bar{r}>1$ with which both of $\cale(b*W)$ and $\cale(-b*W^\mbb{Q})$
satisfy the reverse H\"older inequality and $\bar{q}=\frac{\bar{r}}{\bar{r}-1}$ as its dual.
Let us put $D:=\max\bigl(||\cale(b*W)||_{\mbb{L}^{\bar{r}}(\mbb{P})},||\cale(-b*W^{\mbb{Q}})||_{\mbb{L}^{\bar{r}}(\mbb{Q})}\bigr)$,
which is dominated by some constant depending only on $||H||_{\mbb{H}^2_{BMO}(\mbb{P})}$.

It is clear that the BSDE satisfies the global Lipschitz properties under the measure $\mbb{Q}$.
Furthermore, the following inequality is satisfied due to (reverse) H\"older inequalities:
\bea
&&\mbb{E}^{\mbb{Q}}\Bigl[|\xi|^p+\Bigl(\int_0^T |f(s,0,0,0)|ds\Bigr)^p\Bigr]
=\mbb{E}\Bigl[\cale(b*W)\Bigl(|\xi|^p+\Bigl(\int_0^T |f(s,0,0,0)|ds\Bigr)^p\Bigr)\Bigr] \nn \\
&&\qquad \leq C_{\bar{q}} D \mbb{E}\Bigl[ |\xi|^{p\bar{q}}+\Bigl(\int_0^T |f(s,0,0,0)|ds\Bigr)^{p\bar{q}}\Bigr]^{\frac{1}{\bar{q}}}<\infty~,\nn
\eea
with some positive constant $C_{\bar{q}}$.
Thus, by Lemma B.2 in \cite{FT-BSDEJ}, one concludes that 
there exists a unique solution $(Y,Z,\psi)$ to (\ref{eq-BMO-BSDE-Q}) in $\mbb{Q}$ and hence also 
to (\ref{eq-BMO-BSDE}) in $\mbb{P}$.
Furthermore, it also satisfies by the same lemma,
\bea
||(Y,Z,\psi)||^p_{\calk^p(\mbb{Q})}\leq C_{p,K,T}\mbb{E}^{\mbb{Q}}\Bigl[|\xi|^p+\Bigl(\int_0^T|f(s,0,0,0)|ds\Bigr)^p\Bigr]~.\nn
\eea
We thus have
\bea
&&\bigl|\bigl|(Y,Z,\psi)\bigr|\bigr|^p_{\calk^p(\mbb{P})}\leq C_{\bar{q}}D~\bigl|\bigl|(Y,Z,\psi)
\bigr|\bigr|^{p}_{\calk^{p\bar{q}}(\mbb{Q})}\nn \\
&&\quad \leq C_{p,\bar{q},K,T}D^{1+\frac{1}{\bar{q}}}\Bigl(\mbb{E}
\Bigl[|\xi|^{p\bar{q}^2}+\Bigl(\int_0^T |f(s,0,0,0)|ds\Bigr)^{p\bar{q}^2}\Bigr]\Bigr)^{\frac{1}{\bar{q}^2}}\nn ~,
\eea
which proves the second part of the claim.
\end{proof}
\end{theorem}

Now, we gives the stability result which is required to show the uniqueness of the quadratic-exponential growth BSDE.
Consider the two BSDEs with $i\in\{1,2\}$ satisfying Assumption~\ref{assumption-BMO-BSDE};
\bea
Y_t^i=\xi^i+\int_t^T f^i(s,Y_s^i,Z_s^i,\psi_s^i)ds-\int_t^T Z_s^i dW_s-\int_t^T \int_E \psi_s^i(x)\wt{\mu}(ds,dx)
\label{eq-BMO-BSDE-2}
\eea
and put
$\del Y:=Y^1-Y^2,~\del Z:=Z^1-Z^2,~\del\psi:=\psi^1-\psi^2, ~\del f(s):=(f^1-f^2)(s,Y_s^1,Z_s^1,\psi_s^1)$. 

\begin{lemma}
\label{lemma-BMO-stability}
The unique solutions $(Y^i,Z^i,\psi^i), i\in\{1,2\}$ to the BSDEs (\ref{eq-BMO-BSDE-2}) under Assumption~\ref{assumption-BMO-BSDE}
satisfy 
\bea
\bigl|\bigl|(\del Y,\del Z, \del \psi)\bigr|\bigr|^p_{\calk^p[0,T]}\leq C\Bigl(\mbb{E}\Bigl[|\del \xi|^{p\bar{q}^2}+\Bigl(\int_0^T 
|\del f(s)|ds\Bigr)^{p\bar{q}^2}\Bigr]\Bigr)^{\frac{1}{\bar{q}^2}}\nn
\eea
with a positive constant $q_*\leq \bar{q}<\infty$ whose lower bound $q_*>1$ is controlled only by $||H||_{\mbb{H}^2_{BMO}}$,
and some positive constant $C$ depending only on $(p,\bar{q},T,K,||H||_{\mbb{H}^2_{BMO}})$.
\begin{proof}
Let us introduce a process $(b_s,s\in[0,T])$ defined by
\bea
b_s:=\frac{f^2(s,Y_s^1,Z_s^1,\psi_s^1)-f^2(s,Y_s^1,Z_s^2,\psi_s^1)}{|\del Z_s|^2}\bold{1}_{\del Z_s\neq 0} \del Z_s\nn
\eea
and also a map $\wt{f}:\Omega\times [0,T]\times \mbb{R}\times \mbb{L}^2(E,\nu;\mbb{R}^k)\rightarrow \mbb{R}$ by
\be
\wt{f}(\omega,s,\wt{y},\wt{\psi}):=\del f(\omega,s)+f^2(\omega,s,\wt{y}+Y^2_s,Z_s^2,\wt{\psi}+\psi_s^2)-
f^2(\omega,s,Y^2_s,Z^2_s,\psi_s^2)~. \nn
\ee
Then, $(\del Y, \del Z, \del \psi)$ can be interpreted as the solution to the BSDE
\be
\del Y_t=\del \xi+\int_t^T \Bigl(\wt{f}(s,\del Y_s,\del \psi_s)+b_s\cdot \del Z_s\Bigr)ds
-\int_t^T \del Z_s dW_s -\int_t^T \int_E \del \psi_s(x)\wt{\mu}(ds,dx)~.\nn
\ee
Since $|b_s|\leq H_s\in \mbb{H}^2_{BMO}$ and $\wt{f}$ has the linear-growth property with respect to $(\wt{y},\wt{\psi})$,
Lemma~\ref{lemma-bmolike-apriori} with $g=|\del f|$ gives the desired result.
\end{proof}
\end{lemma}

%%%%%%%%%%%%%%%%%%%%%%%%%%%%%%%%%%%%%%%%%%%%%%%%%%%%%%%%%%%%%%%%%%%%%%%%%%%%%
\section{Some remarks on the comparison principle}
%%%%%%%%%%%%%%%%%%%%%%%%%%%%%%%%%%%%%%%%%%%%%%%%%%%%%%%%%%%%%%%%%%%%%%%%%%%%%
%Firstly, we check the boundedness of the solution for the BSDE with data $(\xi,f^{n,m,k})$
\begin{lemma} 
\label{lemma-ynmk-bound}
If $(Y,Z,\psi)$ is the square integrable solution of the BSDE with data $(\xi,f^{n,m,k})$, then $Y\in\mbb{S}^\infty$.
\begin{proof}
Consider a sequence of the BSDEs with $l\in\mbb{N}$,
\bea
Y^l_t=\xi+\int_t^T F^l(s, Y^l_s,Z^l_s,\psi_s^l)ds-\int_t^T Z_s^l dW_s-\int_t^T \int_E \psi_s^l(x)\wt{\mu}(ds,dx), t\in[0,T]
\label{eq-nmk-l}
\eea
where $F^l(s,y,z,\psi):=f^{n,m,k}(s,y,z,\psi\circ \zeta_l)$ and 
$(\psi_s\circ \zeta_l)(x):=\psi_s(x)\bold{1}_{\{|x|\geq 1/l\}}$. $F^l$ is globally Lipschitz
and satisfy $Q_{\exp}$-structure condition uniformly in $l$.
Since $|f^{n,m}|\leq |\ol{f}^{n}|\vee|\ul{f}^{m}|\leq |f|$,
one sees that $|F^l(s,y,0,\psi)| \leq |f(s,\varphi_k(y),0,\varphi_k(\psi\circ\zeta_l))|$, which is clearly bounded
for all $s,y,\psi$. Thus, by absorbing the $Z$ argument by the measure change, one sees $Y^l\in\mbb{S}^\infty$. 
One can now apply the universal bounds of Lemmas~\ref{lemma-BMO-bound}
and \ref{lemma-universal-bound} to conclude $||Y^l||_{\mbb{S}^\infty}, ||Z^l||_{\mbb{H}^2_{BMO}}, ||\psi^l||_{\mbb{J}^2_{BMO}}$
are bounded uniformly in $l$.  It now suffices to prove $(Y^l,Z^l,\psi^l)$ converges to the 
solution $(Y,Z,\psi)$ of the BSDE with data $(\xi,f^{n,m,k})$.

Since (\ref{eq-nmk-l}) is globally Lipschitz uniformly in $l$, the standard stability formula gives
\bea
||(Y^l-Y^{l^\prime},Z^l-Z^{l^\prime}, \psi^l-\psi^{l^\prime})||^2_{\calk^2}\leq C\mbb{E}\Bigl[\Bigl(\int_0^T |\del f(s)|ds\Bigr)^2\Bigr]
\leq CT\mbb{E}\Bigl[\int_0^T |\del f(s)|^2 ds\Bigr]\nn
\eea
where $C$ is independent of $l$ and $\del f(s):=(F^l-F^{l^\prime})(s,Y^l,Z^l,\psi^l)$. Let suppose $l\leq l^\prime$. 
For any $(s,y,z,\psi)\in[0,T]\times \mbb{R}\times \mbb{R}^d\times \mbb{L}^2(E,\nu;\mbb{R}^k)$, $A_\Gamma$-condition for  $f^{n,m}$ gives
\bea
&&|F^l(s,y,z,\psi)-F^{l^\prime}(s,y,z,\psi)|=|f^{n,m}(s,\varphi_k(y),z,
\varphi_k(\psi\circ \zeta_l))-f^{n,m}(s,\varphi_k(y),z,\varphi_k(\psi\circ\zeta_{l^\prime}))|\nn \\ 
&&\qquad \leq \int_E \Gamma_s^{l,l^\prime}(x)|\varphi_k(\psi(x))|\bold{1}_{\{|x|<1/l\}}\nu(dx)\nn 
\eea
with some non-negative $\calp\otimes\cale$-measurable process $\Gamma^{l,l^\prime}$ satisfying $\Gamma^{l,l^\prime}(x)\leq C(1\wedge |x|)$.
Here, the constant $C$ depends only on $k$.
Noticing the fact that $||\psi^l||_{\mbb{J}^2}$ is bounded uniformly in $l$, the dominated convergence theorem gives
\bea
\mbb{E}\Bigl[\int_0^T |\del f(s)|^2 ds\Bigr]\leq C\Bigl(\int_E |x|^2\bold{1}_{\{|x|<1/l\}}\nu(dx)\Bigr)\mbb{E}\Bigl[\int_0^T\int_E |\psi_s^l(x)|^2\nu(dx)ds\Bigr]\rightarrow 0 \nn
\eea
as $l$ (and hence also $l^\prime$)$\rightarrow \infty$.  This proves $(Y^l,Z^l,\psi^l)_{l\geq 1}$ converges to 
some $(\wt{Y},\wt{Z},\wt{\psi})$ in $\calk^2$. Since $(Y^l)_{l\geq 1}$ are uniformly bounded, so is $\wt{Y}$.
It is straightforward to check $(\wt{Y},\wt{Z},\wt{\psi})$ actually gives a solution to
the BSDE with data $(\xi,f^{n,m,k})$, but it is unique and hence equal to $(Y,Z,\psi)$ due to the global Lipschitz continuity.
\end{proof}
\end{lemma}
The remaining two lemmas are on the comparison principle.
\begin{lemma}
\label{lemma-fnm-BSDE}
With Assumptions~\ref{assumption-Qexp},\ref{assumption-LLC} and \ref{assumption-AGamma}, if there exists a solution $(Y^{n,m},Z^{n,m},\psi^{n,m})\in \mbb{S}^\infty\times \mbb{H}^2\times \mbb{J}^2$
to the BSDE
\bea
Y^{n,m}_t=\xi+\int_t^T f^{n,m}(s,Y_s^{n,m},Z_s^{n,m},\psi^{n,m}_s)ds-\int_t^T Z_s^{n,m}dW_s-\int_t^T\int_E \psi^{n,m}_s(x)
\wt{\mu}(ds,dx)~,\nn 
\eea
then it is unique. Moreover, if the relevant solutions exist for the pairs of $(n,m)$, they satisfy 
$Y^{n,m+1}_t\leq Y^{n,m}_t\leq Y^{n+1,m}_t$ for $\forall t\in[0,T]$ a.s.
\begin{proof}
Since $f^{n,m}$ satisfies the structure condition in Assumption~\ref{assumption-Qexp} uniformly in $(n,m)$, if there exists a bounded solution, then we have
$(Y^{n,m},Z^{n,m},\psi^{n,m})\in\mbb{S}^\infty\times \mbb{H}^2_{BMO}\times \mbb{J}^2_{BMO}$ and the same
universal bounds in Lemmas~\ref{lemma-BMO-bound} and \ref{lemma-universal-bound}. 
Hence, from Assumption~\ref{assumption-LLC}, one can choose a constant $K_M$ as the Lipschitz constant
with regard to $y,\psi$ arguments. Since the driver is $(n\vee m)$-Lipschitz with respect to $z$,
one obtains the same stability condition as the globally Lipschitz BSDE.
The uniqueness of the solution then follows.
Since the driver $f^{n,m}$ satisfies Assumption~\ref{assumption-AGamma}, one has for bounded solutions $(\psi,\psi^\prime)$, 
\bea
\mbb{E}\Bigl[\int_\tau^T |\Gamma_t^{\psi,\psi^{\prime}}(x)|^2\nu(dx)ds\Bigr|\calf_\tau\Bigr] \leq (C_M^2\vee|C^1|) \int_\tau^T |x|^2\nu(dx)ds 
\leq C_0 T
\eea
for any $\tau\in\calt^T_0$ with some constant $C_0$ depending only on the universal bounds. 
This implies $\Gamma^{\psi,\psi^\prime}.\wt{\mu}$ is a BMO-martingale.  Moreover $\cale(\Gamma^{\psi,\psi^\prime}. \wt{\mu})$ 
is a uniformly integrable martingale by Lemma~\ref{lemma-R-Holder}. The comparison principles now follows
in the same way as the Lipschitz case. See, for example, Theorem 2.5 of Royer (2006)~\cite{Royer}.
\end{proof}
\end{lemma}

\begin{lemma}
\label{lemma-fn-tilde}
With Assumptions~\ref{assumption-Qexp},\ref{assumption-LLC} and \ref{assumption-AGamma}, if there exists a solution $(\wt{Y}^{n},\wt{Z}^{n},\wt{\psi}^{n})\in \mbb{S}^\infty\times \mbb{H}^2\times \mbb{J}^2$
to the BSDE with $\wt{f}^n=\ol{f}^n+\ul{f}$
\bea
\wt{Y}^{n}_t=\xi+\int_t^T \wt{f}^{n}(s,\wt{Y}_s^{n},\wt{Z}_s^{n},\wt{\psi}^{n}_s)ds-\int_t^T \wt{Z}_s^{n}dW_s-\int_t^T\int_E \wt{\psi}^{n}_s(x)
\wt{\mu}(ds,dx)~,\nn 
\eea
then it is unique. Moreover, if the relevant solutions exist for $n,n+1$, they satisfy 
$\wt{Y}^{n}_t\leq \wt{Y}^{n+1}_t$ for $\forall t\in[0,T]$ a.s.
\begin{proof}
Since $\wt{f}^n$ satisfies the structure condition in Assumption~\ref{assumption-Qexp},
if there exists a bounded solution it satisfies the universal bounds. Thus the driver is $K_M$-Lipschitz continuous with respect to $y,\psi$
as in the previous lemma. For $z$ argument, the driver is local Lipschitz continuous whose coefficient 
is given by the sum of $n$ and that given in Assumption~\ref{assumption-LLC}. Thanks to the universal bounds,
it has a bounded $H^2_{BMO}$-norm for each $n$. 
It is also easy to confirm that $\wt{f}^n$ satisfies $A_\Gamma$-condition uniformly in $n$ 
as in the proof of Lemma~\ref{lemma-nmk}.
Thus the measure change used in Theorem 2.5 of Royer~\cite{Royer} is still valid
and hence the comparison principle follows. The uniqueness follows from Proposition~\ref{prop-uniqueness} or 
from the comparison principle as \cite{Royer}.
\end{proof}
\end{lemma}

\section{Malliavin differentiability for Lipschitz BSDEs with jumps}
%%%%%%%%%%%%%%%%%%%%%%%%%%%%%%%%%%%%%%%%%%%%%%%%%%%%%%%%%%%%%%%%%%%%%%%%%%%%%%%%
In order to show Malliavin's differentiability of $Q_{\exp}$-growth BSDEs, we have to establish 
the differentiability for Lipschitz BSDEs with slightly more general setup than 
what was proved in \cite{Delong-Imkeller} and \cite{Delong}.
For convenience of the readers, we give the detailed proof in this section.
We closely follow the arguments used in El Karoui et.al. (1997)~\cite{ElKaroui}.
The complication relative to a diffusion case is the treatment of small jumps.
The difference from the work~\cite{Delong-Imkeller} is a local Lipschitz condition instead of 
the global Lipschitz condition for the Malliavin derivative of the 
driver.
%%%%%%%%%%%%%%%%%%%%%%%%%%%%%%%%%%
%~\footnote{In addition, we think that the result of \cite{Delong-Imkeller} misses
%the one condition for the driver 
%\bea
%\lim_{\ep\downarrow 0} \int_0^T\int_{|z|\leq \ep}\mbb{E}\left[\Bigl(\int_0^T |(D_{s,z}f)(r,0)|dr\Bigr)^2\right]z^2\nu(dz)ds=0~.\nn
%\eea
%This stems from an error in the estimate (4.15) and (4.16) (and hence (4.21),(4.22)) in the proof of Theorem 4.1 of 
%\cite{Delong-Imkeller}.
%Note that if one choose $\wt{f}=f$, then $(\wt{Y}^{s,z}, \wt{Z}^{s,z},\wt{U}^{s,z})$ cannot be chosen as zero.
%The left hand side of (4.16), for example, should still be $||Y^{s,z}-\wt{Y}^{s,z}||_{\mbb{S}^2}+\cdots$.
%Adding the contributions from $\wt{Y}^{s,z}$ etc would yield the consistent result to the analysis given here.
%}.

We consider a BSDE defined by
\bea
\hspace{-3mm}Y_t=\xi+\int_t^T f\Bigl(s,Y_s,Z_s,\int_{\mbb{R}_0}\rho(x)G(s,\psi_s(x))\nu(dx)\Bigr)ds-\int_t^T Z_s dW_s-\int_t^T \int_E \psi_s(x)\wt{\mu}(ds,dx),~
\label{eq-Lipschitz-BSDE}
\eea
where $\xi:\Omega\rightarrow \mbb{R}$, $f:\Omega\times [0,T]\times \mbb{R} \times  \mbb{R}^{d}\times 
\mbb{R}^{k}\rightarrow \mbb{R}$.
Here, $\int_{\mbb{R}_0}\rho(x)G(s,\psi_s(x))\nu(dx)$ denotes a $k$-dimensional vector 
whose $i$-th element is given by $\int_{\mbb{R}_0}\rho^i(x)G^i(s,\psi_s^i(x))\nu^i(dx)$
where $\rho^i:\mbb{R}\rightarrow \mbb{R}$, $\G^i:[0,T]\times \mbb{R}\rightarrow \mbb{R}$.
With slight abuse of notation, we use
$\Theta_r:=\Bigl(Y_r,Z_r,\int_{\mbb{R}_0}\rho(x)G(r,\psi_r(x))\nu(dx)\Bigr)$
as a collective argument in this section. The results in this section can be straightforwardly 
extended to multi-dimensional Lipschitz BSDEs.
\begin{assumption}
\label{assumption-Lipschitz}
(i) For every $i\in\{1,\cdots,k\}$, $\rho^i(s)$ and $G^i(s,v)$ are continuous functions in $s\in[0,T]$ and $(s,v)\in[0,T]\times \mbb{R}$,
respectively.
We set without loss of generality that $G^i(\cdot,0)=0$ . In addition
$\int_{\mbb{R}_0}|\rho^i(x)|^2\nu^i(dx)<\infty$, and
with some positive constant $K$, $G^i$ satisfies
\bea
&&|G^i(s,v)-G^i(s,v^\prime)|\leq K|v-v^\prime|, {\text{~~~for every $s\in[0,T]$ and $v,v^\prime\in \mbb{R}$.}}\nn
\eea
(ii)~ The map $(\omega,t)\mapsto f(\omega,t,\cdot)$ is $\mbb{F}$-progressively measurable, and  for every $(y,z,u),(y^\prime,z^\prime,u^\prime)\in\mbb{R}\times \mbb{R}^{d}\times \mbb{R}^{k}$,
there exists some positive constant $K$ such that
\bea
|f(\omega, t,y,z,u)-f(\omega, t,y^\prime,z^\prime, u^\prime)|\leq K(|y-y^\prime|+|z-z^\prime|+|u-u^\prime|)\nn
\eea
$d\mbb{P}\otimes dt$-a.e. $(\omega,t)\in\Omega\times[0,T]$.\\
(iii)~$\xi\in\mbb{L}^4(\Omega,\calf_T,\mbb{P})$ and $\bigl(f(t,0),t\in[0,T]\bigr)\in\mbb{H}^4[0,T]$. 
\end{assumption}
\begin{remark}
Due to the property of $G$ and $\rho$, it is easy to see that
\bea
&&\Bigl|\int_{\mbb{R}_0}\rho(x)G(s,\psi_s(x))\nu(dx)-\int_{\mbb{R}_0}\rho(x)G(s,\psi_s^\prime(x))\nu(dx)\Bigr| 
\leq K^\prime ||\psi_s-\psi_s^\prime||_{\L2nu} \nn
\eea
with some constant $K^\prime >0$. Thus, Assumption~\ref{assumption-Lipschitz} yields the standard global Lipschitz conditions.
By Lemma B.2 in \cite{FT-BSDEJ},  the BSDE (\ref{eq-Lipschitz-BSDE}) has a
unique solution $(Y,Z,\psi)\in\calk^4[0,T]$. In order to show the Malliavin's differentiability, 
we need additional assumptions.
\end{remark}

\begin{assumption}
\label{assumption-M-Lipschitz}
(i) For every $i\in\{1,\cdots,k\}$, $G^i$ is one-time continuously differentiable with respect to its spacial variable $v$ with a uniformly bounded and continuous derivative. \\
(ii) The terminal value is Malliavin differentiable $\xi\in\mbb{D}^{1,2}$
and satisfies 
\be
\mbb{E}\Bigl[\int_{\wt{E}}|D_{s,z}\xi|^2q(ds,dz)\Bigr]<\infty. \nn
\ee
(iii) The driver $f(\cdot,y,z,u)$ is one-time continuously differentiable with respect to $(y,z,u)$ with uniformly bounded 
and continuous derivatives.
For every $(y,z,u)\in \mbb{R}\times \mbb{R}^{d}\times \mbb{R}^{k}$, 
the driver $\bigl(f(t,y,z,u),t\in[0,T]\bigr)$ belongs to $\mbb{L}^{1,2}$ and
its Malliavin derivative is denoted by $(D_{s,z}f)(t,y,z,u)$.  \\
(iv) For every Wiener as well as jump direction, and
for every $(y,z,u),(y^\prime,z^\prime,u^\prime) \in \mbb{R} \times \mbb{R}^{d}\times \mbb{R}^{k}$
and $d\mbb{P}\otimes dt$-a.e. $(\omega,t)\in\Omega\times [0,T]$, 
the Malliavin derivative of the driver satisfies the following local Lipschitz conditions~\footnote{Delong \& Imkeller (2010)~\cite{Delong-Imkeller} has treated a special case where $(K_{s,0},~K_{s,z})$ are positive constants.
The current generalization is necessary when one introduces a Markovian process $X$ driven by a FSDE
to create a forward-backward SDE system, which is the subject of interests in many applications.};
\bea
&&|(D_{s,0}^i f)(t,y,z,u)-(D_{s,0}^if)(t,y^\prime,z^\prime,u^\prime)|\leq K_{s,0}^i(t)\bigl(|y-y^\prime|+|z-z^\prime|+|u-u^\prime|\bigr), \nn
\eea
for $ds$-a.e. $s\in[0,T]$ with $i\in\{1,\cdots,d\}$, and
\bea
&&|(D_{s,z}^i f)(t,y,z,u)-(D_{s,z}^if)(t,y^\prime,z^\prime,u^\prime)|\leq K_{s,z}^i(t)\bigl(|y-y^\prime|+|z-z^\prime|+|u-u^\prime|\bigr), \nn 
\eea
for $m^i(dz)ds$-a.e. $(s,z)\in[0,T]\times \mbb{R}_0$ with $i\in\{1,\cdots,k\}$.
Here, $\bigl(K_{s,0}^i(t),t\in[0,T]\bigr)_{i\in\{1,\cdots,d\}}$
and $\bigl(K_{s,z}^i(t),t\in[0,T]\bigr)_{i\in\{1,\cdots,k\}}$ are $\mbb{R}_+$-valued $\mbb{F}$-progressively measurable processes 
satisfying $\int_{\wt{E}}||K_{s,z}(\cdot)||^4_{\mbb{S}^4[0,T]}q(ds,dz)<\infty$.
%(v) The following equality is supposed to hold;
%\bea
%\lim_{\ep\downarrow 0} \sum_{i=1}^k \int_0^T \int_{|z|\leq \ep}\mbb{E}
%\left[ |D^i_{s,z}\xi|^2+\Bigl(\int_0^T |(D_{s,z}^if)(r,0)|dr\Bigr)^2+||K^i_{s,z}||_T^4\right]z^2\nu^i(dz)ds=0~.\nn
%\eea
\end{assumption}
\begin{remark}
\label{dominated-Lip}
It follows from the conditions (ii), (iii) and (iv) that
\bea
&&\sum_{i=1}^k \int_0^T \int_{|z|\leq \ep}\mbb{E}
\Bigl[ |D^i_{s,z}\xi|^2+\Bigl(\int_0^T |(D_{s,z}^if)(r,0)|dr\Bigr)^2+||K^i_{s,z}||_T^4\Bigr]m^i(dz)ds\rightarrow 0 \nn 
\eea
as $\ep\downarrow 0$ by the dominated convergence. 
\end{remark}

%%%%%%%%%%%%%%%%%%%%%%%%%%%%%%
\begin{theorem}
\label{theorem-Malliavin-Lipschitz}
Suppose that Assumptions~\ref{assumption-Lipschitz} and \ref{assumption-M-Lipschitz} hold true and denote the 
solution to the BSDE (\ref{eq-Lipschitz-BSDE}) as $(Y,Z,\psi)\in\calk^4[0,T]$. Then, the following statements hold:\\
(a)~For each Wiener direction $i\in\{1,\cdots,d\}$ and $ds$-a.e. $s\in[0,T]$, there exists a unique solution 
$(Y^{s,0,i},Z^{s,0,i},\psi^{s,0,i})\in \calk^2[0,T]$ to the BSDE
\bea
Y^{s,0,i}_t=D_{s,0}^i\xi+\int_t^T f^{s,0,i}(r)dr-\int_t^T Z^{s,0,i}_r dW_r-\int_t^T \int_E \psi^{s,0,i}_r(x)\wt{\mu}(dr,dx)
\label{eq-L-W-direction}
\eea
for $0\leq s\leq t\leq T$, where 
\bea
f^{s,0,i}(r)&:=&(D_{s,0}^if)(r,\Theta_r)+\part_\Theta f(r,\Theta_r)\Theta_r^{s,0,i} \nn \\
&=&(D_{s,0}^if)(r,\Theta_r)+\part_y f(r,\Theta_r)Y^{s,0,i}_r+\part_z f(r,\Theta_r)Z^{s,0,i}_r\nn \\
&&+\part_u f(r,\Theta_r) \int_{\mbb{R}_0}\rho(x)\part_v G(r,\psi_r(x))\psi_r^{s,0,i}(x)\nu(dx)\nn~.
\eea
(b)~For each jump direction $i\in\{1,\cdots,k\}$ and $m^i(dz)ds$-a.e. $(s,z)\in[0,T]\times \mbb{R}_0$,
there exists a unique solution $(Y^{s,z,i},Z^{s,z,i},\psi^{s,z,i})\in \calk^2[0,T]$ to the BSDE
\bea
Y^{s,z,i}_t=D_{s,z}^i\xi+\int_t^T f^{s,z,i}(r)dr-\int_t^T Z^{s,z,i}_r dW_r-\int_t^T \int_E \psi^{s,z,i}_r(x)\wt{\mu}(dr,dx)
\label{eq-L-J-direction}
\eea 
for $0\leq s\leq t\leq T$ and $z\neq 0$, where 
\bea
f^{s,z,i}(r)&:=&\frac{1}{z}\Bigl(f(\omega^{s,z},r, \Theta_r+z\Theta_r^{s,z,i})-f(\omega,r,\Theta_r)\Bigr)\nn \\
&=&\frac{1}{z}\Bigl\{ f\Bigl(\omega^{s,z},r,Y_r+z Y_r^{s,z,i}, Z_r+z Z_r^{s,z,i} \nn \\
&&\qquad ,\int_{\mbb{R}_0}\rho(x)G\bigl(r,\psi_r(x)+z\psi^{s,z,i}_r(x)\bigr)\nu(dx)\Bigr)-f(\omega,r,\Theta_r)\Bigr)\Bigr\}\nn.
\eea
(c)~Solution of the BSDE (\ref{eq-Lipschitz-BSDE}) is Malliavin differentiable $(Y,Z,\overline{\psi})\in \mbb{L}^{1,2}\times \mbb{L}^{1,2}\times \mbb{L}^{1,2}$. Put, for every $i$, $Y^{s,\cdot,i}_t=Z^{s,\cdot,i}_t=\psi^{s,\cdot,i}_t(\cdot)\equiv 0$ for $t<s\leq T$, then $\bigl((Y^{s,z,i}_t, Z^{s,z,i}_t, \psi^{s,z,i}_t(x)), 0\leq s,t\leq T, x\in\mbb{R}_0, z\in\mbb{R}\bigr)$
is a version of the Malliavin derivative $\bigl((D_{s,z}^iY_t, D_{s,z}^iZ_t, D_{s,z}^i\psi_t(x)), 0\leq s,t\leq T, x\in\mbb{R}_0, z\in\mbb{R}\bigr)$ for every Wiener and jump direction.
\begin{proof}

%%%%%%%%%%%%%%%%%%%%%%%%%%%%%%%%%%%%%%%%
For notational simplicity, we omit $i$ denoting the direction of derivative
by assuming that we consider each direction separately.
\\\\
{\it{Proof for (a) and (b)}}\\
It is easy to see that both of the BSDEs (\ref{eq-L-W-direction}) and 
(\ref{eq-L-J-direction}) satisfy the standard global Lipschitz conditions.
We have $|f^{s,0}(r)|\leq |(D_{s,0}f)(r,0)|+K_{s,0}(r)|\Theta_r|+K|\Theta^{s,0}_r|$.
Since
\bea
f^{s,z}(r)&=&\frac{f(\omega^{s,z},r,\Theta_r)-f(\omega,r,\Theta_r)}{z}
+\frac{f(\omega^{s,z},r,\Theta_r+z\Theta_r^{s,z})-f(\omega^{s,z},r,\Theta_r)}{z} \nn \\
&=&(D^{s,z}f)(r,\Theta_r)+\frac{f(\omega^{s,z},r,\Theta_r+z\Theta_r^{s,z})-f(\omega^{s,z},r,\Theta_r)}{z}, \nn
\eea
we also have $|f^{s,z}(r)|\leq |(D_{s,z}f)(r,0)|+K_{s,z}(r)|\Theta_r|+K|\Theta_r^{s,z}|~$ for $z\in\mbb{R}_0$.
Thus,  Lemma B.2 in \cite{FT-BSDEJ} tells us that  for all $(s,z)\in[0,T]\times \mbb{R}$ (thus including $\Theta^{s,0}$)
there exists a unique solution $\Theta^{s,z}\in \calk^2[0,T]$ satisfying 
\bea
&&||(Y^{s,z},Z^{s,z},\psi^{s,z})||^2_{\calk^2[0,T]}\leq
C_{K,T}\mbb{E}\Bigl[|D_{s,z}\xi|^2+\Bigl(\int_0^T \Bigl[|(D_{s,z}f)(r,0)|+K_{s,z}(r)|\Theta_r|\Bigr]dr\Bigr)^2\Bigr]\nn \\
&&\leq C_{K,T} \mbb{E}\Bigl[|D_{s,z}\xi|^2+\Bigl(\int_0^T |(D_{s,z}f)(r,0)|dr\Bigr)^2
+||K_{s,z}||^4_T +\Bigl(\int_0^T |\Theta_r|^2dr\Bigr)^2 \Bigr]<\infty.\nn
\eea
Note here that $\Theta \in \calk^4[0,T]$.  By Assumption~\ref{assumption-M-Lipschitz} (ii), (iii) and (iv), it also follows that
\be
\int_{\wt{E}}||(Y^{s,z},Z^{s,z},\psi^{s,z})||^2_{\calk^2[0,T]}q(ds,dz)<\infty~.\nn
\ee
{\it{Proof for (c)}}\\
%%%%%%%%%%%%%%%%%%%%%%%%%%%%%%%%%%%
We consider a sequence of solution $(Y^n,Z^n,\psi^n)_{n\geq 1}$ of the following
BSDEs that converges to $(Y,Z,\psi)$ of (\ref{eq-Lipschitz-BSDE}) in $\calk^4[0,T]$;
\bea
Y_t^{n+1}=\xi+\int_t^T f^n(r)-\int_t^T Z_r^{n+1}dW_r-\int_t^T \int_E \psi_r^{n+1}(x)\wt{\mu}(dr,dx), 
\label{eq-Piccard}
\eea
for $t\in[0,T]$ and $n\in\mbb{N}$, where
$f^n(r):=f\Bigl(r,Y_r^n,Z_r^n,\int_{\mbb{R}_0}\rho(x)G(r,\psi_r^n(x))\nu(dx)\Bigr)$.
The convergence can be proven by
the standard arguments of contraction mapping for the Lipschitz BSDEs. See, for example,
Lemma B.2 in \cite{FT-BSDEJ} and its proof. 
\\\\
%%%%%%%%%%%%%%%%%%%%%%%%%%%%%%%%%%%%%%%%%%%%%%%%%%%%%%%%%%%%%%%%%%%%%%%%%%%%%%%%%%%%
{\bf{~~[First step: Showing $(Y^{n+1},Z^{n+1},\overline{\psi}^{n+1})\in (\mbb{L}^{1,2})^3$]}}\\
%%%%%%%%%%%%%%%%%%%%%%%%%%%%%%%%%%%%%%%%%%%%%%%%%%%%%%%%%%%%%%%%%%%%%%%%%%%%%%%%%%%
We first suppose that $(Y^n,Z^n,\overline{\psi}^n)\in (\mbb{L}^{1,2})^{3}$
and are going to prove that $(Y^{n+1},Z^{n+1},\overline{\psi}^{n+1})\in (\mbb{L}^{1,2})^{3}$.
Then, we can inductively show  $(Y^n,Z^n,\overline{\psi}^n)\in (\mbb{L}^{1,2})^3$ for every $n\in\mbb{N}$.
Firstly, the {\it chain} rules (Theorem 3.5 and Theorem 12.8 in \cite{Nunno} with the division by the jump size in the 
current convention) and Lemma 3.2 in \cite{Delong-Imkeller} show that
\bea
\int_{\mbb{R}_0}\rho(x)G(r,\psi_r^n(x))\nu(dx)dr\in \mbb{D}^{1,2}~. 
\label{eq-G-1}
\eea
In particular, this is because 
\bea
&&\int_{\wt{E}}\bigl|\bigl|D_{t,z}G(\cdot,\psi_\cdot^n)\bigr|\bigr|^2_{\mbb{J}^2[0,T]}q(dt,dz)\leq K^2\int_{\wt{E}}\bigl|\bigl|D_{t,z} \psi_\cdot^n\bigr|\bigr|^2_{\mbb{J}^2[0,T]}q(dt,dz)<\infty\nn,
\eea
where we have used the bounded derivative and the Lipschitz condition for $G$ and the 
assumption that $\overline{\psi}^n\in\mbb{L}^{1,2}$.
This also shows that $G(\cdot,\psi^n_\cdot)\in \mbb{L}^{1,2}$.

By (\ref{eq-G-1}) and by the  general chain rule for random functions (See, Theorem 3.12~\cite{Geiss}
for Wiener directions and Proposition 5.5~\cite{Sole} for jump directions in a canonical Levy space, respectively),
we see $f^n(r)=f(r,\Theta^n_r)\in\mbb{D}^{1,2}$ for every $r\in[0,T]$.
It is easy to check $||f^n(\cdot)||_{\mbb{H}^2[0,T]}^2<\infty$. 
Next, Assumption~\ref{assumption-M-Lipschitz}, the hypothesis
$(Y^n,Z^n,\overline{\psi}^n) \in \calk^4[0,T]\cap (\mbb{L}^{1,2})^{3}$ and the estimate $|D_{s,z}f^n(r)|\leq |(D_{s,z}f)(r,0)|+K_{s,z}(r)|\Theta_r^n|+K|D_{s,z}\Theta_r^n|$ imply
\bea
&&\hspace{-3mm}\int_{\wt{E}}||D_{t,z}f^n(\cdot)||^2_{\mbb{H}^2[0,T]}q(dt,dz)\nn \\
%&&\leq C_K \int_{\wt{E}}\mbb{E}\Bigl[\int_0^T \Bigl( |(D_{t,z}f)(r,0)|^2+|K_{t,z}(r)|^2|\Theta^n_r|^2+|D_{t,z}\Theta_r^n|^2\Bigr)dr
%\Bigr]q(dt,dz)\nn\\
&&\leq C_K\int_{\wt{E}}\mbb{E}\Bigl[\int_0^T\bigl(
|(D_{t,z}f)(r,0)|^2+|D_{t,z}\Theta^n_r|^2\bigr)dr+||K_{t,z}||^4_T+
\bigl(\int_0^T |\Theta^n_r|^2 dr\bigr)^2\Bigr]q(dt,dz) <\infty \nn
\eea
with some positive constant $C_K$.
Thus, Lemma 3.2 \cite{Delong-Imkeller} shows that $\int_t^T f^n(r)dr\in \mbb{D}^{1,2}$ for every $t\in[0,T]$.
As a result, we have 
$\xi+\int_t^T f^n(r)\in \mbb{D}^{1,2}$ for each $t\in[0,T]$.
Thus, by Lemma 3.1 \cite{Delong-Imkeller}, we conclude that $Y^{n+1}_t=\mbb{E}
\left[\xi+\int_t^T f^n(r)\Bigr|\calf_t\right]\in \mbb{D}^{1,2}$,
which then implies
\bea
\int_t^T Z_r^{n+1}dW_r+\int_t^T \int_E \psi^{n+1}_r(x)\wt{\mu}(dr,dx)=-Y_t^{n+1}+\xi+\int_t^T f^n(r)dr\in \mbb{D}^{1,2}~, \nn
\eea
which, together with Lemma 3.3~\cite{Delong-Imkeller}, shows $Z^{n+1},\overline{\psi}^{n+1} \in \mbb{L}^{1,2}$.

We are now going to prove $Y^{n+1}\in \mbb{L}^{1,2}$.
For a Wiener $(z=0)$ as well as a jump $(z\neq 0)$ direction, we have, 
\bea
&&D_{s,z}Y_t^{n+1}=D_{s,z}\xi+\int_t^T D_{s,z}f^n(r)dr-\int_t^T D_{s,z}Z_r^{n+1}dW_r
-\int_t^T \int_E D_{s,z}\psi_r^{n+1}(x)\wt{\mu}(dr,dx),\nn \\
&& \text{for} ~0\leq s\leq t\leq T \quad \text{and} \quad z\in\mbb{R}^k, \nn
\eea
by Lemma 3.3~\cite{Delong-Imkeller}.
By Lemmas B.2 in \cite{FT-BSDEJ}, one obtains
\bea
&&\int_{\wt{E}}||D_{s,z}Y^{n+1}||^2_{\mbb{S}^2[0,T]}q(ds,dz) \leq C_{K,T}\int_{\wt{E}}\mbb{E}\Bigl[|D_{s,z}\xi|^2
+\Bigl(\int_0^T
|D_{s,z}f^n(r)|dr\Bigr)^2\Bigr]q(ds,dz)\nn \\
&&\leq C_{K,T}\int_{\wt{E}}\mbb{E}\Bigl[
|D_{s,z}\xi|^2+\Bigl(\int_0^T|(D_{s,z}f)(r,0)|+|D_{s,z}\Theta^n_r|dr\Bigr)^2 \nn \\
&&\hspace{20mm}+||K_{s,z}||^4_T+\Bigl(\int_0^T |\Theta^n_r|^2dr\Bigr)^2\Bigr]q(ds,dz)<\infty~, 
\label{eq-Dsz-Yn}
\eea
where $D_{s,z}Y^{n+1}_t\equiv 0$ for $t<s$ is used. Hence $(Y^{n+1},Z^{n+1},\overline{\psi}^{n+1})\in(\mbb{L}^{1,2})^3$ is proved.
\\\\
%%%%%%%%%%%%%%%%%%%%%%%%%%%%%%%%%%%%%%%%%%%%%%%%%%
{\bf{~~[Second step: convergence of $D_{s,0}\Theta^{n}\rightarrow \Theta^{s,0}$]}}\\
%%%%%%%%%%%%%%%%%%%%%%%%%%%%%%%%%%%%%%%%%%%%%%%%%%
Let us set the difference process as follows:
\bea
\Del^{s,0} Y^{n}:=Y^{s,0}-D_{s,0}Y^{n}, \quad \Del^{s,0} Z^{n}:=Z^{s,0}-D_{s,0}Z^{n}, \quad \Del^{s,0} \psi^{n}:=\psi^{s,0}-D_{s,0}\psi^{n}.\nn
\eea
and denote $\Del^{s,0} \Theta^n:=(\Del^{s,0} Y^n,\Del^{s,0} Z^n,\Del^{s,0}\psi^n)$ for every $n\in\mbb{N}$. 
We claim 
\be
\lim_{n\rightarrow \infty} \int_0^T ||(\Del^{s,0}\Theta^n)||^2_{\calk^2[0,T]}ds=0~.
\label{eq-2nd-claim}
\ee
Since 
$|f^{s,0}(r)-D_{s,0}f^n(r)|\leq K_{s,0}(r)|\Theta_r-\Theta_r^n|+|\part_\Theta f(r,\Theta_r)-\part_\Theta f(r,\Theta_r^n)|
|\Theta_r^{s,0}|+K|\Del^{s,0}\Theta^n_r|$,
the a priori estimate given in Lemma B.2~\cite{FT-BSDEJ} gives
\bea
&&\hspace{-3mm}\int_0^T\bigl|\bigl|(\Del^{s,0} Y^{n+1},\Del^{s,0} Z^{n+1},\Del^{s,0}\psi^{n+1})||^2_{\calk^2[0,T]}ds
\leq C_T \int_0^T \mbb{E}\Bigl[\Bigl(\int_0^T |f^{s,0}(r)-D_{s,0}f^n(r)|dr\Bigr)^2\Bigr]ds\nn\\
&&\leq C_T\int_0^T \mbb{E}\Bigl[\Bigl(\int_0^T\Bigr[
K_{s,0}(r)|\Theta_r-\Theta^n_r|+|\part_\Theta f(r,\Theta_r)-\part_\Theta f(r,\Theta_r^n)||\Theta_r^{s,0}|\Bigr]dr\Bigr)^2\Bigr]ds\nn \\
&&\quad+C_{T,K}\int_0^T\mbb{E}\Bigl[\Bigl(\int_0^T |\Del^{s,0} \Theta_r^n|dr\Bigr)^2\Bigr]ds\nn~.
\eea
One sees that the first line converges to zero because $\Theta^n\rightarrow \Theta \in \calk^4[0,T]$.
Thus, by using a sequence of small positive constants $(\ep_n)_{n\geq 1}$ converging to zero, one can write
\bea
&&\int_0^T\bigl|\bigl|(\Del^{s,0} Y^{n+1},\Del^{s,0} Z^{n+1},\Del^{s,0} \psi^{n+1})||^2_{\calk^2[0,T]}ds
\leq \ep_n+C_{T,K}\int_0^T \mbb{E}\left[\Bigl(\int_0^T |\Del^{s,0} \Theta_r^n|dr\Bigr)^2\right]ds\nn~ \nn \\
&&\leq \ep_n+C_{T,K}^\prime \max(T^2,T)\int_0^T\bigl|\bigl|(\Del^{s,0} Y^n,\Del^{s,0} Z^n,\Del^{s,0} \psi^n)\bigr|\bigr|^2_{\calk^2[0,T]}ds.\nn
\eea

For a sufficiently {\bf{small}} $T(>0)$ so that
$\alpha:=C^\prime_{T,K}\max(T^2,T)<1$,
one obtains
$\int_0^T ||(\Del^{s,0}\Theta^{n+1})||^2_{\calk^2[0,T]}ds\leq \ep_n+\alpha\int_0^T||(\Del^{s,0}\Theta^{n})||^2_{\calk^2[0,T]}ds$.
Then, by fixing some $n_0\in\mbb{N}$, 
\bea
&&\int_0^T ||(\Del^{s,0}\Theta^{n+n_0})||^2_{\calk^2[0,T]}ds\leq \frac{\ep_{n_0}}{1-\alpha}+\alpha^n\int_0^T ||(\Del^{s,0}\Theta^{n_0})||^2_{\calk^2[0,T]}ds.\nn
\eea
Thus, by passing $n$ and then $n_0$ to $\infty$, (\ref{eq-2nd-claim}) is proved for small $T$.

For {\bf{general}} $T>0$, one can use a time partition $0=T_0<T_1<\cdots<T_N=T$ that is
fine enough so that $\alpha<1$ in every time interval.
Due to the uniqueness of the solution, by setting $Y^{s,0}_{T_i}$ as the terminal condition for the interval $[T_{i-1},T_i]$,
one can prove (\ref{eq-2nd-claim}) for the interval. Repeating the procedures from $i=N$ to $i=1$ proves the claim.
\\\\
%%%%%%%%%%%%%%%%%%%%%%%%%%%%%%%%%%%%%%%%%%%
{\bf{~~[Third step: convergence of $D_{s,z}\Theta^{n}\rightarrow \Theta^{s,z}$ $(z\neq 0)$]}}\\
%%%%%%%%%%%%%%%%%%%%%%%%%%%%%%%%%%%%%%%%%%%
Choosing one direction of jump (omit $i$ for simplicity) and put
\bea
\Del^{s,z} Y^{n}:=Y^{s,z}-D_{s,z}Y^{n}, \quad \Del^{s,z} Z^{n}:=Z^{s,z}-D_{s,z}Z^{n}, \quad \Del^{s,z} \psi^{n}:=\psi^{s,z}-D_{s,z}\psi^{n}.\nn
\eea
and denote $\Del^{s,z} \Theta^n:=(\Del^{s,z} Y^n,\Del^{s,z} Z^n,\Del^{s,z}\psi^n)$ for every $n\in\mbb{N}$. 
In this step, our final goal is to show the convergence
\be
\lim_{n\rightarrow \infty} \int_0^T\int_{\mbb{R}_0}||(\Del^{s,z}\Theta^{n})||^2_{\calk^2[0,T]}m(dz)ds=0~.
\label{eq-3rd-claim}
\ee

Before discussing  (\ref{eq-3rd-claim}), we have to prove first that the convergence 
\bea
\hspace{-5mm}\lim_{\ep\downarrow 0}\int_0^T\int_{|z|>\ep}\bigl|\bigl|(\Del^{s,z}\Theta^{n+1})\bigr|\bigr|^2_{\calk^2[0,T]}m(dz)ds
=\int_0^T\int_{\mbb{R}_0}\bigl|\bigl|(\Del^{s,z}\Theta^{n+1})\bigr|\bigr|^2_{\calk^2[0,T]}m(dz)ds
\label{eq-uniform-ep}
\eea
occurs {\it uniformly} in (sufficiently large) $n$.  As the proof of Theorem 4.1~\cite{Delong-Imkeller}, 
it suffices to show that, for each $\ep>0$,  there exists a positive constant $C$ and $\bar{\ep}>0$ independent of $n$ 
such that
\bea
\int_0^T\int_{|z|\leq \bar{\ep}}\bigl|\bigl|(\Del^{s,z}\Theta^{n+1})\bigr|\bigr|^2_{\calk^2[0,T]}m(dz)ds<C\ep~. \nn
\eea
By Remark~\ref{dominated-Lip},  for a given arbitrary $\ep>0$, there exists
$\bar{\ep}>0$ such that
\bea
\label{eq-ep-1}
&&\hspace{-7mm}\bullet~\int_0^T\int_{|z|\leq \bar{\ep}}\mbb{E}\Bigl[
|D_{s,z}\xi|^2+\Bigl(\int_0^T |(D_{s,z}f)(r,0)|dr\Bigr)^2+||K_{s,z}||^4_T\Bigr]m(dz)ds<\ep \\
&&\hspace{-7mm}\bullet~\int_0^T\int_{|z|\leq \bar{\ep}}m(dz)ds<\ep. 
\label{eq-ep-2}
\eea
Let us fix $\bar{\ep}>0$ as above. By Lemma B.2~\cite{FT-BSDEJ}, 
we have $\||(\Del^{s,z}\Theta^{n+1})||^2_{\calk^2[0,T]} \leq C_T
\mbb{E}\bigl[\bigl(\int_0^T |f^{s,z}(r)-D_{s,z}f^n(r)|dr\bigr)^2\bigr]$.
Using the (local) Lipschitz properties, it is easy to show that
\be
|f^{s,z}(r)-D_{s,z}f^n(r)|\leq K_{s,z}(r)|\Theta_r-\Theta_r^n|+K|\Theta_r^{s,z}|+K|D_{s,z}\Theta_r^{n}|\nn 
\ee
and hence
\bea
&&\int_0^T\int_{|z|\leq \bar{\ep}}\bigl|\bigl|(\Del^{s,z}\Theta^{n+1})\bigr|\bigr|^2_{\calk^2[0,T]}m(dz)ds \leq C_{T,K}\int_0^T \int_{|z|\leq \bar{\ep}}\mbb{E}\Bigl[\Bigl(\int_0^T K_{s,z}(r)|\Theta_r-\Theta_r^n|dr\Bigr)^2 \nn \\
&&\qquad +\Bigl(\int_0^T |\Theta^{s,z}_r|dr\Bigr)^2+\Bigl(\int_0^T |D_{s,z}\Theta_r^n|dr\Bigr)^2\Bigr]m(dz)ds~.
\label{eq-uniform}
\eea

We are now going to discuss each term of (\ref{eq-uniform}).
For the first term, it is straightforward to see that there exists $n$ independent constant $C$ such that
\bea
&&C_{T,K}\int_0^T \int_{|z|\leq \bar{\ep}}\mbb{E}\Bigl[\Bigl(\int_0^T K_{s,z}(r)|\Theta_r-\Theta_r^n|dr\Bigr)^2\Bigr]\nn \\
&&\leq C_{T,K}\int_0^T \int_{|z|\leq \bar{\ep}}\mbb{E}\Bigl[
||K_{s,z}||^4_T+\Bigl(\int_0^T |\Theta_r-\Theta_r^n|^2 dr\Bigr)^2\Bigr]m(dz)ds<C\ep\nn
\eea
where the last inequality follows from (\ref{eq-ep-1}), (\ref{eq-ep-2}) and the fact that 
$||\Theta-\Theta^n||^4_{\mbb{H}^4[0,T]}$ is bounded due to the convergence $\Theta^n\rightarrow \Theta$ in $\calk^4[0,T]$.
For the second term of (\ref{eq-uniform}), one can show
\bea
&&\hspace{-6mm}C_{T,K}\int_0^T\int_{|z|\leq \bar{\ep}}\mbb{E}\Bigl[\Bigl(\int_0^T |\Theta_r^{s,z}|dr\Bigr)^2\Bigr]m(dz)ds 
\leq C_{T,K} \int_0^T\int_{|z|\leq \bar{\ep}}||(\Theta^{s,z})||^2_{\calk^2[0,T]}m(dz)ds\nn\\
&&\hspace{-3mm}\leq C_{T,K}\int_0^T\int_{|z|\leq \bar{\ep}}\mbb{E}\Bigl[|D_{s,z}\xi|^2+\Bigl(\int_0^T|(D_{s,z}f)(r,0)|dr\Bigr)^2+||K_{s,z}||^4_T
+\Bigl(\int_0^T |\Theta_r|^2dr\Bigr)^2\Bigr]m(dz)ds\nn \\
&&\hspace{-3mm}<C\ep 
\label{eq-Ysz-secondterm}
\eea
where the last inequality follows from (\ref{eq-ep-1}), (\ref{eq-ep-2}) and the fact that $\Theta\in\calk^4[0,T]$.
Finally, the third term of (\ref{eq-uniform}) can be evaluated as
\bea
C_{T,K}\int_0^T\int_{|z|\leq \bar{\ep}}\mbb{E}\Bigl[\Bigl(\int_0^T |D_{s,z}\Theta_r^n|dr\Bigr)^2\Bigr]m(dz)ds \leq C_{T,K}\int_0^T\int_{|z|\leq \bar{\ep}}||(D_{s,z}\Theta^n)||^2_{\calk^2[0,T]}m(dz)ds~.\nn
\eea
Here, by the same a priori estimate used in (\ref{eq-Dsz-Yn}),
\bea
&&C_{T,K}||(D_{s,z}\Theta^n)||^2_{\calk^2[0,T]}\leq C_{K,T}\mbb{E}\left[
|D_{s,z}\xi|^2+\Bigl(\int_0^T |(D_{s,z}f)(r,0)|dr\Bigr)^2+||K_{s,z}||^4_T\right. \nn \\
&&\left.\quad+\Bigl(\int_0^T |\Theta^{n-1}_r|^2 dr\Bigr)^2\right]+C_{K,T}\mbb{E}\left[\Bigl(\int_0^T |D_{s,z}\Theta^{n-1}_r|dr\Bigr)^2\right]
\nn 
\eea
\bea
&&\leq C_{K,T}\mbb{E}\left[\ep_{n-1}+|D_{s,z}\xi|^2+\Bigl(\int_0^T|(D_{s,z}f)(r,0)|dr\Bigr)^2+
||K_{s,z}||^4_T+\Bigl(\int_0^T |\Theta_r|^2dr\Bigr)^2\right]\nn \\
&&+C_{K,T}\max(T^2,T)||(D_{s,z}\Theta^{n-1})||^2_{\calk^2[0,T]}~,
\label{eq-Dsz-Yn-2}
\eea
where  $(\ep_n)_{n\geq 1}$ is a sequence of positive constants with $\ep_n:=\bigl|||\Theta^{n}||^4_{\mbb{H}^4[0,T]}-||\Theta||^4_{\mbb{H}^4[0,T]}\bigr|$. 
It is bounded $(\sup_{n\in\mbb{N}}\bigl(\ep_n\bigr)\leq \del)$ with some $n$-independent constant $\del$ due to the convergence of $\Theta^n\rightarrow \Theta$ in $\calk^4[0,T]$.
Choosing the terminal time $T$ small enough so that $\alpha:=C_{K,T}\max(T^2,T)<1$, (\ref{eq-Dsz-Yn-2}) yields
\bea
&&C_{T,K}\int_0^T\int_{|z|\leq \bar{\ep}}||(D_{s,z}\Theta^{n})||^2_{\calk^2[0,T]}m(dz)ds\nn \\
&&\leq \frac{C_{K,T}}{1-\alpha}\int_0^T\int_{|z|\leq \bar{\ep}}\mbb{E}\left[\del+|D_{s,z}\xi|^2+\Bigl(\int_0^T|(D_{s,z}f)(r,0)|dr\Bigr)^2+
||K_{s,z}||^4_T\right.\nn \\
&&\left.+\Bigl(\int_0^T |\Theta_r|^2dr\Bigr)^2\right]m(dz)ds+\alpha^{n-1}\int_0^T\int_{|z|\leq \bar{\ep}}||(D_{s,z}\Theta^{1})||^2_{\calk^2[0,T]}
m(dz)ds.\nn
\eea
It is free to choose $\Theta^1 \equiv 0$ in the fixed point iteration (\ref{eq-Piccard}). 
Thus, the right hand side is dominated by $C\ep$ with some $n$ independent constant $C$
due to (\ref{eq-ep-1}) and (\ref{eq-ep-2}).

%%%%%%%%%%%%%%%%%%%%%
%\subsubsection*{Small terminal time $T$}
%%%%%%%%%%%%%%%%%%%%%
By the previous arguments, we have shown that the convergence of (\ref{eq-uniform-ep})
is uniform in $n$, at least for {\bf{sufficiently small}} $T$.
In this case, one can exchange the order of limit operations;
\bea
\lim_{n\rightarrow \infty}\lim_{\ep\downarrow 0} \int_0^T\int_{|z|>\ep}\bigl|\bigl|(\Del^{s,z}\Theta^{n+1})\bigr|\bigr|^2_{\calk^2}m(dz)ds\nn=\lim_{\ep\downarrow 0}\lim_{n\rightarrow \infty}\int_0^T\int_{|z|>\ep}\bigl|\bigl|(\Del^{s,z}\Theta^{n+1})\bigr|\bigr|^2_{\calk^2}m(dz)ds~.
\eea
Therefore, in order to show the convergence (\ref{eq-3rd-claim}),
it is enough to prove
\bea
\lim_{n\rightarrow \infty}\int_0^T\int_{|z|>\ep}\bigl|\bigl|(\Del^{s,z}\Theta^{n+1})\bigr|\bigr|^2_{\calk^2[0,T]}m(dz)ds=0\nn
\eea
for each $\ep>0$.
An inequality from the Lipschitz property of the driver
\bea
&&|f^{s,z}(r)-D_{s,z}f^n(r)|\leq \frac{1}{|z|}\bigl| f(\omega^{s,z},r,\Theta_r+z\Theta_r^{s,z})-f(\omega^{s,z},r,\Theta_r^n+zD_{s,z}\Theta_r^n)\bigr|\nn \\
&&\qquad +\frac{1}{|z|}\bigl|f(\omega,r,\Theta_r)-f(\omega,r,\Theta_r^n)\bigr|  \leq  \frac{2K}{|z|}|\Theta_r-\Theta_r^n|+K|\Del^{s,z}\Theta^n_r|\nn
\eea
implies
\bea
&&\int_0^T\int_{|z|>\ep}\bigl|\bigl|(\Del^{s,z}\Theta^{n+1})\bigr|\bigr|^2_{\calk^2[0,T]}m(dz)ds\nn \\
&&\leq C_{T,K}\int_0^T\int_{|z|>\ep}\mbb{E}\left[
\frac{1}{|z|^2}\Bigl(\int_0^T |\Theta_r-\Theta_r^n|dr\Bigr)^2+\Bigl(\int_0^T |\Del^{s,z}\Theta^n_r|dr\Bigr)^2\right]m(dz)ds\nn \\
&&\leq \ep_n+C_{T,K}\max(T^2,T)\int_0^T\int_{|z|>\ep}||(\Del^{s,z}\Theta^n)||^2_{\calk^2[0,T]}m(dz)ds\nn
\eea
where $\ep_n\rightarrow 0$ as $n\rightarrow 0$ due to the convergence of $\Theta^n\rightarrow \Theta$.
If necessary by re-choosing $T$ small enough so that $\alpha:=C_{T,K}\max(T^2,T)<1$,  one gets
\bea
\int_0^T\int_{|z|>\ep}\bigl|\bigl|(\Del^{s,z}\Theta^{n+n_0})\bigr|\bigr|^2_{\calk^2[0,T]}m(dz)ds\leq \frac{\ep_{n_0}}{1-\alpha}+\alpha^{n}\int_0^T\int_{|z|>\ep}\bigl|\bigl|(\Del^{s,z}\Theta^{n_0})\bigr|\bigr|^2_{\calk^2[0,T]}m(dz)ds\nn.
\eea
By passing to the limit $n,n_0\rightarrow \infty$, (\ref{eq-3rd-claim}) is proved for small $T$.

For {\bf{general}} $T>0$, one can construct a partition $0=T_0<T_1<\cdots<T_N=T$
fine enough so that one can conclude by the previous arguments
\bea
\lim_{n\rightarrow 0}\int_{T_{N-1}}^T\int_{|z|>\ep}\bigl|\bigl|(\Del^{s,z}\Theta^{n})\bigr|\bigr|^2_{\calk^2[0,T]}m(dz)ds=0~.\nn
\eea
Note that (\ref{eq-Ysz-secondterm}) implies 
$\lim_{\ep\downarrow 0} \int_0^T\int_{|z|<\bar{\ep}}\mbb{E}|Y^{s,z}_{T_{N-1}}|^2 m(dz)ds=0$, in particular.
Therefore, by the same procedures with a new terminal value $Y^{s,z}_{T_{N-1}}$ instead of $D_{s,z}\xi$,
the convergence (\ref{eq-3rd-claim}) in $[T_{N-2},T_{N-1}]$ is proved. Repeating the same arguments
proves (\ref{eq-3rd-claim}) for general $T$. Hence, one can conclude $(Y^n,Z^n,\overline{\psi}^n)$ converges to
$\bigl((Y,Z,\overline{\psi}),(Y^{s,z},Z^{s,z},\overline{\psi}^{s,z})\bigr)$ in $(\mbb{L}^{1,2})^3$.
Finally, thanks to the closability of the Malliavin derivatives in $\mbb{L}^{1,2}$
(See Theorem 12.6 in \cite{Nunno}.), one concludes $(Y,Z,\overline{\psi})\in\mbb{L}^{1,2}$
and that $(Y^{s,z},Z^{s,z},\psi^{s,z})$ is a version of $(D_{s,z}Y,D_{s,z}Z,D_{s,z}\psi)$.
\end{proof}
\end{theorem}

%%%%%%%%%%%%%%%%%%%%%%%%%%%%%%%%%%%%%%%%%%%%%%%%%%%%%%%%%%%%%%%%%%%%%%
%%%%%%%%%%%%%%%%%%%%%%%%%%%%%%%%%%%%%%%%%%%%%%%%%%%%%%%%%%%%%%%%%%%%%%
\section{Technical details omitted in the proof of Theorem~\ref{theorem-Qexp-MD}}
%%%%%%%%%%%%%%%%%%%%%%%%%
\subsection{Proof for (\ref{eq-Ds0Theta-conv})}
\label{app-QMD}
%%%%%%%%%%%%%%%%%%%%%%%%%%%%%
By (\ref{eq-Dsz-Ym-L12-bound}) and the dominated convergence theorem,  
it suffices to show 
\be
\lim_{m\rightarrow \infty} \bigl|\bigl|(\Del^{s,0}Y^m, \Del^{s,0}Z^m, \Del^{s,0}\psi^m)\bigr|\bigr|^p_{\calk^p[0,T]}=0
\nn
\ee 
for $ds$-a.e. $s\in[0,T]$.
Since
\bea
&&\bullet~f^{s,0}(r)-D_{s,0}f_m(r)=f^{s,0}(r)-\bigl((D_{s,0}f_m)(r,\Theta_r^m)+\part_\Theta f_m(r,\Theta_r^m)\Theta_r^{s,0}\bigr)\nn \\
&&\hspace{40mm} +\part_\Theta f_m(r,\Theta_r^m)(\Theta_r^{s,0}-D_{s,0}\Theta_r^m)~, \nn 
\eea
and
\bea
&&\bullet~\Bigl|f^{s,0}(r)-\bigl((D_{s,0}f_m)(r,\Theta_r^m)+\part_\Theta f_m(r,\Theta_r^m)\Theta_r^{s,0}\bigr)\Bigr|
\leq |(D_{s,0}f)(r,\Theta_r)-(D_{s,0}f)(r,\Theta_r^m)|\nn \\
&&\qquad +|(D_{s,0}f)(r,\Theta_r^m)-(D_{s,0}f_m)(r,\Theta_r^m)|+|\part_\Theta f(r,\Theta_r)-\part_\Theta f_m(r,\Theta_r^m)||\Theta_r^{s,0}|~, \nn
\eea
Lemma~\ref{lemma-BMO-stability} implies that
{\small
\bea
&&\bigl|\bigl| (\Del^{s,0}Y^m, \Del^{s,0}Z^m, \Del^{s,0}\psi^m)\bigr|\bigr|^p_{\calk^p[0,T]}ds \leq C \mbb{E}\Bigl[ 
\Bigl(\int_0^T|(D_{s,0}f)(r,\Theta_r)-(D_{s,0}f)(r,\Theta^m_r)|dr\Bigr)^{p\bar{q}^2}\nn \\
&&\hspace{-5mm}+\Bigl(\int_0^T |(D_{s,0}f)(r,\Theta_r^m)-(D_{s,0}f_m)(r,\Theta_r^m)|dr\Bigr)^{p\bar{q}^2}+\Bigl(\int_0^T |\part_\Theta f(r,\Theta_r)-\part_\Theta f_m(r,\Theta^m_r)||\Theta_r^{s,0}|dr\Bigr)^{p\bar{q}^2}
\Bigr]^{\frac{1}{\bar{q}^2}} \nn
%\label{eq-DelS0}
\eea}
where, as before, $C>0$ and $\bar{q}>1$ are constants independent of $m$.

Let us check each term. By the local Lipschitz property, the first term yields
\bea
&&\mbb{E}\Bigl[ \Bigl(\int_0^T|(D_{s,0}f)(r,\Theta_r)-(D_{s,0}f)(r,\Theta^m_r)|dr\Bigr)^{p\bar{q}^2} \Bigr] \nn \\
&&\quad\qquad \leq C\mbb{E}\Bigl[||K_{s,0}^M||^{2p\bar{q}^2}\Bigr]^{\frac{1}{2}}
\mbb{E}\Bigl[||\del Y^m||_T^{2p\bar{q}^2}+\Bigl(\int_0^T ||\del \psi^m_r||^2_{\L2nu}dr\Bigr)^{p\bar{q}^2}\Bigr]^{\frac{1}{2}}\nn \\
&&\quad \qquad +C\mbb{E}\Bigl[||K_{s,0}^M||^{2p\bar{q}^2}_T\Bigl(\int_0^T |H^m(r)|^2dr\Bigr)^{p\bar{q}^2}\Bigr]^{\frac{1}{2}}
\mbb{E}\Bigl[\Bigl(\int_0^T |\del Z_r^m|^2 dr\Bigr)^{p\bar{q}^2}\Bigr]^{\frac{1}{2}}, 
\label{eq-DelS0-1st}
\eea
where the process $H^m$ is defined by $H^m(r):=1+|Z_r|+|Z_r^m|+||\psi_r||_{\L2nu}+||\psi^m_r||_{\L2nu}$ 
and $(\del Y^m,\del Z^m,\del\psi^m):=(Y-Y^m,Z-Z^m,\psi-\psi^m)$. Since $H^m\in\mbb{H}^2_{BMO}$ with the norm dominated by constant independent of $m$,
the convergence of $\Theta^m\rightarrow \Theta$ in $\mbb{S}^\infty\times \mbb{H}^2_{BMO}\times \mbb{J}^2_{BMO}$
implies that (\ref{eq-DelS0-1st}) converges to zero as $m\rightarrow \infty$.

Secondly, by definition of the truncated driver,
$(D_{s,0}f_m)(r,\Theta_r^m)=(D_{s,0}f)\bigl(r,\varphi_m(\Theta^m)\bigr)$.
Since both $\Theta^m$ and $\varphi_m(\Theta^m)$ converge to $\Theta$ in $\mbb{S}^\infty\times \mbb{H}^2_{BMO}\times \mbb{J}^2_{BMO}$, 
the convergence of the second term
can be shown in the same way as the first term.

Finally, by the Cauchy-Schwartz inequality,
\bea
&&\mbb{E}\Bigl[\Bigl(\int_0^T |\part_\Theta f(r,\Theta_r)-\part_\Theta f_m(r,\Theta^m_r)||\Theta_r^{s,0}|dr\Bigr)^{p\bar{q}^2}\Bigr] \nn \\
&&\quad \leq \mbb{E}\Bigl[\Bigl(\int_0^T |\part_\Theta f(r,\Theta_r)-\part_\Theta f_m(r,\Theta^m_r)|^2dr\Bigr)^{p\bar{q}^2}\Bigr]^{\frac{1}{2}}
\mbb{E}\Bigl[\Bigl(\int_0^T |\Theta^{s,0}_r|^2dr\Bigr)^{p\bar{q}^2}\Bigr]^{\frac{1}{2}}.
\label{eq-w-appendix}
\eea
By (\ref{eq-deriv-estimate}), there exists a constant $C_M$ depends only on the universal bounds such that
$|\part_\Theta f_m (r,\Theta_r^m)|\leq C_M(1+|Z_r^m|+||\psi_r^m||_{\mbb{L}^2(\nu)})$. Since $Z^m\rightarrow Z$ (resp. $\psi^m\rightarrow \psi$) in $\mbb{H}^2_{BMO}$
(resp. $\mbb{J}^2_{BMO}$), the energy inequality of Lemma~\ref{lemma-energy} gives the convergence in $\mbb{H}^{p^\prime}$ (resp. $\mbb{J}^{p\prime}$)
with $\forall p^\prime\geq 2$. Thus, by extracting subsequence if necessary, one sees 
$\sup_m |Z^m|, \sup_m||\psi^m||_{\mbb{L}^2(\nu)}$ are in $\mbb{H}^{p^\prime}$ for any $p^\prime\geq 2$
from Lemma 2.5 of \cite{Kobylanski}.
Since $\part_\Theta f_m(r,\Theta_r^m)\rightarrow \part_\Theta f(r,\Theta_r)$ $dt\otimes d\mbb{P}$-a.e.,
the dominated convergence shows the RHS of (\ref{eq-w-appendix}) tends to 0 as $m\rightarrow \infty$.
One can confirm the convergence actually occurs in the entire sequence, since otherwise there exists a
subsequence $(m_j)$ such that the RHS must be bounded from below by some positive constant.
However, one can once again choose a further subsequence from $(m_j)$ 
so that the RHS converges to zero by the dominated convergence as the last discussion, which is a contradiction.
This proves (\ref{eq-Ds0Theta-conv}).

%%%%%%%%%%%%%%%%%%%%%%%%%%%%%%%%%%%%%%%%%%%%%%%%%%
\subsection{Proof for (\ref{eq-DszTheta-conv})}
\label{app-QMD2}
%%%%%%%%%%%%%%%%%%%%%%%%%%%%%%%%%%%%%%%%%%%%%%%%%%
Let us define a $d$-dimensional $\mbb{F}$-progressively measurable process $(b_{s,z}^m(r),r\in[0,T])$ by
\bea
b^m_{s,z}(\omega, r)&:=&\frac{f_m(\omega^{s,z},r,\check{\Xi}^m_{s,z}(r))-f_m(\omega^{s,z},r,\Xi^m_{s,z}(r))}
{z|\Del^{s,z}Z^m_r|^2}\bold{1}_{\Del^{s,z}Z^m_r\neq 0} \Del^{s,z}Z_r^m\nn 
\eea
where $\check{\Xi}^m_{s,z}:=(\caly^m_{s,z},Z^m+zZ^{s,z},\int_{\mbb{R}_0}\rho(x)G_m(r,\Psi^m_{s,z}(\cdot,x))\nu(dx))$ and \\
$\Xi^m_{s,z}:=(\caly^m_{s,z},\calz^{m}_{s,z},\int_{\mbb{R}_0}\rho(x)G_m(r,\Psi^m_{s,z}(\cdot,x))\nu(dx))$.
Noticing the fact $\calz^{s,z}=Z+zZ^{s,z}$, one sees $(\caly^m_{s,z},Z^m+zZ^{s,z},\Psi^m_{s,z})\rightarrow (\caly^{s,z},\calz^{s,z},\Psi^{s,z})$ in $\mbb{S}^\infty\times \mbb{H}^2_{BMO}\times \mbb{J}^2_{BMO}$. 
Let us also introduce a map $\wt{f}^m_{s,z}:\Omega\times[0,T]\times \mbb{R}\times \mbb{L}^2(E,\nu;\mbb{R}^k)\rightarrow \mbb{R}$ by
\bea
\wt{f}^m_{s,z}(\omega,r,\wt{y},\wt{\psi})&:=&(D_{s,z}f)(r,\Theta_r)-(D_{s,z}f_m)(r,\Theta_r^m)
-\frac{1}{z}\bigl[f(\omega^{s,z},r,\Theta_r)-f_m(\omega^{s,z},r,\Theta_r^m)\bigr]\nn \\
&&\hspace{-10mm} +\frac{1}{z}\Bigl\{f\Bigr(\omega^{s,z},r,z\wt{y}+\caly^{m}_{s,z}(r)+\del Y^m_r,\calz^{s,z}_r \nn \\
&&\hspace{-10mm} ,\int_{\mbb{R}_0}\rho(x)G(r,z\wt{\psi}(x)+\Psi^{m}_{s,z}(r,x)+\del\psi^m_r(x))\nu(dx)\Bigr)  -f_m(\omega^{s,z},r,\check{\Xi}^m_{s,z}(r))\Bigr\}~.\nn
\eea
Then, $(\Del^{s,z}Y^m, \Del^{s,z}Z^m,\Del^{s,z}\psi^m)$ is the solution to the BSDE
\bea
&&\Del^{s,z}Y^m_t=\int_t^T\Bigl(\wt{f}^m_{s,z}(r,\Del^{s,z}Y_r^m,\Del^{s,z}\psi_r^m)+
b_{s,z}^m(r)\cdot \Del^{s,z}Z_r^m\Bigr)dr\nn \\
&&\qquad -\int_t^T \Del^{s,z}Z_r^m dW_r-\int_t^T \int_E \Del^{s,z}\psi_r^m(x)\wt{\mu}(dr,dx).\nn
\eea
By denoting an $\mbb{F}$-progressively measurable process $H^m_{s,z}$ as
\bea
H^m_{s,z}(r):=K_M\Bigl(1+|\calz_{s,z}^m(r)|+|\calz^{s,z}_r|+|\del Z^m|+2||\rho||_{\L2nu}G_M^\prime
||\Psi^m_{s,z}(r,\cdot)||_{\L2nu}\Bigr), \nn
\eea
one obtains $|b^m_{s,z}(r)|\leq H^m_{s,z}(r)$ for $\forall r\in[0,T]$.
Here, $H^m_{s,z}\in\mbb{H}^2_{BMO}$ and for $m(dz)ds$-a.e. $(s,z)\in[0,T]\times \mbb{R}_0$,
its norm $||H^m_{s,z}||_{\mbb{H}^2_{BMO}}$ is bounded by some $m$-independent constant thanks to the universal bounds.
Furthermore, the new driver satisfies the linear growth property
$\displaystyle~|\wt{f}^m_{s,z}(r,\wt{y},\wt{\psi})|\leq |\wt{f}^m_{s,z}(r,0,0)|+K_M\bigl(|\wt{y}|+||\rho||_{\L2nu}G_M^\prime 
||\wt{\psi}||_{\L2nu}\bigr)$
and 
\bea
&&\hspace{-5mm}|\wt{f}^m(s,z)(r,0,0)|\leq |(D_{s,z}f)(r,\Theta_r)-(D_{s,z}f_m)(r,\Theta_r^m)|+\frac{1}{|z|}|f(\omega^{s,z},r,\Theta_r)
-f_m(\omega^{s,z},r,\Theta_r^m)| \nn \\
&&+\frac{1}{|z|}|f(\omega^{s,z},r,\check{\Xi}^m_{s,z}(r))-f_m(\omega^{s,z},r,\check{\Xi}^m_{s,z}(r))|+C K_M \frac{1}{|z|}\Bigl(|\del Y^m_r|+||\del \psi^m_r||_{\L2nu}+\calh^m_{s,z}(r)|\del Z^m_r|\Bigr) \nn
\eea
where $C$ is a positive constant depending only on $||\rho||_{\L2nu}, G_M^\prime$ and
\bea
\calh^m_{s,z}(r):=1+2|\calz^{s,z}_r|+|\del Z_r^m|+2||\Psi^m_{s,z}(r,\cdot)||_{\L2nu}+||\del \psi^m_r||_{\L2nu}~.\nn
\eea
$\calh^m_{s,z}\in \mbb{H}^2_{BMO}$ and its norm is bounded by some $m$-independent constant $m(dz)ds$-a.e. $(s,z)\in[0,T]\times \mbb{R}_0$.
By applying Lemma~\ref{lemma-bmolike-apriori}, one obtains
\bea
&&\bigl|\bigl|(\Del^{s,z}Y^m,\Del^{s,z}Z^m,\Del^{s,z}\psi^m)\bigr|\bigr|^p_{\calk^p[0,T]}\nn \\
&&\qquad \leq  C\mbb{E}\Bigl[ \Bigl(\int_0^T |(D_{s,z}f)(r,\Theta_r)-(D_{s,z}f_m)(r,\Theta_r^m)|dr\Bigr)^{p\bar{q}^2}
\Bigr]^{\frac{1}{\bar{q}^2}}\nn \\
&&\qquad +\frac{C}{|z|^p}\mbb{E}\Bigl[\Bigl(\int_0^T |f(\omega^{s,z},r,\Theta_r)-f_m(\omega^{s,z},r,\Theta_r^m)|dr\Bigr)^{p\bar{q}^2}
\Bigr]^{\frac{1}{\bar{q}^2}} \nn \\
&&\qquad+\frac{C}{|z|^p}\mbb{E}\Bigl[\Bigl(\int_0^T |f(\omega^{s,z},r,\check{\Xi}^m_{s,z}(r))-f_m(\omega^{s,z},r,\check{\Xi}^m_{s,z}(r))|dr
\Bigr)^{p\bar{q}^2}\Bigr]^{\frac{1}{\bar{q}^2}}\nn \\
&&\qquad+\frac{C}{|z|^p}\mbb{E}\Bigl[\Bigl(\int_0^T \bigl[ |\del Y^m_r|
+||\del \psi^m_r||_{\L2nu}+\calh^m_{s,z}(r)|\del Z_r^m|\bigr]dr\Bigr)^{p\bar{q}^2}\Bigr]^{\frac{1}{\bar{q}^2}},
\label{eq-Delsz-estimate}
\eea
where the positive constants $C$ and $\bar{q}>1$ are $m$-independent as before.

Due to (\ref{eq-Dsz-Ym-L12-bound}) and (\ref{eq-Dsz-ep-conv}), the convergence in $\lim_{\ep\downarrow 0}$
is uniform in $m$ and hence the order of limit operations can be exchanged. By the dominated convergence from (\ref{eq-Dsz-Ym-L12-bound}),
\bea
&&\lim_{m\rightarrow \infty}\lim_{\ep\downarrow 0}\int_0^T\int_{|z|>\ep}\bigl|\bigl| (\Del^{s,z} Y^m,\Del^{s,z} Z^m, \Del^{s,z} \psi^m)\bigr|\bigr|^p_{\calk^p[0,T]}m(dz)ds\nn\\
%&&\qquad =\lim_{\ep\downarrow 0}\lim_{m\rightarrow \infty} \int_0^T \int_{|z|>\ep}\bigl|\bigl| (\Del^{s,z} Y^m,\Del^{s,z} Z^m, \Del^{s,z} %\psi^m)\bigr|\bigr|^p_{\calk^p[0,T]}m(dz)ds \nn \\
&&\qquad =\lim_{\ep\downarrow 0}\int_0^T \int_{|z|>\ep}
\lim_{m\rightarrow \infty}\bigl|\bigl| (\Del^{s,z} Y^m,\Del^{s,z} Z^m, \Del^{s,z} \psi^m)\bigr|\bigr|^p_{\calk^p[0,T]}m(dz)ds.\nn
\eea
Therefore, in order to prove the convergence (\ref{eq-DszTheta-conv})
it suffices to show, for $m(dz)ds$-a.e. $(s,z)\in[0,T]\times \mbb{R}_0$, 
$\lim_{m\rightarrow \infty}\bigl|\bigl| (\Del^{s,z} Y^m,\Del^{s,z} Z^m, \Del^{s,z} \psi^m)\bigr|\bigr|^p_{\calk^p[0,T]}=0$.
This can be easily confirmed from (\ref{eq-Delsz-estimate}) by using the local Lipschitz continuity and the fact that
$\Theta^m$ and $\varphi_m(\Theta^m)\rightarrow \Theta$ and $\check{\Xi}^m_{s,z}$ and $\varphi_m(\check{\Xi}^m_{s,z})\rightarrow 
\Xi^{s,z}$ converge in $\mbb{S}^\infty\times \mbb{H}^2_{BMO}\times \mbb{J}^2_{BMO}$.
This finishes the proof for (\ref{eq-DszTheta-conv})
\end{appendix}
\small
%%%%%%%%%%%%%%%%%%%%%%%%%%%%%%%%%%%%%%%%%
\section*{Acknowledgement}
We thank the anonymous referees for valuable comments which greatly improve the presentation of the paper.
The research is partially supported by Center for Advanced Research in Finance (CARF) and
JSPS KAKENHI (Grant Number 25380389).

%%%%%%%%%%%%%%%%%%%%%%%%%%%%%%%%%%%%%%%%%%%%%%%%%%%%%%%%%%%%%%%%%%%%%

%%%%%%%%%%%%%%%%%%%%%%%%%%%%%%%%%%%%%%%%%%%%%%%%%%%%%%%%%%%%%%%%%%%%

\end{document}